\documentclass[a4paper, 10pt]{article}
\usepackage[latin1]{inputenc}

\pdfoutput=1

\usepackage{jcappub} % for details on the use of the package, please
\usepackage{amssymb, amsmath}
\usepackage[below]{placeins}
\usepackage{graphicx}
\usepackage{verbatim}
\usepackage{amsmath, amssymb, amsfonts }
\usepackage{helvet}
\usepackage{footnote}
\usepackage{subfigure}

\usepackage{microtype}

\graphicspath{{figures/}, 
              {figures/initial_conditions_plot/}, 
              {figures/February_et_al_comparison_plot/},
              {figures/February_et_al_comparison_plot/l_2/},
              {figures/February_et_al_comparison_plot/l_10/}
              {figures/void_size/},
              {figures/void_depth/},
              {figures/PowerSpectra/},
              {figures/initial_params_test/}
             }

\renewcommand{\d}{\mathrm{d}}
\newcommand{\p}{_{\|}}
\renewcommand{\o}{_{\perp}}
\newcommand{\st}{_{\ast}}
\newcommand{\I}{\int_{\Omega}}

\newcommand{\lm}{^{(\ell m)}}
\newcommand{\hpel}[1]{\hat{\phi}^{#1}_{e,l}(\mu_e^{-1}(x))}

\newcommand{\hpelh}[1]{\hat{\phi}^{#1}_{e,l}(\hat{x})}
\newcommand{\hpemh}[1]{\hat{\phi}^{#1}_{e,m}(\hat{x})}
\newcommand{\pvec}{\hat{\phi}_{e,l} = (\hat{\phi}^{(1)}_{e,l}, \hat{\phi}^{(2)}_{e,l}, \hat{\phi}^{(3)}_{e,l}, \hat{\phi}^{(4)}_{e,l}, \hat{\phi}^{(5)}_{e,l})}
\newcommand{\avol}{\alpha_{h,e} }
\newcommand{\avoltemp}[2]{ \left[\alpha_{h,e}(#2)\right]^{\mathrm{temp},#1} }
\newcommand{\avolspat}[2]{ \left[\alpha_{h,e}(#2)\right]^{\mathrm{spat},#1} }

\title{Evolution of linear perturbations in Lemaître-Tolman-Bondi void models  }
\author[a]{Sven Meyer}
\author[a]{, Matthias Redlich}
\author[a]{, and Matthias Bartelmann}

\affiliation[a]{Zentrum f\"ur Astronomie der Universit\"at Heidelberg, Institut f\"ur 
Theoretische Astrophysik, Albert-Ueberle-Str.~2, 69120 Heidelberg, Germany}

\emailAdd{ sven.meyer@uni-heidelberg.de }

\abstract{
We study the evolution of linear perturbations in a Lemaître-Tolman-Bondi (LTB) void model with realistic cosmological initial conditions. Linear perturbation theory in LTB models is substantially more complicated than in standard Friedmann universes as the inhomogeneous background causes gauge-invariant perturbations couple at first order. As shown by Clarkson et al. (2009) (\cite{clarkson_perturbation_2009}), the evolution is constrained by a system of linear partial differential equations which need to be integrated numerically. We present a new numerical scheme using finite element methods to solve this equation system and generate scalar initial conditions based on Gaussian random fields with an underlying power spectrum for the Bardeen potential. After spherical harmonic decomposition, the initial fluctuations are propagated in time and estimates of angular power spectra of each gauge invariant variable are computed as functions of redshift. This allows to analyse the coupling strength in a statistical way. We find significant couplings up to $25\%$ for large and deep voids of Gpc scale as required to fit the distance redshift relations of SNe.    
}

\begin{document}

\maketitle
\flushbottom

%***************************************************************************************************************************
\section{Introduction} 
\label{ltb:introduction}
%***************************************************************************************************************************

The Copernican principle states that, averaged over sufficiently large scales ($\ge 100 \mathrm{Mpc}$), there is no distinct position in the Universe. As one of the theoretical foundations of the standard Friedmann-Lemaître-Robertson-Walker (FLRW) models, it is a well-established concept in modern cosmology, though very difficult to test observationally. Since we only have direct access to observational data on our past lightcone, we are not able to distinguish temporal evolution from spatial variations which are neglected by assumption in the standard model. One possible approach to test homogeneity has been to study Lemaître-Tolman-Bondi (LTB) models (see \cite{lemaitre_expansion_1931, tolman_effect_1934, bondi_spherically_1947} for original works and \cite{enqvist_lemaitre_2008, bolejko_testing_2009, marra_observational_2011, clarkson_establishing_2012} for detailed reviews) which are a class of cosmological models based on a radially inhomogeneous solution of Einstein's field equations. Their spatial hypersurfaces are spherically symmetric about a distinct central worldline. If our galaxy is assumed to be located in a large underdense region of Gpc scale (\cite{february_rendering_2010, celerier_we_2000}), cosmological models built upon this class of solutions are able to fit distance-redshift relations of type Ia supernovae without any assumption of dark energy. However, an off-centre observer would see a significant dipole amplitude in the cosmic microwave background (CMB) which has to be in agreement with the observed signal. Our position in the void region is therefore constrained to be close to the center within a few tens of Mpc (\cite{alnes_cmb_2006, foreman_spatial_2010} for corresponding CMB analyses) requiring a high degree of spatial fine tuning. LTB models have been rigorously confronted with observations (see \cite{biswas_testing_2010, garcia-bellido_confronting_2008, bull_kinematic_2012, celerier_we_2000, zibin_can_2008, moss_precision_2011, valkenburg_testing_2014, zumalacarregui_tension_2012, zhang_confirmation_2011, caldwell_test_2008, garcia-bellido_looking_2008, zibin_linear_2011}) and their applicability to describe the local universe around our position has recently been reanalysed in detail in \cite{redlich_probing_2014}. \\

However, current analyses of these models are limited to observables that do not depend on the details of linear structure formation, since linear perturbation theory in LTB models is quite a challenge and therefore still under development. The reason is that the evolution of perturbations on inhomogeneous backgrounds is substantially more complicated than in spatially homogeneous FLRW models, because decomposition into gauge invariant 3-scalar-, 3-vector- and 3-tensor modes (SVT variables) is no longer straightforwardly possible. In addition, anisotropic and position-dependent structure growth causes gauge-invariant perturbation variables to couple already at first order which is described by a system of coupled partial- instead of ordinary differential equations. However, significant progress has been made in the past years on several approaches. Alonso et al (2010) (\cite{alonso_large_2010}) managed to set-up a Newtonian N-body simulation in the gravitational potential of a large Gpc void and studied Newtonian perturbations by comparing simulations with the theoretically predicted void profile of \cite{garcia-bellido_confronting_2008}. Nishikawa et al. (2012) (see \cite{nishikawa_evolution_2012} and also \cite{nishikawa_two-point_2013, nishikawa_comparison_2014}) studied the linear density evolution in void models by applying secondary linear perturbations on a primary non-linearly perturbed FLRW model that accounts for the void. Zibin (2008) (see \cite{zibin_scalar_2008} and further application in \cite{dunsby_how_2010}) used a covariant 1+1+2 formalism for scalar perturbations in LTB spacetimes and obtained evolution equations and matter transfer functions in the, so-called, \textit{silent approximation} by neglecting the magnetic part of the Weyl tensor and effectively the coupling of scalar to tensor modes. \\

Clarkson et al. (2009) (see \cite{clarkson_perturbation_2009} for a remarkable paper in this context) obtained a full set of gauge-invariant perturbation variables in spherically symmetric dust spacetimes. They derived first order evolution equations given by linear partial differential equations containing the full coupling of the perturbations on inhomogeneous backgrounds. Their results are built upon earlier studies of gauge-invariant perturbations in general spherically symmetric spacetimes using a 2+2 split of the background spacetime (see \cite{gerlach_relativistic_1978, gerlach_homogeneous_1978, gerlach_gauge-invariant_1979, seidel_gravitational_1990} and \cite{tomita_perturbations_1997} for a cosmological application). These results have been evaluated in a covariant perfect fluid frame by Gundlach \& Martin-García (\cite{gundlach_gauge-invariant_2000, martin-garcia_gauge-invariant_2001}) in the context of perturbed stellar collapse. Clarkson et al. specified these evolution equations for dust solutions and even performed the FLRW limit showing the complicated mixing of FLRW scalar, vector, and tensor degrees of freedom in each LTB gauge-invariant perturbation. However, numerics of the corresponding partial differential equation system is very challenging. In case of negligible coupling, February et al. (2010) (\cite{february_galaxy_2013}) managed to predict two-point density correlation function in LTB void universes. A first test run of a full numerical solution of the system has been performed by February et al. (2014) (\cite{february_evolution_2014}). Their numerical scheme is based on combined second-order finite differencing in space and fourth-order time integration. Starting with void profile of Gaussian shape, they ran several test cases by initialising only a single perturbation variable by five Gaussian peaks that are placed at equidistant positions in the void. \\

However, estimating the relevance of the coupling strength in a realistic cosmological environment is still an open issue. We are therefore going to extend the approach of February et al. (2014) by generating realistic initial conditions and evolve the system from an initial FLRW state. By assuming the universe to be initially homogeneous and isotropic, we sample initial conditions from a power spectrum of the Bardeen potential $\Psi$ in the matter dominated era. This enables us to study the spacetime evolution of perturbations in LTB models in a statistical way by comparing angular power spectra of each LTB gauge invariant at different redshifts on the past null cone. We developed a new numerical scheme to solve the underlying partial differential equation system based on a finite element technique for the spatial discretization in each timestep. This approach has proven to be more flexible than finite differences, since the grid structure can easily be adapted to the problem itself. In addition, we use a numerical implementation of the LTB background model (as also applied in \cite{redlich_probing_2014}) such that we are not limited to strictly hyperbolic background evolution models and very flexible in the choice of the void density profile. \\

The paper is structured as follows:  In Sects. \ref{ltb:background} and \ref{ltb:perturbation}, we give an overview on the background model implementation and perturbation theory in LTB models using a 2+2 split of the spacetime. We outline the sampling technique and spherical harmonics decomposition of the initial Bardeen potential field in Sect. \ref{ltb:initial}. The numerical scheme to solve the LTB perturbation equations is outlined in Sect. \ref{ltb:numerics} and presented in detail in Appendix \ref{ltb:dune}. In the following, we show results of different test runs by first confirming the results of \cite{february_evolution_2014} (see Sect. \ref{ltb:comparison}) with our numerical scheme and then applying it to the realistically sampled initial conditions by showing angular power spectra and coupling strengths at different redshifts (Sects. \ref{ltb:powerspectra} and \ref{ltb:coupling}). We conclude with a short discussion of the results and plans for future work (Sects. \ref{ltb:discussion} and \ref{ltb:conclusion}).

%***************************************************************************************************************************
\section{Background LTB Model} 
\label{ltb:background}
%***************************************************************************************************************************

The LTB solution is a radially inhomogeneous dust solution of Einstein's field equations with spatial hypersurfaces spherically symmetric about a central worldline. Consequently, these models are isotropic around this center and anisotropic everywhere else. The absence of pressure contributions in the particularly simple energy momentum tensor  

\begin{equation}
T_\mathrm{\mu\nu} = \rho(t,r) u_\mu u_\nu
 \label{ltb:background:1}
\end{equation}

and the resulting geodesic motion of dust-comoving observers allow us to choose comoving-synchronous coordinates. The line element then takes the form 

\begin{equation}
  ds^2 = -\d t^2 + \frac{a\p^2(t,r)}{1- \kappa(r) r^2} \d r^2 + r^2 a\o^2(t,r) \d \Omega^2
 \label{ltb:background:2}
\end{equation}

with $a\p(t,r) = (r a\o(t,r))_{,r}$. $a\o(t,r)$ and $\kappa(r)$ are free functions of the coordinate time and radius.

Using this notation proposed by Clarkson (2012) (\cite{clarkson_establishing_2012}), we directly see the differences to the spatially homogeneous FLRW models. The off-center anisotropy causes the FLRW scale factor $a(t)$ to be replaced by two scale factors $a\p(t,r)$ and $a\o(t,r)$ expressing the expansion parallel and perpendicular to the radial direction. The two scale factors induce two Hubble rates 

\begin{equation}
 H\o(t,r) = \frac{\dot{a}\o(t,r)}{a\o(t,r)}, \quad H\p(t,r) = \frac{\dot{a}\p(t,r)}{a\p(t,r)}
 \label{ltb:background:3}
\end{equation}

that depend on time and radial position. In addition, these models allow for a radial dependence of the density $\rho(t,r)$ and spatial curvature $\kappa(r)$. It is convenient to define an active gravitational mass $M(r)$ inside a spherical shell of radius $r$ which is fixed by the field equations to 

\begin{equation}
 \frac{ (M(r)r^3)_{,r}}{ r^2 a\o^2(t,r) a\p(t,r) } = 8 \pi G \rho(t,r). 
 \label{ltb:background:4}
\end{equation}

The time evolution of the scale factor $a\o(t,r)$ in terms of mass and curvature profile is also given by Einstein's field equations which, in addition to Eq. (\ref{ltb:background:4}), become   

\begin{equation}
 H\o^2(t,r) = \frac{M(t,r)}{a\o^3(t,r)} - \frac{\kappa(r)}{a\o^2(t,r)}
 \label{ltb:background:5}
 \end{equation}

and can be integrated to 

\begin{equation}
 t -t_B(r)= \int_0^{r a\o}{ \frac{\d(r \tilde{a}\o)}{\sqrt{-2\kappa(r) r^2 + \frac{2M(r)}{r \tilde{a}\o} } }}.
 \label{ltb:background:6}
\end{equation}

For consistency with the standard inflationary paradigm, we assume a synchronous big bang by setting $t_B(r) = 0$ for all radii. It can be shown (see \cite{silk_large-scale_1977, zibin_can_2008}) that fluctuations in the bang time function correspond to decaying modes in linear perturbation theory resulting in large inhomogeneities at early times. This would be highly inconsistent with the observed remarkable uniformity of the last scattering surface. For this reason, the void solution will be asymptotically embedded into a FLRW model at large radii and early times. \\

As well-known and widely used in the literature, Eq. (\ref{ltb:background:6}) can be solved parametrically depending on the sign of the curvature profile $\kappa(r)$: \\

$\kappa(r) > 0$ (elliptic evolution):

\begin{eqnarray}
 a\o(t,r) &=& \frac{M(r)}{2\kappa(r) } \left( 1- \cos(\eta) \right) \\
 \eta - \sin(\eta) &=& \frac{ 2[\kappa(r)]^{3/2}}{M(r)} \left( t  - t_0 \right)
  \label{ltb:background:7}
\end{eqnarray}

$\kappa(r) = 0$ (parabolic evolution):

\begin{equation}
 a\o(t,r) = \left[ \frac{9}{4} M(r) \left( t  - t_0 \right)^2 \right]^{1/3}
   \label{ltb:background:8}
\end{equation}

$\kappa(r) < 0$ (hyperbolic evolution):

\begin{eqnarray}
 a\o(t,r) &=& \frac{M(r)}{-2\kappa(r) } \left( \cosh(\eta) -1  \right) \\
 \sinh(\eta) -\eta  &=& \frac{ 2[-\kappa(r)]^{3/2}}{M(r)} \left( t  - t_0 \right) 
   \label{ltb:background:9}
\end{eqnarray}

The line element and all previous formulae are invariant under the coordinate transformation $r = f(r')$ which is a gauge freedom in this context. For consistency with standard FLRW models, we fix this gauge such that $a\o(t_0,r) = 1$. \\

The mass profile is then simply given by

\begin{equation}
 M(r) = \frac{8\pi}{r^3} \int_0^r{r'^2 \rho(t_0,r') \d r'}.
 \label{ltb:background:10}
\end{equation}

In general, an observer in LTB spacetime can only access information from her past null cone (PNC).  We will assume throughout this analysis that observers are moving on the central worldline of the spacetime which is supported by the small dipole signal of the CMB. Inward radial null geodesics are then described by the following system of equations 

\begin{align}
 \label{ltb:background:11}
 \frac{\d t(r)}{\d r} &= - \frac{a\p(t(r),r)}{\sqrt{1-\kappa(r) r^2}},\\
 \label{ltb:background:12}
 \frac{1}{ 1 + z(r)} \frac{\d z(r)}{\d r} &=  \frac{\dot{a}\p(t(r),r)}{\sqrt{1-\kappa(r) r^2}},
 \end{align}

 which can be integrated numerically. By interpolation, we can effectively invert the result in order to transform arbitrarily between redshift and LTB coordinates on the PNC.
 
 For a practical implementation of the background model, we adapted the algorithm outlined in Redlich et al. (2014) (see \cite{ redlich_probing_2014}) which shall be shortly summarised here. For a more detailed discussion of the background model implementation, the interested reader is referred to the corresponding paper\footnote{Note that in the context of perturbation theory, the notation of Clarkson (2012) (\cite{clarkson_establishing_2012}) using  $\{ a\o, a\p, M, \kappa\}$ turns out to be more appropriate then the standard notation $\{ R, R', \tilde{M}, E\}$ applied in Redlich et al. (2014) \cite{ redlich_probing_2014}. For better comparison, these quantities are related via $a\o(t,r) = R(t,r)/r$, $a\p(t,r) = R'(t,r)$, $M(r) = 2\tilde{M}(r)/r^3$ and $\kappa(r) = -2E(r)/r^2$. }.\\ 
 
 \begin{enumerate}
  \item Since we neglect fluctuations in the bang time function and assume an initial homogeneous universe, the background FLRW solution and the LTB patch have the same global age $t_0$ given by 
 
  \begin{equation}
    t_0 = \frac{1}{H_0} \int_0^1{ \frac{\sqrt{a} }{ \sqrt{ \Omega_m + \Omega_k a } } \d a  }
    \label{ltb:background:13}
  \end{equation}

  \item We fix a density profile $\rho(t_0,r)$ at present time which sets the mass profile according to Eq. (\ref{ltb:background:4}). The curvature profile $\kappa(r)$ is implicitly defined by Eq. (\ref{ltb:background:3}) and has to be computed numerically. Using Eq. (\ref{ltb:background:6}) in combination with a proper root finder (see \cite{ redlich_probing_2014}), we solve for $\kappa(r)$ as function of $t_0$, $r$ and $M(r)$.   
  
  \item With the mass and curvature profiles at hand, the time evolution of the background model is completely determined by Eq. ({\ref{ltb:background:3}}) which is integrated backwards to an initial hypersurface of constant time $t_\mathrm{ini}$. All necessary coefficients of the LTB perturbation equations are then fixed on the full domain of interest and can be accessed by 2d linear interpolation. 
  
  \item Eqs. (\ref{ltb:background:11}) and  (\ref{ltb:background:12}) can be integrated in a similar way.
 \end{enumerate}

 As can be seen from Eqs. (\ref{ltb:background:4}) - (\ref{ltb:background:6}), LTB models are generally characterised by three free radial functions $t_B(r)$, $M(r)$, and $\kappa(r)$ that have to be specified individually. By demanding a spatially homogeneous universe at early times, we have already fixed the bang time function. In addition, we set a proper gauge for the tangential scale factor ($a\o(t_0,r) =1$) which allows to integrate  Eq. (\ref{ltb:background:4}) and determines the mass profile as functional of the density profile $\rho(t_0,r)$ on the $t_0$-hypersurface.  Since the gauge choice also specifies the upper integral boundary in Eq.  (\ref{ltb:background:6}), the curvature profile is also fixed as function of $t_0$ and $M(r)$. Consequently, the void density profile at $t=t_0$ is sufficient to characterise the LTB model at all times and spatial scales. \\
 
 For the purpose of this study, we assume a Gaussian shaped void of the form $\rho(t_0,r) = f(r) \bar{\rho}(t_0)$ with radial profile
 
 \begin{equation}
  f(r) =  1 + \left( \frac{\Omega_\mathrm{in}}{\Omega_\mathrm{out}} - 1 \right) \exp\left( -\frac{r^2}{L^2} \right).
  \label{ltb:background:14}
 \end{equation}

  The void extent $L$ as well as the central- and asymptotic density parameters $\Omega_\mathrm{in}$ and $\Omega_\mathrm{out}$ can be varied individually. As standard example for our study, we choose a $2$ Gpc void with $\Omega_\mathrm{in}=0.2$ that is asymptotically embedded into an Einstein-deSitter (EdS) model ($\Omega_\mathrm{out} = 1$). Nonetheless, by the structure of the background implementation, more complicated void shapes can be included without problems. We want to stress at this point that the current analysis is meant as a theoretical test study with no reference to any observable constraints so far. Although the full spectrum of Gaussian initial conditions will be taken into account, we still rely on specific Gaussian void profiles that are certainly not related to any observational constraints on LTB models.

%***************************************************************************************************************************
\section{Linear perturbations} 
\label{ltb:perturbation}
%***************************************************************************************************************************

Perturbations on homogeneous backgrounds have extensively been studied and are well-understood (see \cite{bertschinger_cosmological_2000}, \cite{malik_cosmological_2009}, and \cite{tsagas_relativistic_2008} for detailed discussions). Perturbation variables can be split into scalars, vectors and tensors by their transformation properties on the underlying homogeneous 3-space. The main property of these models is that the high degree of symmetry causes these perturbation types to decouple at first order such that their time evolution can be studied separately. \\

This is no longer the case on general spherically symmetric backgrounds. In fact, radial dependence of structure growth in inhomogeneous models causes all perturbation types to couple which makes their time evolution and physical interpretation highly difficult. One possible approach is a covariant 2+2 split of the spacetime ($\mathcal{M}^4 = \mathcal{M}^2 \times \mathcal{S}^2$) where perturbations can be characterised according to their transformation properties on $\mathcal{S}^2$ which is analogous, but not equivalent, to the homogeneous case. Perturbations then decouple into a polar (even parity) and an axial (odd parity) branch which roughly corresponds to their covariant curl- and divergence-free parts on the two-sphere. In addition, it is convenient to separate the angular parts into spherical harmonics and covariant derivatives thereof (vector and tensor spherical harmonics). Spherical symmetry of the background causes the evolution of perturbations to separate into spherical harmonic modes $\ell, m$. By suitable combinations, sets of gauge-invariant metric and fluid perturbations and corresponding evolution equations can be derived. This formalism has been developed by Gerlach \& Sengupta (1979) (\cite{gerlach_gauge-invariant_1979}) for general spherically symmetric spacetimes and reformulated in perfect fluids frames by Gundlach \& Martin-Garcia (2000) (GMG) (see \cite{gundlach_gauge-invariant_2000, martin-garcia_gauge-invariant_2001}) to study anisotropic stellar collapse. \\

In a remarkable work following up this approach, Clarkson, Clifton and February (CCF) (\cite{clarkson_perturbation_2009}) specified GMG's approach to dust solutions and obtained a full set of LTB gauge-invariant perturbation variables and corresponding master and constraint equations. Their work and recent applications in \cite{february_galaxy_2013, february_evolution_2014} can be considered as the key theoretical background for our analysis. In analogy to the conformal Newtonian gauge in FLRW models, there exists a special gauge, the Regge-Wheeler (RW) gauge (see \cite{regge_stability_1957}), in which the perturbation variables correspond to the gauge-invariants. Throughout this work, we will focus on the polar branch, since it contains the generalized scalar gravitational potential and density perturbation. In addition, we restrict ourselves to the case of $\ell \geq 2$. For $\ell=0,1$, similar gauge invariants variables and evolution equations can be derived (see \cite{gundlach_gauge-invariant_2000} and the appendix of \cite{clarkson_perturbation_2009} for details), but we are not going to consider them here. \\

For the polar sector and $\ell \geq 2$, we can define a set of four metric perturbations $\{\eta\lm, \chi\lm, \varphi\lm, \varsigma\lm \}$ and three fluid perturbations $\{\Delta\lm, w\lm, v\lm \}$ that assemble the general form of the perturbed LTB metric and energy momentum tensor.  In RW gauge, we have (see \cite{clarkson_perturbation_2009})  

 \begin{align}
  \label{ltb:perturbation:1}
  ds^2 &= -\left[ 1 + (2\eta\lm - \chi\lm - \varphi\lm) Y\lm \right]\d t^2 - \frac{2 a\p \varsigma\lm  Y\lm }{\sqrt{1 - \kappa r^2}} \d t \d r  \\
       & \ \ \ \ + \frac{a\p^2}{1 - \kappa r^2} \left[ 1 + (\chi\lm + \varphi\lm) Y\lm \right] \d r^2  + r^2 a\o^2 \left[1 + \varphi\lm  Y\lm \right] \d \Omega^2 \nonumber \\ \nonumber \\ 
  \label{ltb:perturbation:2}
  \rho &= \rho^\mathrm{LTB} \left( 1 + \Delta\lm  Y\lm \right) \\
  \label{ltb:perturbation:3}
  u_\mu &= \left[ u_A + \left( w\lm n_A + \frac{1}{2} k_{AB} u^B  \right) Y\lm , v\lm  Y_b\lm  \right] 
 \end{align}

with sums over $(\ell, m)$ implied and $ Y_b^{(\ell m)} = \nabla_b  Y^{(\ell m)}$.\footnote{There are three types of indices needed for the 2+2 split of the spacetime. By convention of GMG and CCF, we use Greek indices for the full spacetime coordinates, capital Roman letters for the $(t,r)$ part and small Roman letters for the angular parts.}  Each perturbation variable is therefore a spherical harmonic coefficient and a free function of $t$ and $r$. The unit vectors in time and radial direction are given by $u_A = (-1,0)$ and $n_A = (0, a\p/\sqrt{1-\kappa r^2})$. $k_{AB}$ corresponds to the metric perturbation in the $(t,r)$-submanifold. \\ 
   
The evolution equations for the polar metric perturbations are then given by a closed system of master equations \footnote{ $\dot{()}$ denotes a derivative with respect to coordinate time $t$ whereas $()'$ a derivative with respect to LTB radial coordinate $r$ and we dropped the $(\ell, m)$ superscript here.}: 

\begin{align}
  \label{ltb:perturbation:4}
 \ddot{\chi} &= \frac{\chi'' - C \chi'}{Z^2} - 3 H\p \dot{\chi} + \left[A - \frac{(\ell-1) (\ell+2)}{r^2 a\o^2}\right] \chi + \frac{2\sigma}{Z} \varsigma' + \frac{2}{Z} \left[H\p ' - 2 \sigma \frac{a\p}{r a\o}\right] \varsigma - 4 \sigma \dot{\varphi} + A \varphi \\
  \label{ltb:perturbation:5}
 \ddot{\varphi} &= - 4 H\o \dot{\varphi} + \frac{2 \kappa}{a\o^2} \varphi - H\o \dot{\chi} + Z^{-2} \frac{a\p}{r a\o} \chi' - \left[ \frac{1 - 2\kappa r^2}{r^2 a\o^2} - \frac{\ell (\ell +1)}{2r^2 a\o^2}\right] \chi + \frac{2}{Z} \frac{a\p}{r a\o} \sigma \varsigma \\  
  \label{ltb:perturbation:6}
 \dot{\varsigma} &= - 2 H\p \varsigma - \frac{\chi'}{Z} \\
  \label{ltb:perturbation:7}
 \eta &= 0.
\end{align}

The remaining part of the field equations describes the coupling to the fluid perturbations 

\begin{align}
  \label{ltb:perturbation:8}
 \alpha w &= \frac{1}{Z} \dot{\varphi}' - \frac{1}{Z} (\sigma - H\o) \varphi' - \frac{1}{Z} \frac{a\p}{r a\o} \dot{\chi} + \frac{H\o}{Z} \chi' + \left[ \frac{\ell(\ell+1)}{2r^2 a\o^2} + D + \frac{\kappa}{a\o^2} \right] \varsigma \\
  \label{ltb:perturbation:9}
 \alpha  \Delta &= - \frac{1}{Z^2} \varphi'' + \frac{1}{Z^2} \left( C - 4 \frac{a\p}{r a\o} \right) \varphi' + \left( H\p + 2 H\o \right) \dot{\varphi} + \frac{1}{Z^2} \frac{a\p}{r a\o} \chi' + H\o \dot{\chi} \\ \nonumber
                & \quad + \left[ \frac{\ell(\ell+1)}{r^2 a\o^2} + 2 D \right] \left( \chi + \varphi \right) - \frac{(\ell-1)(\ell+2)}{2 r^2 a\o^2} \chi + \frac{2 H\o}{Z} \varsigma' + \frac{2}{Z} \left(H\p + H\o \right) \frac{a\p}{r a\o} \varsigma \\
  \label{ltb:perturbation:10}
\alpha  v &= \dot{\varphi} + \frac{\dot{\chi}}{2} + H\p \left( \chi + \varphi \right) + \frac{1}{2 Z} \varsigma'.
\end{align}

The coefficients are given by the following quantities of the background LTB model:

\begin{align*}
\alpha &= 8\pi G \rho = \frac{\kappa}{a\o^2} \left( 1 + 2 \frac{a\p}{a\o}\right) + H\o \left( H\o + 2 H\p \right) + \frac{\kappa' r}{a\o a\p} \\
A &= 2 \alpha - \frac{6 M}{a\o^3} - 4 H\o \sigma \\
C &= \frac{a\p'}{a\p} + \frac{ \kappa r + \frac{1}{2} \kappa'r^2 }{1 - \kappa r^2} + \frac{2 a\p}{r a\o} \\ 
D &= -\frac{\alpha}{2} + H\o \left( H\o + 2H\p \right) \\ 
\sigma &= H\p - H\o \\
Z &= \frac{a\p}{\sqrt{1-\kappa r^2}}.
\end{align*}
   
Local energy momentum conservation $\nabla_\mu T^\mu_{ \ \nu} = 0$ leads to derivative constraints that the system naturally obeys: 

\begin{align}
\label{ltb:perturbation:11}
\dot{w} &= \frac{1}{2Z} \varphi' - H\p  \left( w + \frac{\varsigma}{2} \right) \\
\label{ltb:perturbation:12}
\dot{\Delta} &= - \frac{\dot{\chi} + 3 \dot{\varphi}}{2} + \frac{\ell (\ell +1)}{r^2 a\o^2} v - \frac{1}{Z} \left[ \left( w + \frac{\varsigma}{2} \right)' + \left( \frac{\alpha'}{\alpha} + \frac{2 a\p}{r a\o} \right)  \left( w + \frac{\varsigma}{2} \right)\right] \\
\label{ltb:perturbation:13}
\dot{v} &= \frac{\chi + \varphi}{2}.
\end{align}

Einstein's field equations to first order show that the metric perturbations $\eta$, $\chi$, $\varphi$ and $\varsigma$ are master variables of the system, i.e., if the solution to the system (\ref{ltb:perturbation:4}) - (\ref{ltb:perturbation:7}) is known for each angular scale $\ell$, Eqs. (\ref{ltb:perturbation:8}) - (\ref{ltb:perturbation:10}) reduce to simple identities for the fluid perturbation variables. The evolution in time and radius is constrained by linear partial differential equations that contain a non-trivial coupling caused by radially-dependent and off-center anisotropic structure growth. \\

Although the master equations will look similar to their FLRW counterparts if the coupling terms are neglected (in fact, $\chi$ obeys a wave equation, $\varphi$ a scalar Bardeen-like equation and $\varsigma$ a vector $a^{-2}$-decay law), the physical interpretation of these gauge-invariants is more subtle. Scalar-vector-tensor decomposition does not naturally exist on spherically symmetric backgrounds and therefore LTB gauge-invariants are a priori completely different from FLRW gauge-invariants. In extensive calculations, CCF managed to perform the FLRW limit and showed that these gauge-invariants are intrinsically a complicated mixture of all perturbation types. In fact, it turns out that only $\chi$ is a genuine gravitational wave mode whereas $\varsigma$ contains vector and tensor degrees of freedom and $\varphi$ as well as $\eta$ contain degrees of freedom of all FLRW-perturbation types. Similar results are obtained for the fluid variables (see appendices of \cite{clarkson_perturbation_2009} for the details). It is therefore hard to disentangle these gauge-invariants aiming at a direct comparison with homogeneous models. There is the possibility of constructing SVT variables that, in the FLRW limit, reduce to pure scalar, vector and tensor modes, but these ones are of complicated structure (see \cite{clarkson_perturbation_2009}). \\

Notwithstanding any physical interpretation or comparison to spatially homogeneous models, the intrinsic coupling of the perturbation variables itself as seen in  Eq. (\ref{ltb:perturbation:4}) - (\ref{ltb:perturbation:7}) can directly be studied if the system is integrated numerically. By carrying out the FLRW limit, CCF showed that the scalar Bardeen potential $\Psi$ is only contained in $\varphi$. Hence, starting from an initial scalar potential perturbation at early times, we can directly quantify the influence of the coupling on the spacetime evolution by comparing the fully coupled with the uncoupled case.\\    

As already pointed out in the introduction, the main intention of this analysis is considering the full evolution of perturbation equations from realistic initial scalar perturbations and investigating the strength of the coupling in a statistical way. We are not yet aiming at a direct comparison of structure formation in LTB and FLRW models which is postponed to a future paper.

%***************************************************************************************************************************
\section{Numerical setup} 
\label{ltb:numerics}
%***************************************************************************************************************************

Numerical integration of a system of partial differential equations involves methods based on, at least, spatial grids. We use the well-established \textit{method of lines} by performing a spatial discretization with finite elements which is then integrated in time using an implicit Alexander S-stable method (see \cite{alexander_diagonally_1977}). All these methods have been preimplemented in the \textit{Distributed Unified Numerics Environment} (DUNE) (see \cite{bastian_towards_2004, bastian_generic_2008, bastian_generic_2008-1, blatt_generic_2008}). In contrast to finite differencing methods applied in \cite{february_evolution_2014}, finite elements, although with slightly more complicated mathematical background, are more flexible and also well-suited for irregular or locally refined grids. In addition, higher accuracy can easily be achieved by increasing the basis polynomial degree (see Appendix \ref{ltb:dune} for details of the implementation). We decided to prefer finite elements to finite differences, since the initial profiles for the spherical harmonic coefficients show considerable small scale fluctuation with increasing spherical harmonic $\ell$-mode and we therefore need a robust scheme for numerical differentiation that also works for on grids that can be adapted to properties of the solution itself. February et al. have shown that their approach using second order finite differences works well for the series of test runs they performed on a regular grid. We have been able to reproduce their results with our finite element setup (see Sect. \ref{ltb:comparison}). \\

\begin{figure}
 \centering
 \includegraphics[width=12cm]{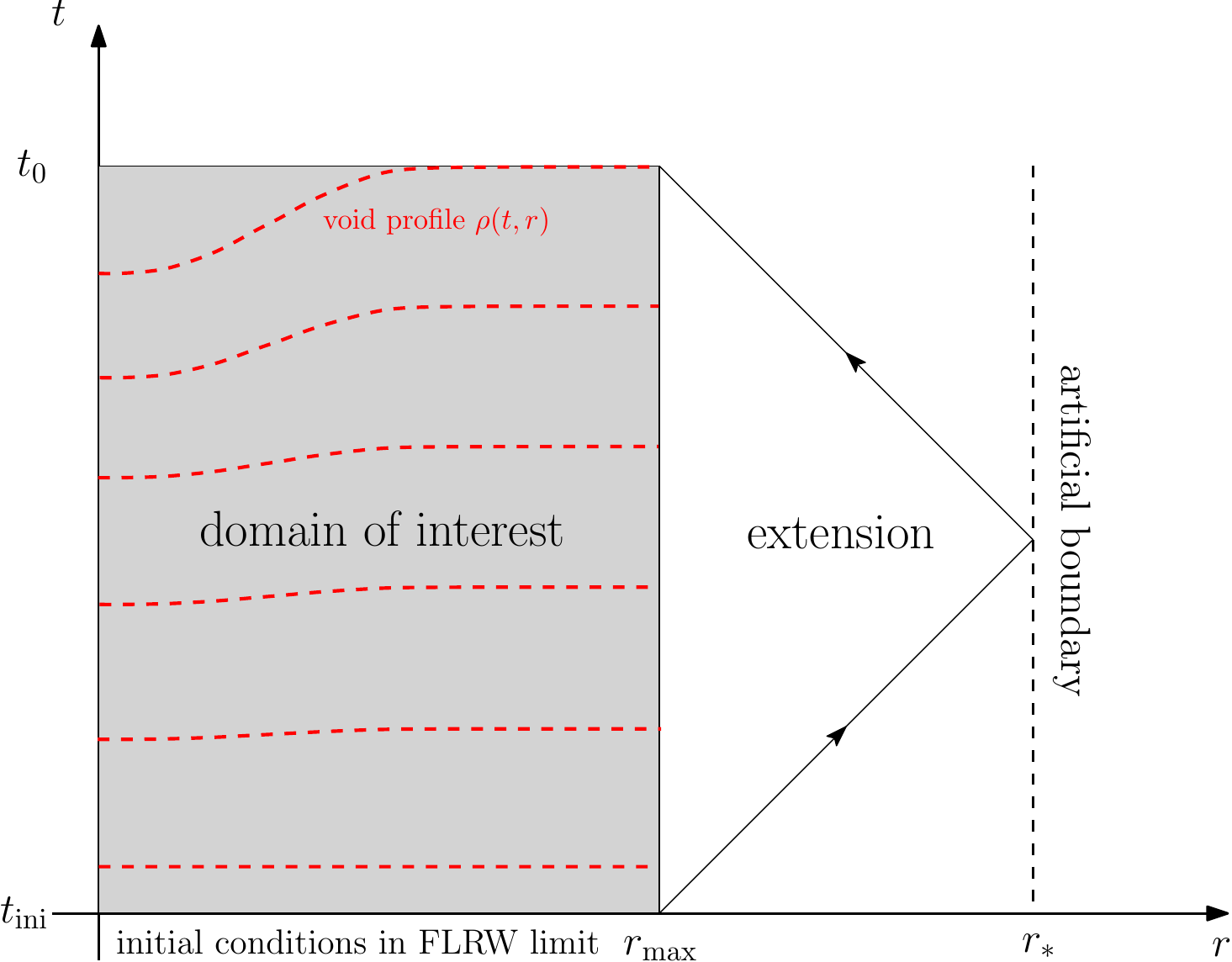}
 \caption{Construction of the artificial boundary condition according to \cite{february_evolution_2014}: $r_\ast$ is chosen to be causally disconnected from the domain of interest by integrating radial null geodesics in the background spacetime (see Eq. \ref{ltb:numerics:2}). The void is fully contained in the domain of interest and deepens with increasing cosmic time. }
 \label{ltb:numerics:fig:1}
\end{figure}

The solution of a system of partial differential equations with time dependence is constrained by initial and boundary conditions. While the setup of initial conditions will be discussed in detail in the next section, boundary conditions can be chosen in a natural way by requiring regularity of the solution. We define a spatial \textit{domain of interest} of (at least) three times the void size in which we want to study the numerical solution. As proposed by \cite{february_evolution_2014}, we impose an artificial spatial boundary condition at $r=r\st$ that is causally disconnected from the domain of interest in order to prevent any artificially reflected propagating modes to reenter (see Fig. (\ref{ltb:numerics:fig:1})). By taking the characteristics of the system into account and following null geodesics in the background spacetime, we can integrate 

\begin{equation*}
 \frac{\d t(r)}{\d r} = \pm \frac{a\p(t,r)}{\sqrt{1-\kappa(r)r^2}} = Z(t,r)
\end{equation*}

with $t(r_\mathrm{max}) = t_\mathrm{ini}$) to obtain\footnote{We assume in addition that the unperturbed LTB metric in the region $[r_\mathrm{max}, r\st]$ is well-described by the background EdS model.}

\begin{equation}
  r_\ast = r_\mathrm{max} + \frac{1}{2} \int_{t_{\mathrm{ini}}}^{t_{0}} { Z^{-1}(t,r_\mathrm{max}) \d t}
 \label{ltb:numerics:1}
\end{equation}

As shown by GMG in \cite{gundlach_gauge-invariant_2000}, a solution to Eqs. (\ref{ltb:perturbation:4}) - (\ref{ltb:perturbation:10}) has to obey certain conditions for regularity at $r=0$:

\begin{equation*}
  \chi = \bar{\chi} r^{\ell +2}, \ \varphi = \bar{\varphi} r^{\ell}, \ \varsigma = \bar{\varsigma} r^{\ell +1}, \ \Delta = \bar{\Delta} r^{\ell}, \ w = \bar{w} r^{\ell -1}, \ v = \bar{v} r^{\ell },
\end{equation*}

where the barred quantities can be expanded in positive even powers of $r$. This enforces a zero Dirichlet boundary condition at the origin for $\ell>2$.  We require the solution to vanish at $r_\ast$ as well such that no propagating modes are generated there. \\

The grid is generally divided into three parts: To fulfill this regularity conditions with sufficient accuracy, we have a fine structured grid at the origin, followed by an equidistant grid covering the full domain of interest and a coarse grid covering the region between the outer boundary of the domain of interest and the artificial boundary. \\

Since we use an implicit time integration scheme, the Courant-Friedrichs-Levy (CFL) condition (\cite{courant_partial_1967}) is not necessary for stability, but nonetheless we assign the timesteps dynamically according to this condition in order to adapt the time resolution to intrinsic timescales of the system. Using the system's characteristics again, we obtain

\begin{equation}
\frac{\Delta t(t)}{\Delta r} = 0.99 \cdot \min_{ r_\mathrm{min}\le r \le r_\mathrm{max}} (Z(t,r)) \le \min_{ r_\mathrm{min} \le r \le r_\mathrm{max} } (Z(t,r)).
\label{ltb:numerics:2}
\end{equation}

\begin{figure}
 \centering
  \includegraphics[width=12cm]{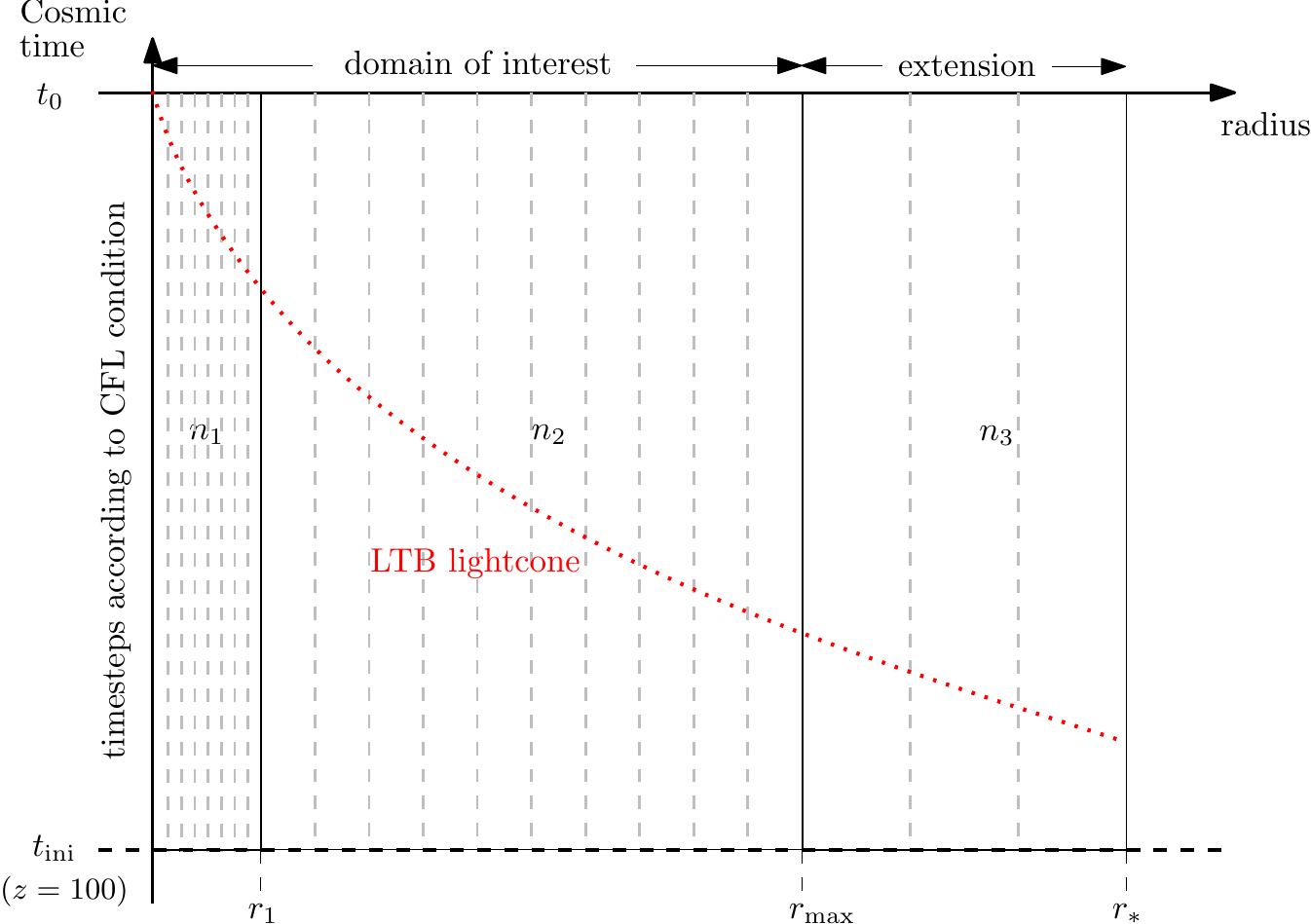}
 \caption{ Schematic structure of the irregular spatial grid: At small radii, we apply a finely resolved grid around $[r_\mathrm{min}, r_1]$ with $n_1$ bins to recover the regularity condition precisely. This part is followed by a regular grid in  $[r_1, r_\mathrm{max}]$ with $n_2$ bins and a very coarse grid in the extension $[r_\mathrm{max}, r_\ast]$ with $n_3$ bins. The parameters $r_1$, $n_1$, $n_2$, and $n_3$ can be adjusted freely. For all runs shown in Sects. (\ref{ltb:powerspectra}) and (\ref{ltb:coupling}), we have chosen $r_1=100 \ \mathrm{Mpc}$, $n_1=50$, $n_2=256$ and $n_3=20$. Final results will be interpolated on the LTB backward lightcone that is shown here schematically for illustration.}
\end{figure}

In the polar sector, Einstein's field equations split into a complete set of master equations for the metric perturbations and corresponding constraint equations for the fluid perturbations. We therefore evolve Eqs. (\ref{ltb:perturbation:4}) - (\ref{ltb:perturbation:7}) in time and use the results in each timestep to constrain the fluid perturbations using the identities given by Eqs. (\ref{ltb:perturbation:8}) - (\ref{ltb:perturbation:10}). 
Grid structure, basic finite element map and the equation systems have to be passed to the DUNE interface that performs a residual calculation to solve the underlying linear equation system in each time step. For details of the implementation, the reader is referred to Appendix \ref{ltb:dune}.  

\begin{comment}
For reasons of stability at the spatial origin $r=0$, we decided to evolve the derivative constraint Eqs. (\ref{ltb:perturbation:11}) - (\ref{ltb:perturbation:13}) and use (\ref{ltb:perturbation:8}) - (\ref{ltb:perturbation:10}) just once to fix proper initial values. \\
\end{comment}

%***************************************************************************************************************************
\section{Initial conditions} 
\label{ltb:initial}
%***************************************************************************************************************************

As we are dealing with a system of second order differential equations in time, we have to specify initial conditions for the variables themselves and their first time derivatives on an initial hypersurface $t=t_\mathrm{ini}$. Overall, we assume vanishing initial rates for the perturbation variables and specify the initial states as radial profiles. In accordance with the standard inflationary paradigm, we assume the early universe to be spatially homogeneous and therefore apply standard FLRW perturbation theory. Hence, initial density seeds for structure formation in the matter-dominated era are described by the power spectrum

\begin{equation}
 P_\delta(k, a) = \frac{D_+^2(a)}{D_+^2(a_\mathrm{ini})} P_\mathrm{ini}(k) T^2(k)
 \label{ltb:initial:1}
\end{equation}

 where $T(k)$ denotes the transfer function of Bardeen et al. (1986) (\cite{bardeen_statistics_1986}). By Poisson's equation, the power spectrum of the Bardeen potential $\Psi$ (\cite{bardeen_gauge-invariant_1980}) yields

 \begin{equation}
 P_\Psi(k, a) = \frac{9}{4} \frac{\Omega_{m,0}^2 H_0^4}{ a^2 k^4} P_\delta(k,a). 
 \label{ltb:initial:2}
\end{equation}

We fix an exemplary initial hypersurface of constant time $t_\mathrm{ini}$ that intersects the LTB past lightcone at redshift $z_\mathrm{ini}=100$. At $t=t_\mathrm{ini}$, we draw a 3d realisation of the Gaussian random field describing the scalar Bardeen potential in a cube of scale $L$ with the following properties:

\begin{enumerate}
 \item $\langle \Psi \rangle = 0$ in configuration space. 
 \item the variance in configuration space is given by the potential power spectrum
 \item the cube is filled in an Hermitian conjugate way: $\Psi^*(k_x, k_y, k_z) = \Psi(-k_x, -k_y, -k_z)$
\end{enumerate}

Precisely, the variance in Fourier space is given by $\hat{\sigma}^2(k) = P_\Psi(k)/L^3$ as we have to take the dimensionless power spectrum into account. In comparison to previous applications in \cite{bertschinger_multiscale_2001}, we assume a prefactor $1/(2\pi)^3$ in the Fourier convention and define the power spectrum as $\langle \Psi(\vec{k}) \Psi(\vec{k}')^\ast\rangle = (2\pi)^3 P_\Psi(k) \delta^{(3)}_D(\vec{k} + \vec{k}')$. In fact, this yields the prefactor $1/L^3$ instead of $(2\pi)^3/L^3$ obtained in \cite{bertschinger_multiscale_2001}. \\

\begin{figure}
 \centering
 \includegraphics[width=0.6\hsize]{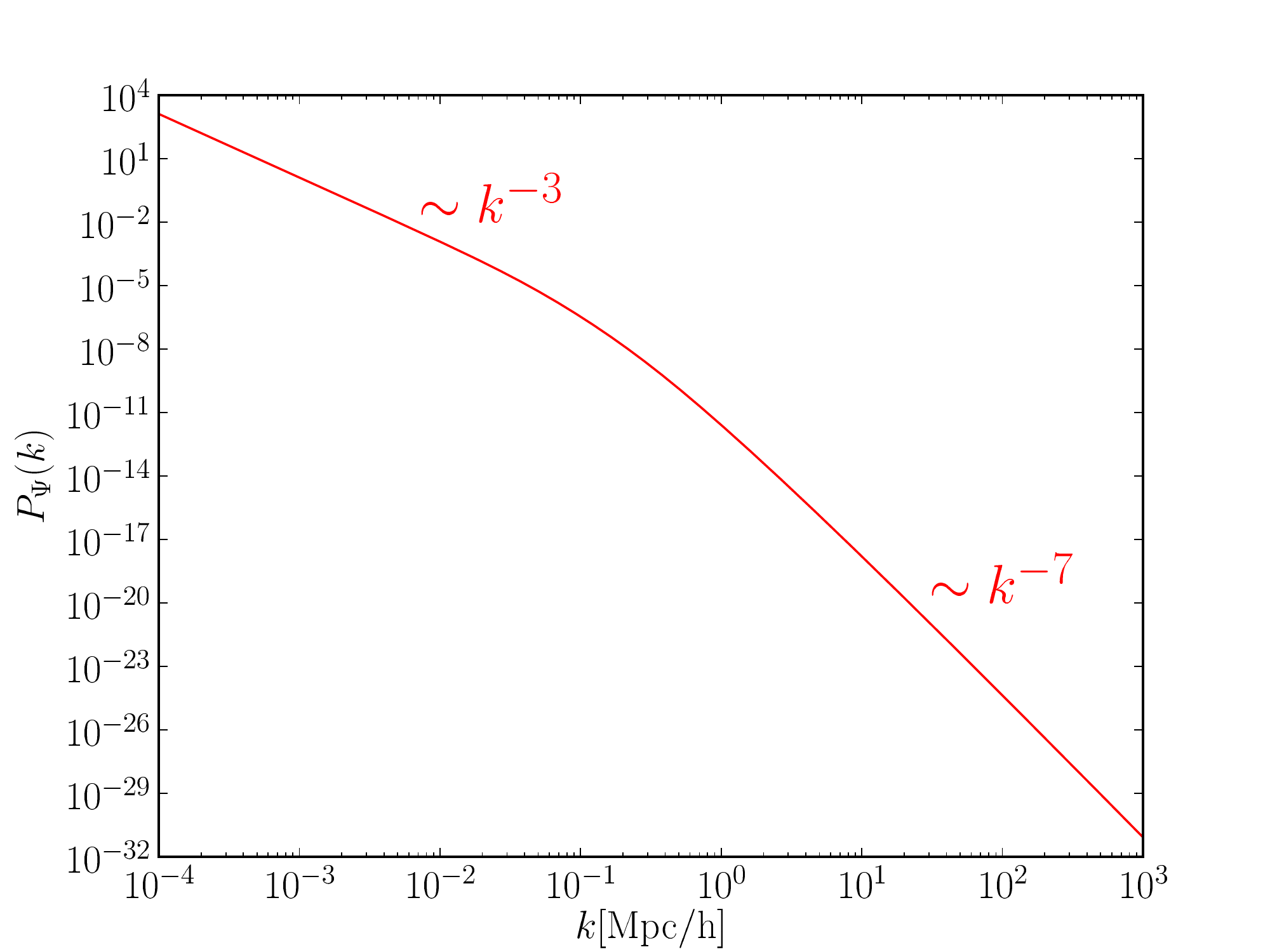}
  \caption{Initial potential power spectrum according to Eq. (\ref{ltb:initial:2}). Due to the asymptotic behaviour of $k^{-7}$ on small scales, the resulting finite sample of the power spectrum is very smooth compared to the directization.}
  \label{ltb:initial:fig:1}
\end{figure}

\begin{figure}
 \subfigure[]{\includegraphics[width=0.5\hsize]{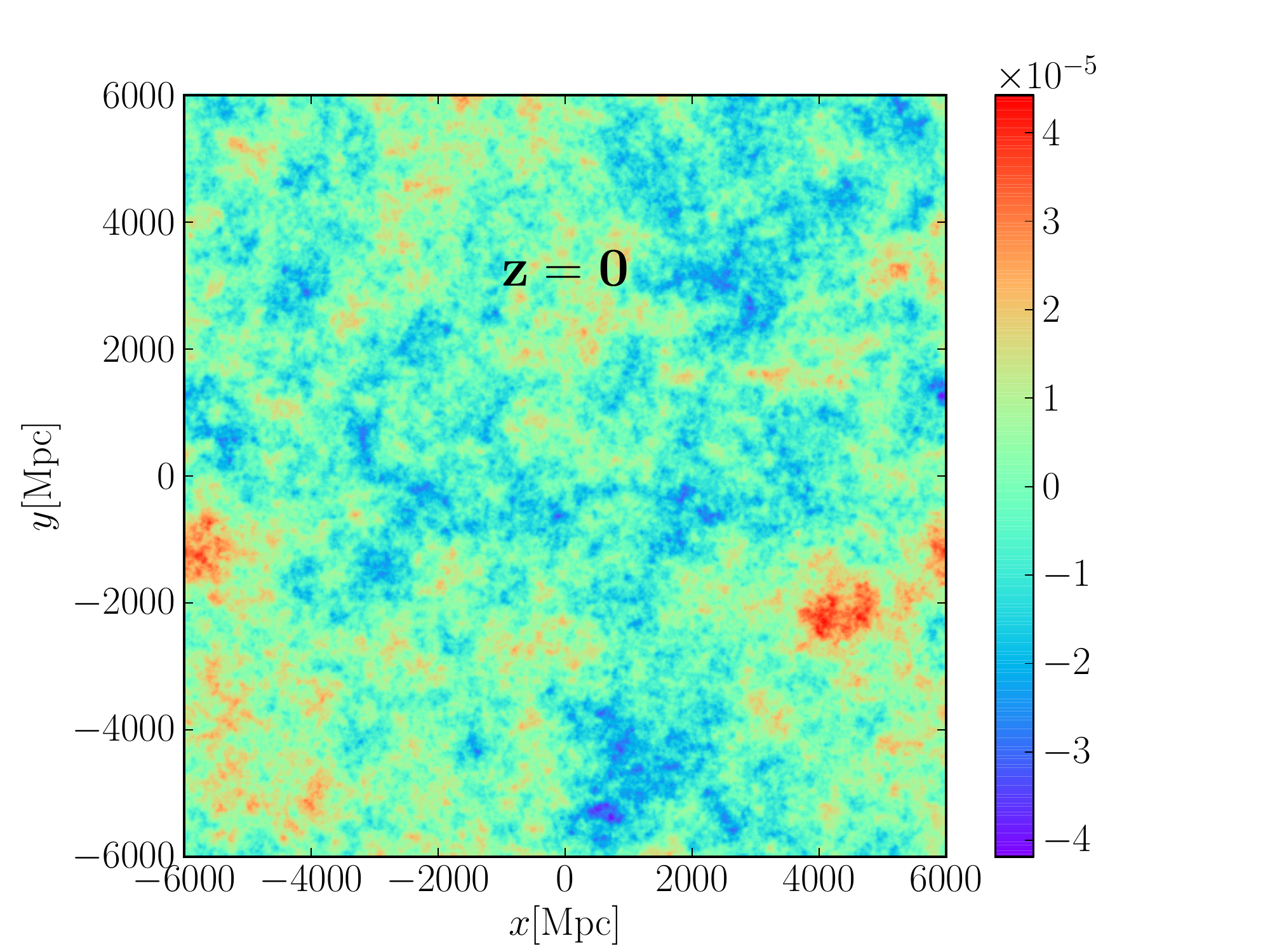}}\hfill
 \subfigure[]{\includegraphics[width=0.5\hsize]{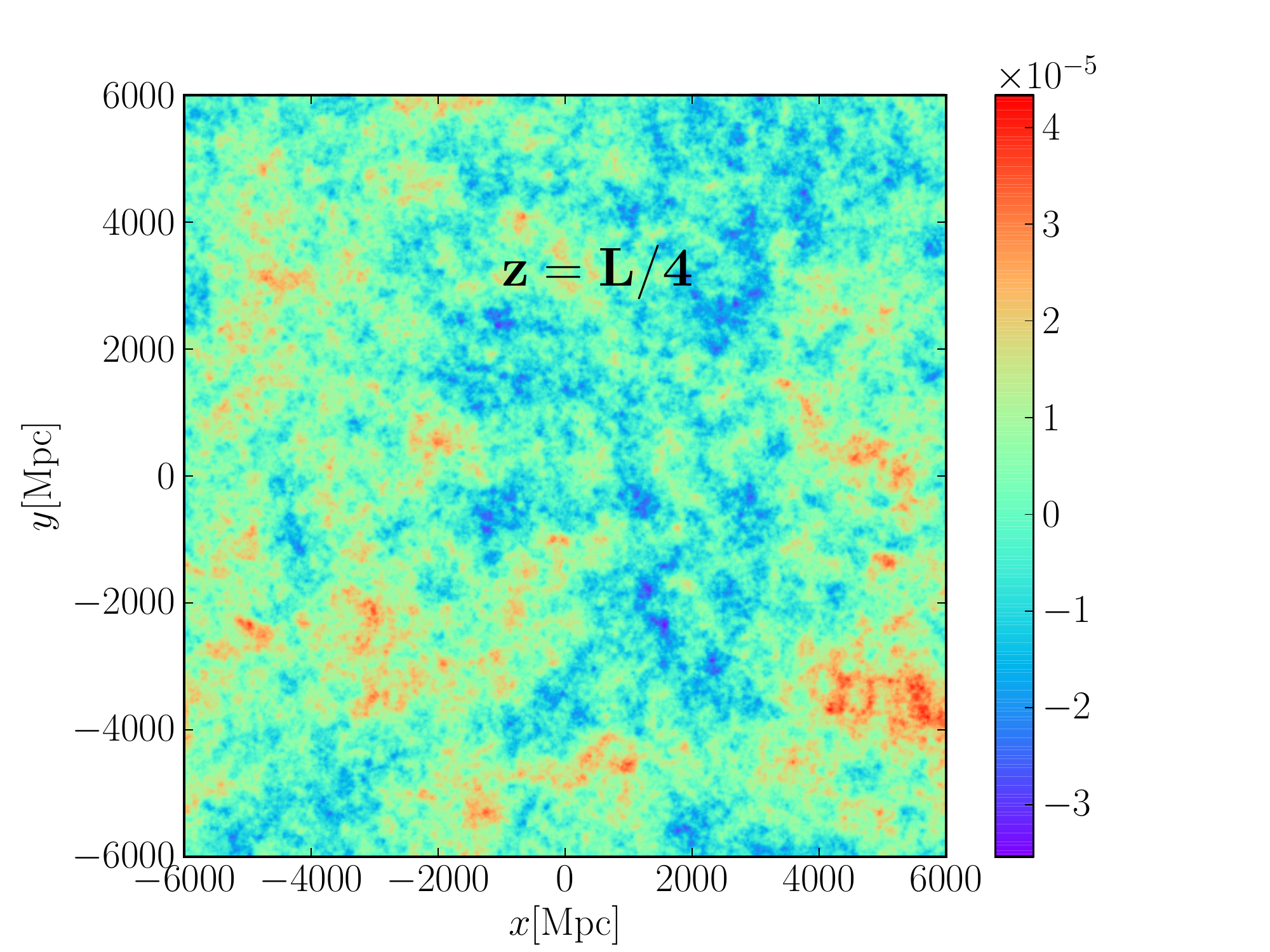}} \\
 \subfigure[]{\includegraphics[width=0.5\hsize]{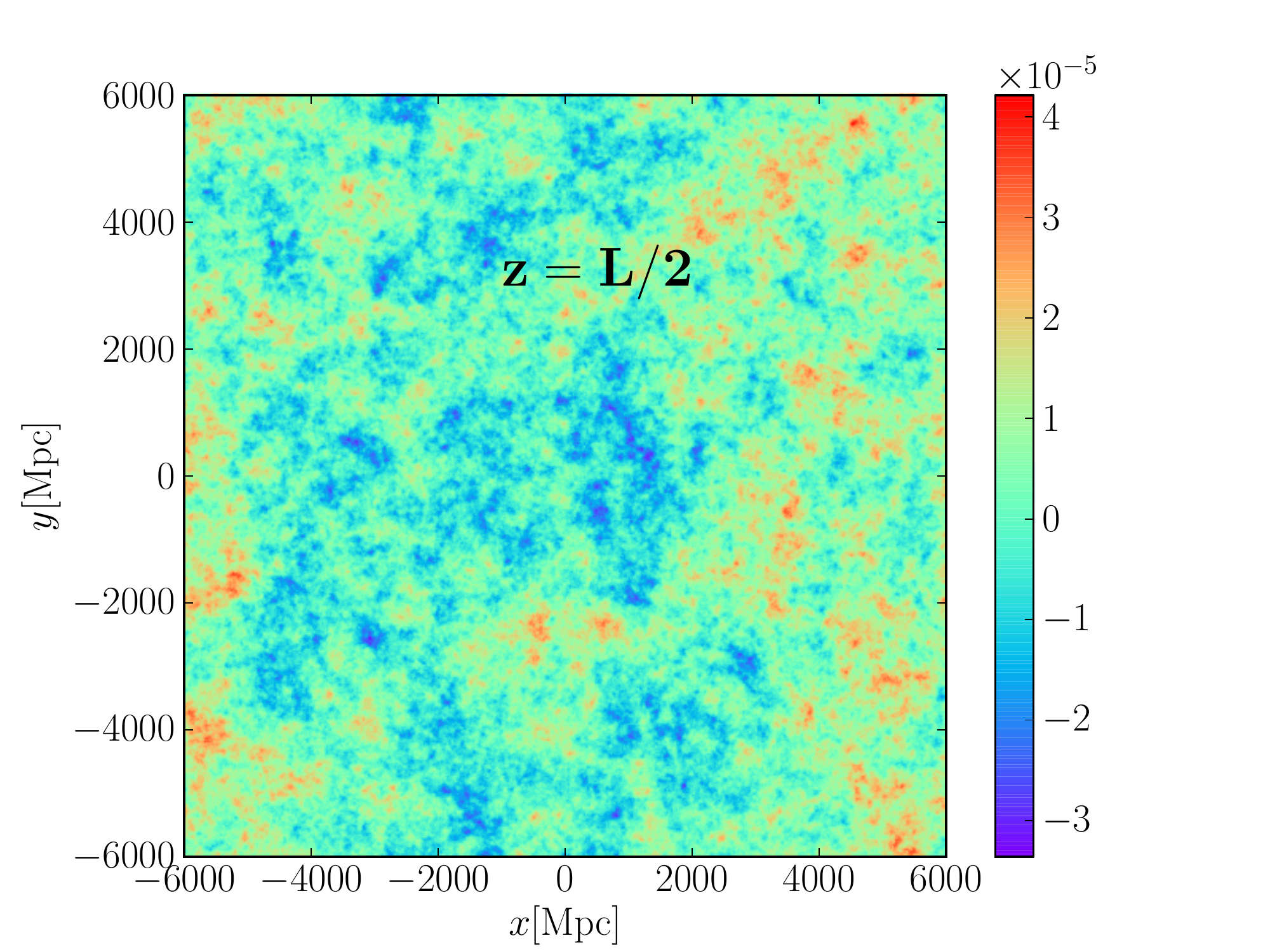}} \hfill
 \subfigure[]{\includegraphics[width=0.5\hsize]{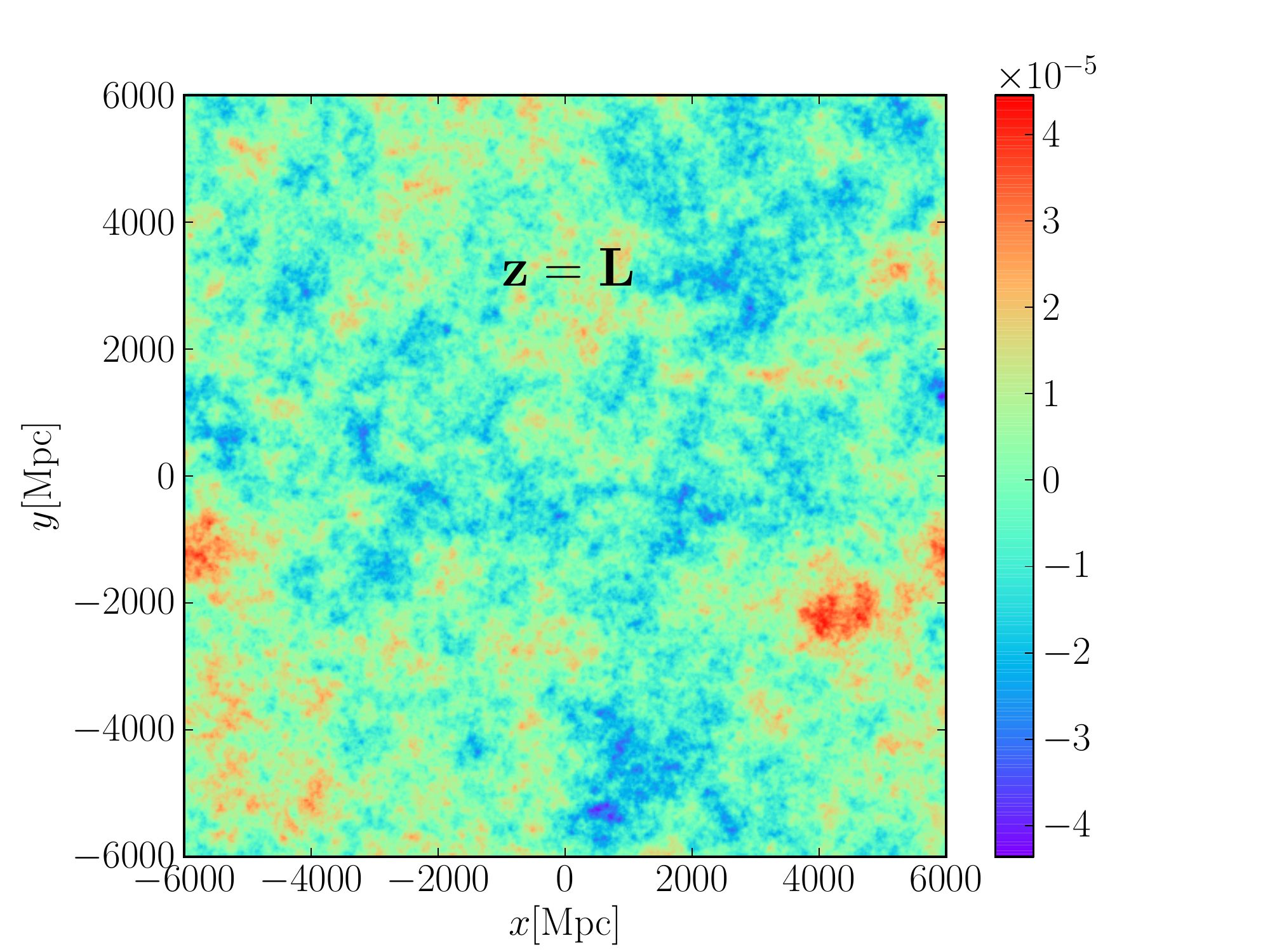}}
 \caption{2d slices of the 3d Gaussian realisation of the Bardeen potential fluctuations. In Cartesian coordinates, the xy surface is plotted for $z=0, z=L/4, z = L/2, \ \text{and} \ z = L$ in the corresponding subfigures (a), (b), (c), (d). We use $N_\mathrm{nodes} = 1024^3$ with full spatial extension of $L=12$ Gpc/h. The obvious similarity of (a) and (d) is caused by periodic boundary conditions applied by the Fourier transform. }
 \label{ltb:initial:fig:2}
\end{figure}

\begin{figure}
 \subfigure[]{\includegraphics[width=0.5\hsize]{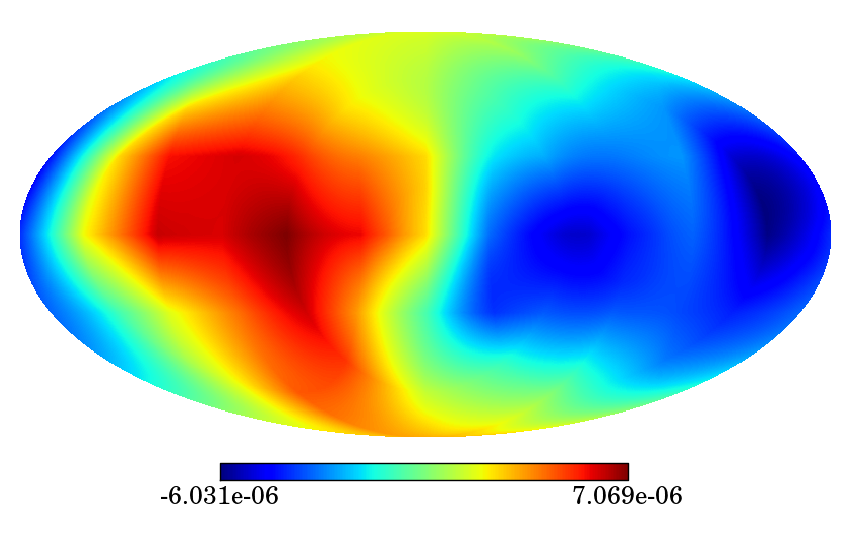}}\hfill
 \subfigure[]{\includegraphics[width=0.5\hsize]{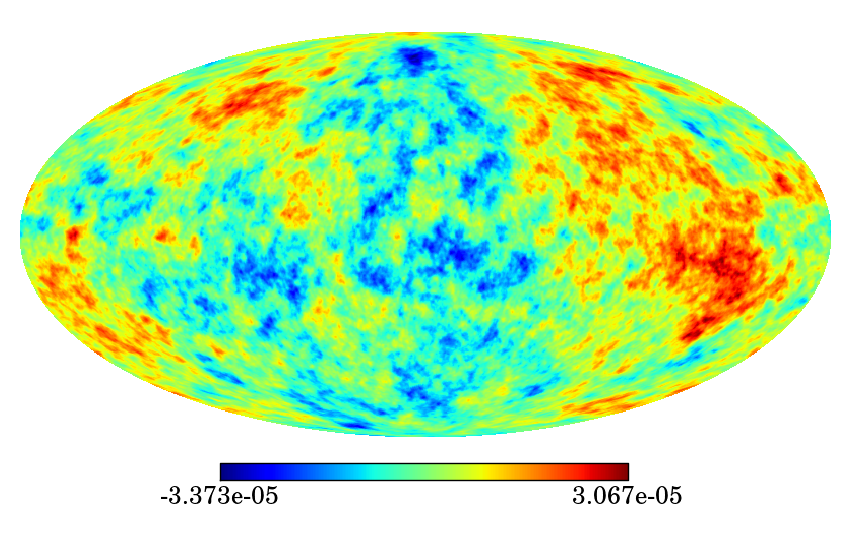}} \\
 \subfigure[]{\includegraphics[width=0.5\hsize]{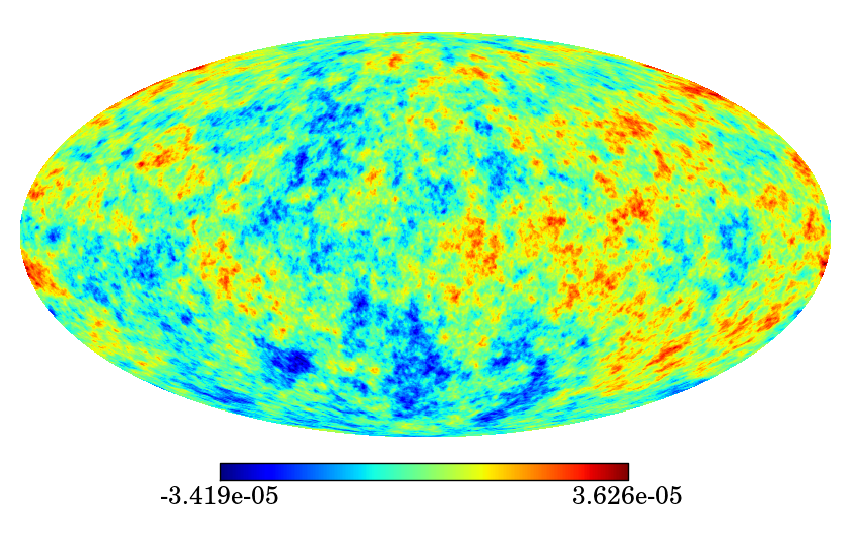}}\hfill
 \subfigure[]{\includegraphics[width=0.5\hsize]{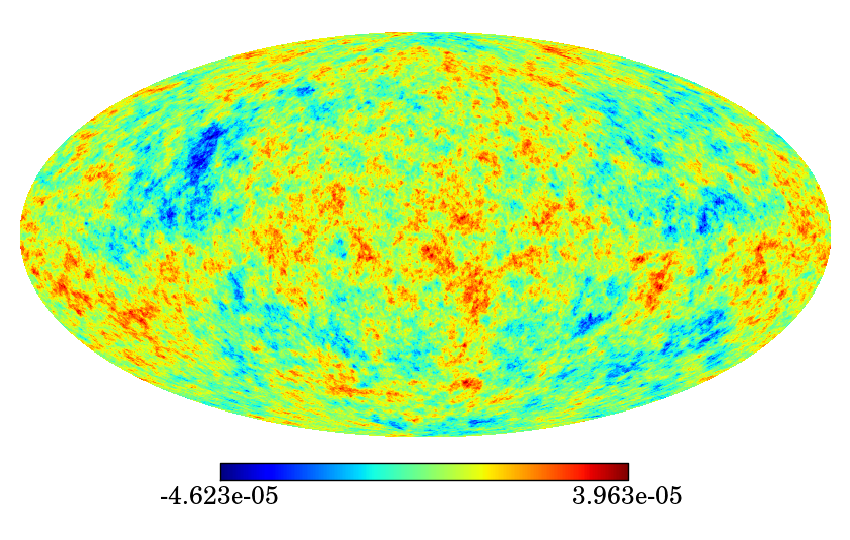}} \\
 \caption{ Healpix maps in Mollweide projection at different radii $r=r_\mathrm{min}$ (a), $r=r_\mathrm{max}/4$ (b), $r=r_\mathrm{max}/2$ (c), and  $r=r_\mathrm{max}= L/2$ (d). By construction of the Healpix scheme (see \cite{gorski_healpix:_2011}), Healpix maps contain $(12 N_\mathrm{side})^2$ pixel with $N_\mathrm{side} = 512$ in this case. We apparently see fluctuations on smaller scales for spheres at increasing radius, since the spherical surface increases with the square of the sphere's radius. }
 \label{ltb:initial:fig:3}
\end{figure}

 Transforming back to real space yields a discretized (real-valued) potential with node-values $\Psi(x_i, y_j, z_k)$ in the 3d cube. Since the potential power spectrum behaves like $k^{-7}$ on small scales, the resulting field is very smooth compared to the discretization scale. Four 2d slices of the 3d initial realisation are shown in Fig. (\ref{ltb:initial:fig:2}). 
 
 \begin{comment}
 \textbf{The sampled potential fluctuations produce, by construction, density fluctuations of the correct amplitude as initial density power spectrum (Eq. (\ref{ltb:initial:1})) entering in Eq. (\ref{ltb:initial:2}) is normalised with an appropriate normalisation factor $\sigma_8$.} (\textit{this comment might be  }) \\
 \end{comment}
  
We can draw concentric spheres around an arbitrary centre and discretize each of them according to the Healpix pixelisation scheme described in \cite{gorski_healpix:_2011}. This scheme allows fast and efficient numerical integration on $\mathcal{S}^2$ needed for spherical harmonic decomposition

\begin{equation}
 \Psi( t_\mathrm{ini}, r ,\theta, \phi) = \sum_{(\ell, m)} \Psi\lm (t_\mathrm{ini},r)Y\lm(\theta, \phi)
\label{ltb:initial:6}
\end{equation}

with spherical harmonic coefficients estimated by quadratures on the two sphere:

\begin{equation}
 \begin{split}
   \Psi\lm(t_\mathrm{ini}, r) &= \int_\Omega { \d\Omega \ \Psi( t_\mathrm{ini}, r ,\theta, \phi) (Y\lm)^*(\theta, \phi) } \\
                              &\approx \frac{4\pi}{N_\mathrm{pix}} \sum_{p=0}^{N_\mathrm{pix}}{ \Psi( t_\mathrm{ini}, r ,\theta_p, \phi_p) (Y\lm)^*(\theta_p, \phi_p) }  
 \end{split}
\label{ltb:initial:7}
\end{equation}

By placing concentric discretized spheres (Healpix maps) at equidistant radial positions in the 3d cube, we finally obtain initial radial profiles for spherical harmonic coefficients of the Bardeen potential $\Psi\lm(t_\mathrm{ini},r)$ for each angular scale $\ell$ and orientation $m$. These profiles will be called \textit{coefficient profiles} in the following. We have to mention here that we effectively measure proper distances $d_p(t_\mathrm{ini},r)$ between the centre and the spherical surfaces that do not reduce to the LTB radial coordinate in general. However, assuming the void depth to be negligible at $t=t_\mathrm{ini}$, we can write (see also \cite{alonso_large_2010, february_galaxy_2013}):

\begin{equation}
 \begin{split}
  d_p(t_\mathrm{ini},r) = a(t_\mathrm{ini}) \cdot r_\mathrm{FLRW} &= \int_0^{r_\mathrm{LTB}} { \frac{a\p(t_\mathrm{ini}, r) }{\sqrt{1-\kappa(r)r^2}} \d r } \\
                                                            &\approx a\o(t_\mathrm{ini}, r_\mathrm{LTB}) \cdot r_\mathrm{LTB}  \\
                                                            &\approx  a(t_\mathrm{ini}) \cdot r_\mathrm{LTB}
 \end{split}
 \label{ltb:initial:8}
\end{equation}

In this particular limit, we can therefore safely identify the LTB and FLRW radial coordinates. \\

Fig. (\ref{ltb:initial:fig:3}) shows examples of Healpix maps at increasing radii using Mollweide projection. As the spheres grow in radius, angular scales of fluctuations seem to decrease due to larger arc lengths with increasing radii. We want to point out at this stage that, in principle, the method can be extended to vector and tensor perturbations once a proper sampling technique is available. The Bardeen potential power spectrum itself can also be replaced correspondingly, if, for example, the primordial curvature perturbation (as applied in \cite{february_galaxy_2013}) turns out to be a more realistic representation of potential fluctuations in the matter dominated FLRW era. \\

\begin{figure}
 \includegraphics[width = 16cm]{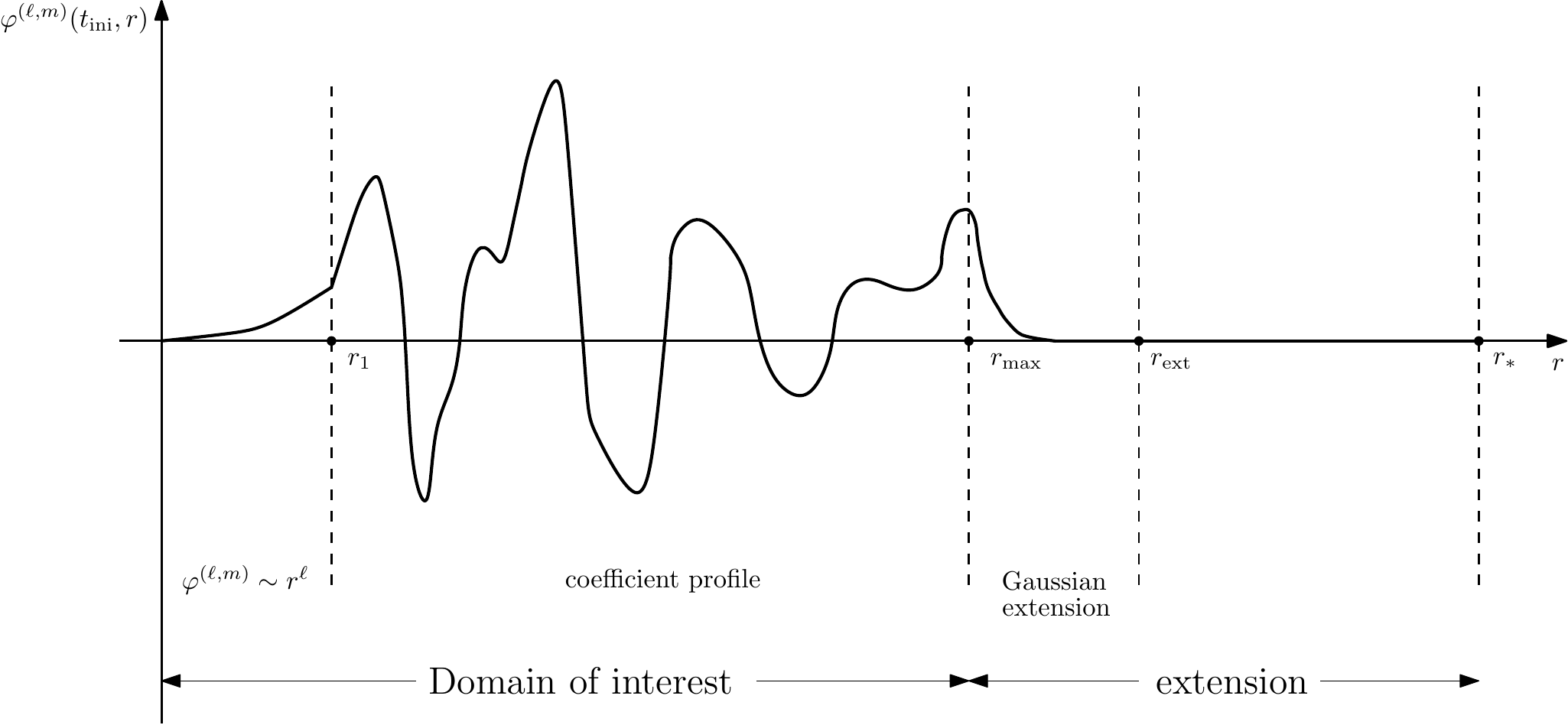}
 \caption{Schematic summary of all modifications performed on the initial profiles. At small radii, we require the regularity condition to hold which is then matched continuously to the spherical harmonic coefficient profiles extracted from the Healpix maps. In order to make the profile dropping to zero in a smooth way, a Gaussian extension is applied such that a vanishing initial profile is achieved in the extension region.  }
 \label{ltb:initial:fig:4}
\end{figure}

As shown by CCF in \cite{clarkson_perturbation_2009}, the only remaining gauge-invariant polar perturbation in the FLRW limit with initial scalar perturbations is $\varphi\lm = -2 \Psi\lm$  while vector and tensor contributions vanish. This turns out to be a very simple initial configuration for our case study.  Nonetheless, initial coefficient profiles for LTB gauge-invariants have to obey certain regularity properties at the radial origin and need to be slightly adapted to the grid structure applied. We therefore have to modify and extend the profiles artificially in the following way (see also Fig. (\ref{ltb:initial:fig:4}) for an illustration):

\begin{enumerate}
\item As shown by GMG in \cite{gundlach_gauge-invariant_2000}, initial profiles have to obey $\varphi\lm \sim r^\ell$ close to $r=0$ in order to be regular solutions. We therefore modify the profile at small radii (typically up to the first radial bin) to obey this functional shape. We then match this region continuously to the exterior profile.
\item The boundary of the domain of interest $r_\mathrm{max}$ and the numerical boundary $r\st$ are different, since the numerical boundary is constructed to be causally disconnected from $r_\mathrm{max}$. In the region $[r_\mathrm{max}, r\st ]$, all perturbation variables have to be zero initially such that no propagating mode can be excited there that influences the results in the domain of interest. 
\item In order to have a smooth transition, we define a transition region $[r_\mathrm{max}, r_\mathrm{ext}]$ where the initial profile is extended by a Gaussian function centered at $r_\mathrm{max}$ and a FWHM of one fifth of the size of the extension region.
\item In order to prevent shot noise due to the pixel size, we smooth the coefficient profile by a smoothing scale of one pixel diagonal corresponding to roughly $20 \mathrm{Mpc/h}$.   
\end{enumerate}

The smoothing- and extension scales are variable quantities and artificial parameters of the initial profiles. Their influence of the final results will be probed and quantified below. \\

Summing up, we have overall initial conditions  

\begin{equation}
  \begin{split}
   &\varphi\lm(t_\mathrm{ini},r) = -2 \Psi\lm(t_\mathrm{ini},r) \\
   &\chi\lm(t_\mathrm{ini},r) = \varsigma\lm(t_\mathrm{ini},r) = 0 \\
   &\dot{\chi}\lm(t_\mathrm{ini},r) = \dot{\varphi}\lm(t_\mathrm{ini},r) =\dot{\varsigma}\lm(t_\mathrm{ini},r) = 0 
  \end{split}
  \label{ltb:initial:9}
\end{equation}

for the metric perturbations which respect the grid structure and boundary conditions posed by the problem itself. The initial fluid perturbations ($\Delta\lm, w\lm, v\lm  $) are then constrained by Eqs. (\ref{ltb:perturbation:8}) - (\ref{ltb:perturbation:10}). \\

We should mention at this point that we assume a vanishing initial rate for the Bardeen potential, because the LTB patch is asymptotically embedded into an EdS model. Hence, $\Psi$ does not depend on time in this limit. Certainly, the setup is inconsistent for a general initial FLRW model and needs to be generalized to void models with different asymptotic limits.  \\

\begin{figure}
 \subfigure[]{\includegraphics[width=0.5\hsize]{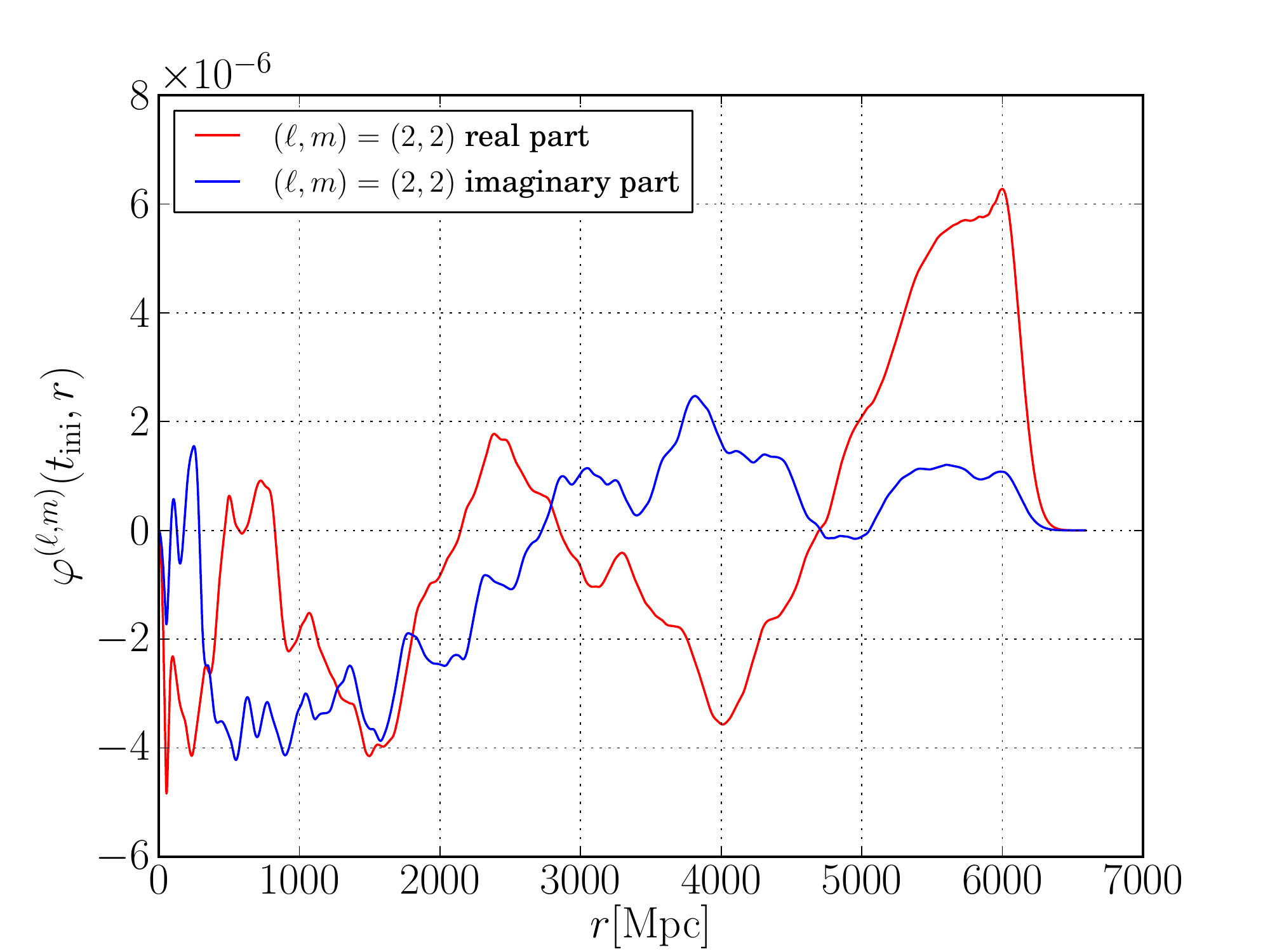}}\hfill
 \subfigure[]{\includegraphics[width=0.5\hsize]{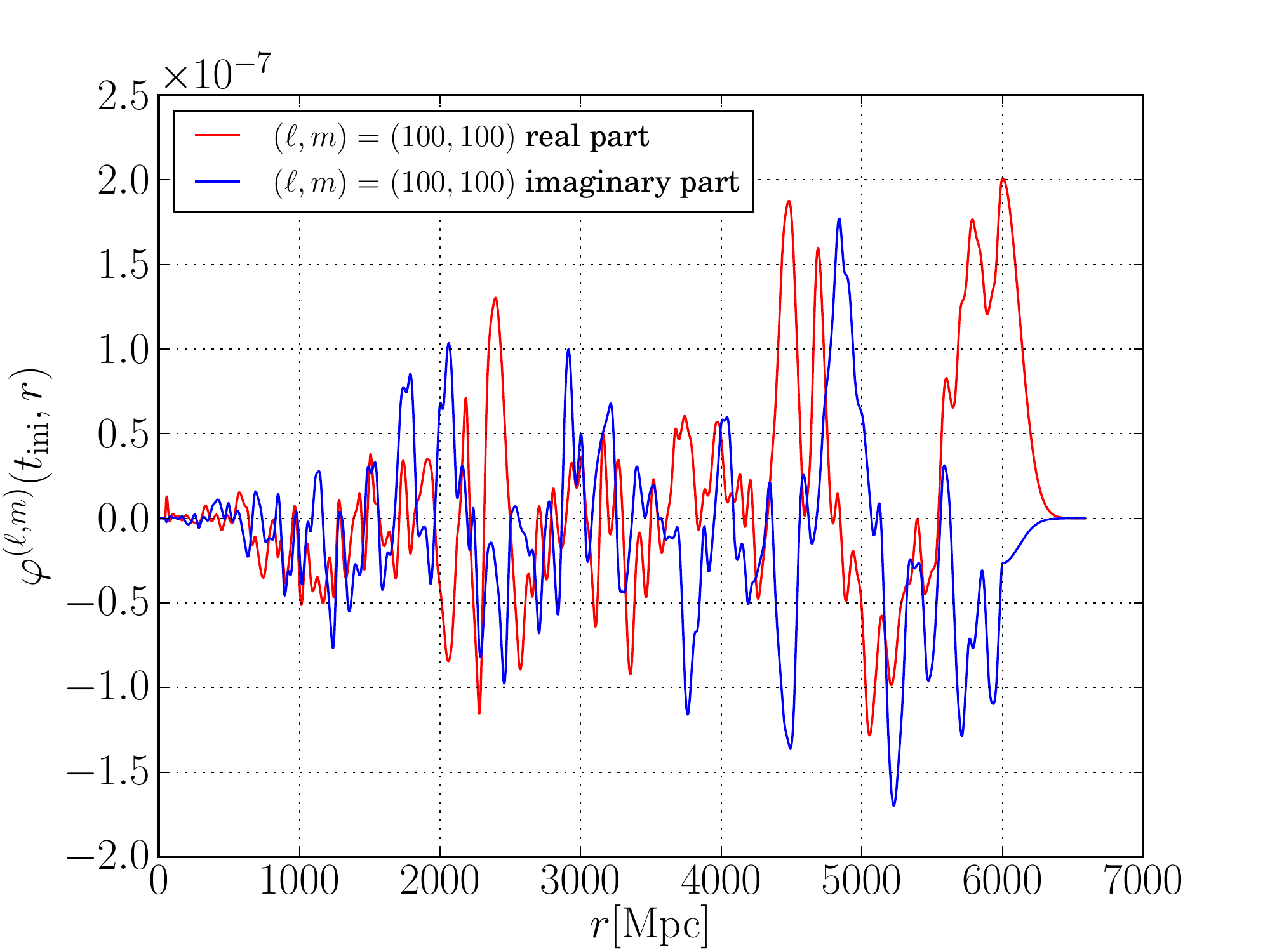}} \\
 \subfigure[]{\includegraphics[width=0.5\hsize]{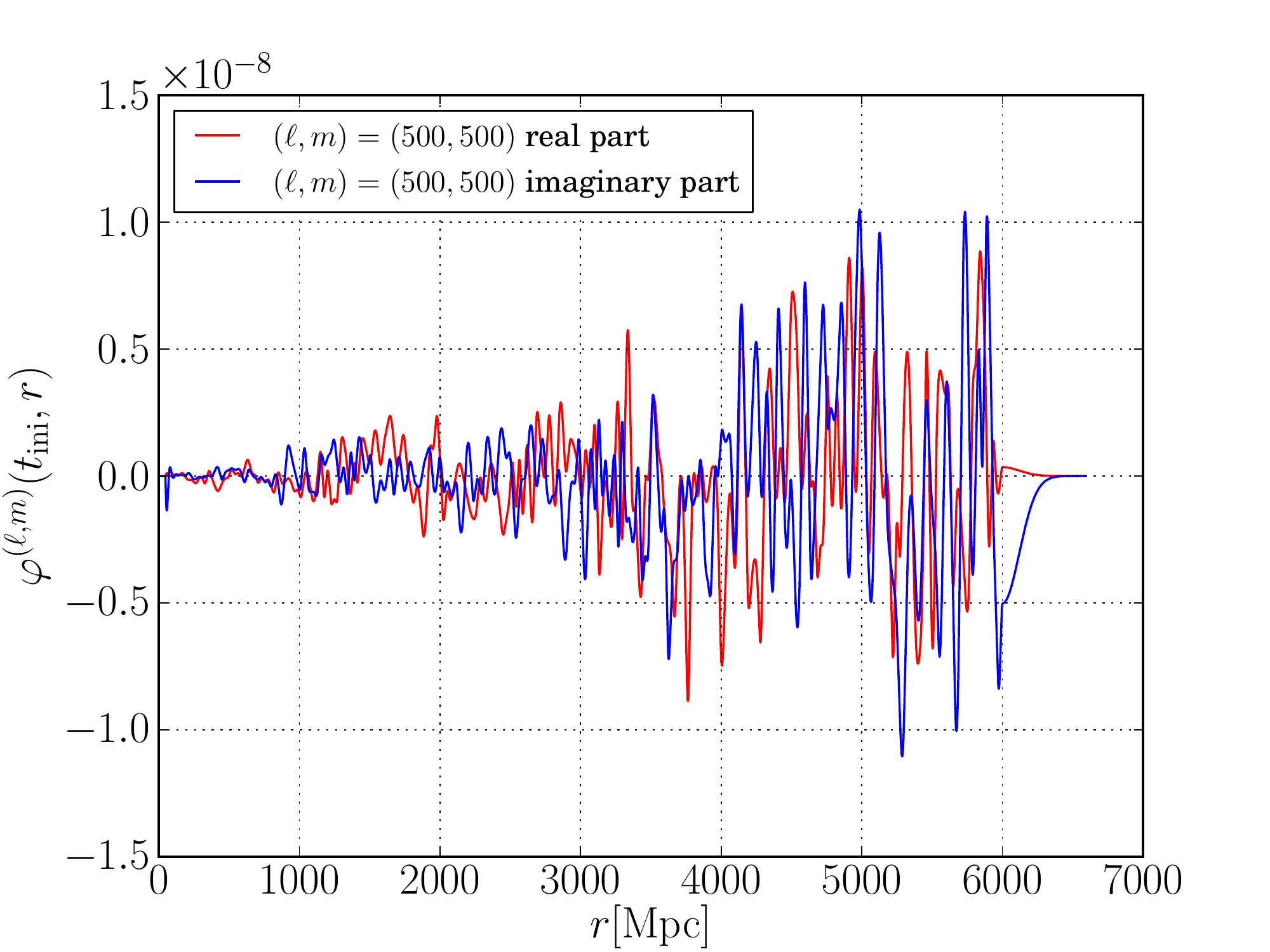}}\hfill
 \subfigure[]{\includegraphics[width=0.5\hsize]{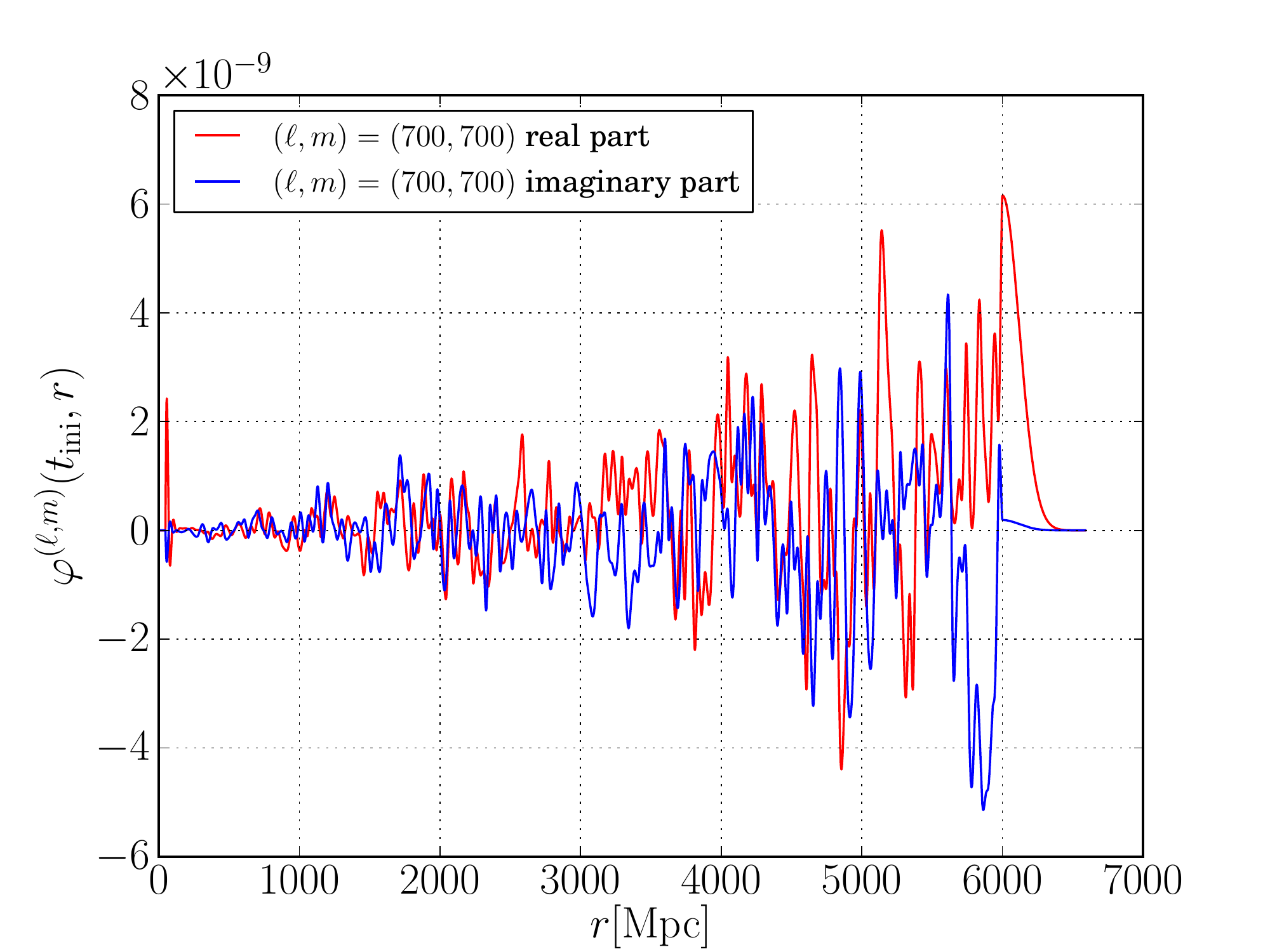}} \\
 \caption{Examples for initial radial profiles for  $\varphi\lm$ for various $\ell$- and $m$-modes. The amplitude of decreasing angular scales (increasing $\ell$-modes) drops by almost three orders of magnitude. This is due to the steepness of the initial potential power spectrum ($P_\Psi(k) \sim k^{-7}$) on small scales that causes high $\ell$-modes to be suppressed. }
 \label{ltb:initial:fig:5}
 \end{figure}

Examples of radial profiles of the spherical harmonic coefficients are shown in Fig. (\ref{ltb:initial:fig:5}). Due to the steepness of the initial potential power spectrum in Eq. (\ref{ltb:initial:2}), small angular scales are strongly suppressed leading to a decrease of coefficient amplitudes of nearly four orders of magnitude from $\ell=2$ to $\ell=1000$. In addition, Healpix maps at small radii do not contain information on small angular scales. This is an effect due to the discretization of the initial profiles which cannot be avoided by our sampling technique.  This becomes problematic in case of small redshifts on the LTB lightcone and will be discussed in detail below.

%***************************************************************************************************************************
\section{Results} 
\label{ltb:results}
%***************************************************************************************************************************

We tested the performance of our numerical scheme in mainly two situations. First, we tried to reproduce the results of February et al. (2014) obtained in \cite{february_evolution_2014}. As a second step, we take realistically sampled initial conditions into account and compute angular power spectra that allow us to extract statistical properties of the evolution of perturbations in LTB spacetimes. We finally try to characterise coupling strengths as functions of void depths and sizes.     

%---------------------------------------------------------------------------------------------------------------------------
\subsection{Reproducing previous results in the field} 
\label{ltb:comparison}
%---------------------------------------------------------------------------------------------------------------------------
 
February et al. (2014) (\cite{february_evolution_2014}) set up a numerical solution to the evolution equations for a simple test case by initializing each variable separately with five Gaussian peaks probing the behavior of the solution at different positions in the void. Starting from a Gaussian shaped void with $\Omega_\mathrm{in} = 0.2$ and $L = 2$ Gpc asymptotically embedded into an EdS model, initial conditions for one variable each are given by

 \begin{figure}
 \centering
 \includegraphics[width=10cm]{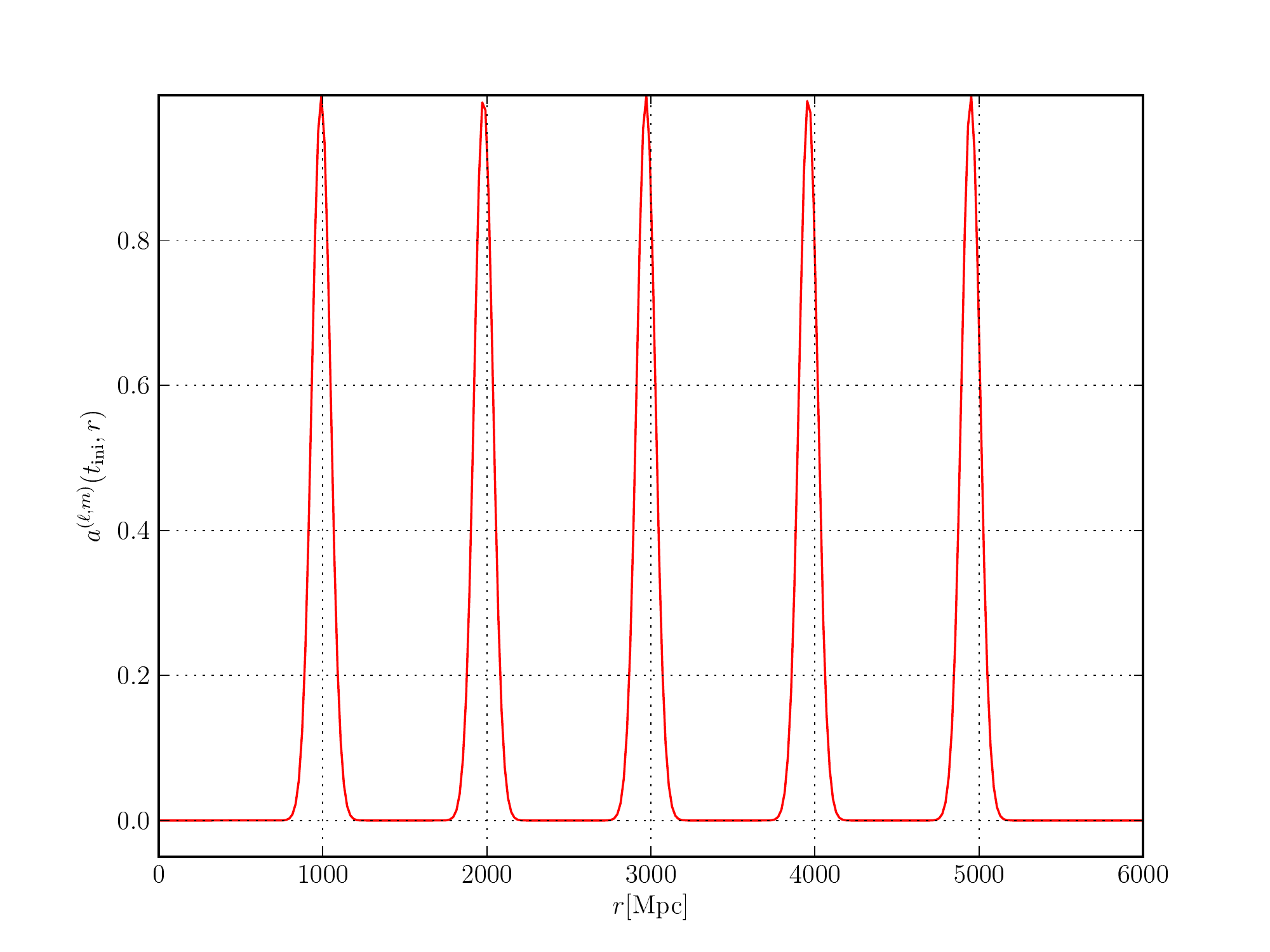}
 \caption{initial Gaussian profile as applied by February et al. (2014) (\cite{february_evolution_2014})}
 \label{ltb:comparison:fig:1}
\end{figure}

\begin{align}
 \label{ltb:comparison:1}
 a\lm(t_\mathrm{ini}, r) &= \sum_{i=1}^5 { \exp{\left( -\frac{(r-r_i)^2}{s^2} \right)} } \\
 \label{ltb:comparison:2}
 \dot{a}\lm(t_\mathrm{ini}, r) &= 0
 \end{align}

with $r_i \in 0.99 \cdot \left\{ 1,2,3,4,5\right\} \ \mathrm{Gpc}$ and $s = 0.08 \ \mathrm{Gpc}$ (see also Fig. (\ref{ltb:comparison:fig:1})). All other variables and their time derivatives are set to zero.

 \begin{figure}[ht]
  \centering
  \subfigure[Case 1: initialise $\varphi$ for $\ell=2$]{ \includegraphics[width=0.48\hsize]{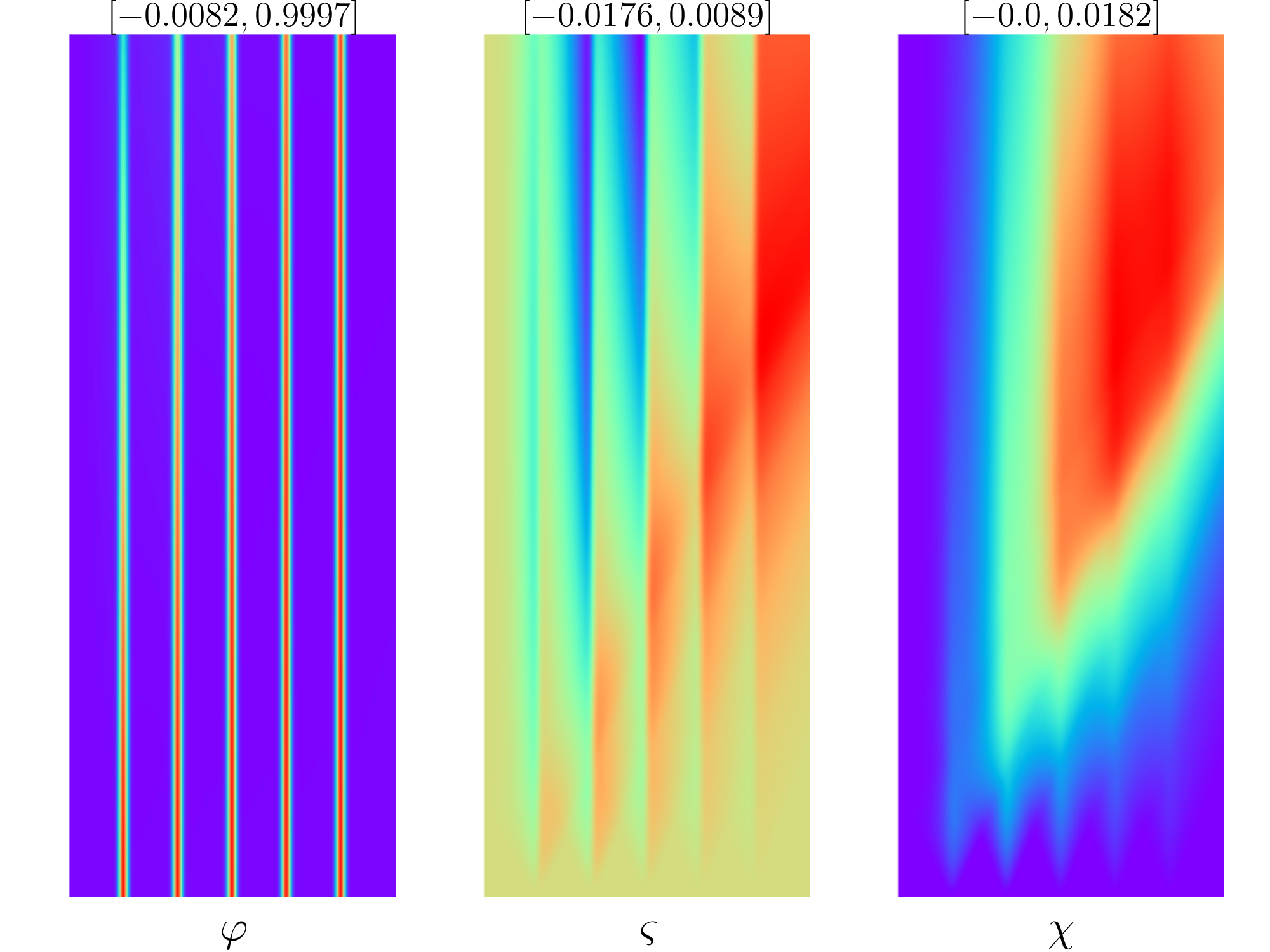} } 
  \subfigure[Case 2: initialise $\varphi$ for $\ell=10$]{ \includegraphics[width=0.48\hsize]{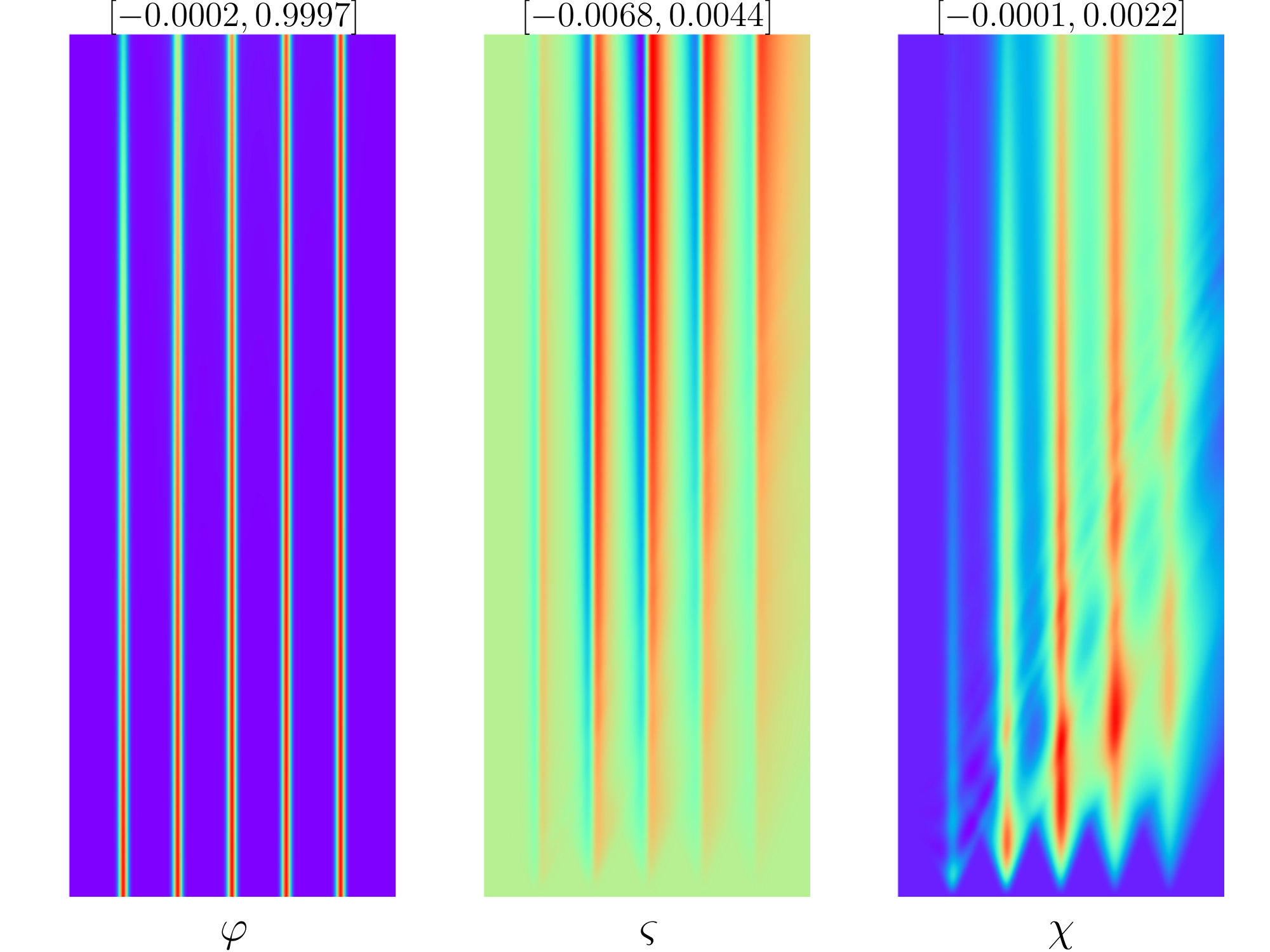} } \\
  \subfigure[Case 3: initialise $\varsigma$ for $\ell=2$]{ \includegraphics[width=0.48\hsize]{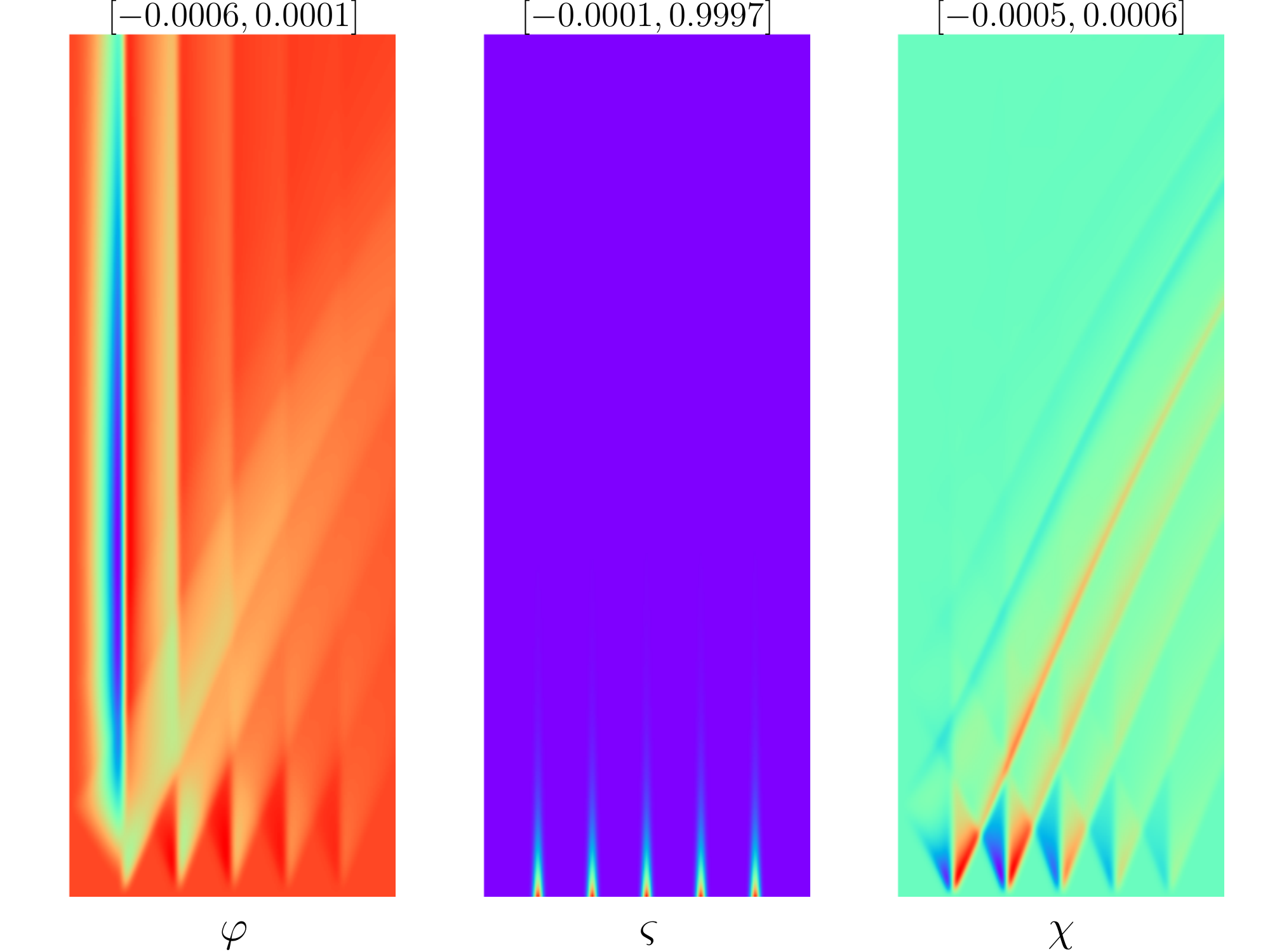} } 
  \subfigure[Case 4: initialise $\varsigma$ for $\ell=10$]{ \includegraphics[width=0.48\hsize]{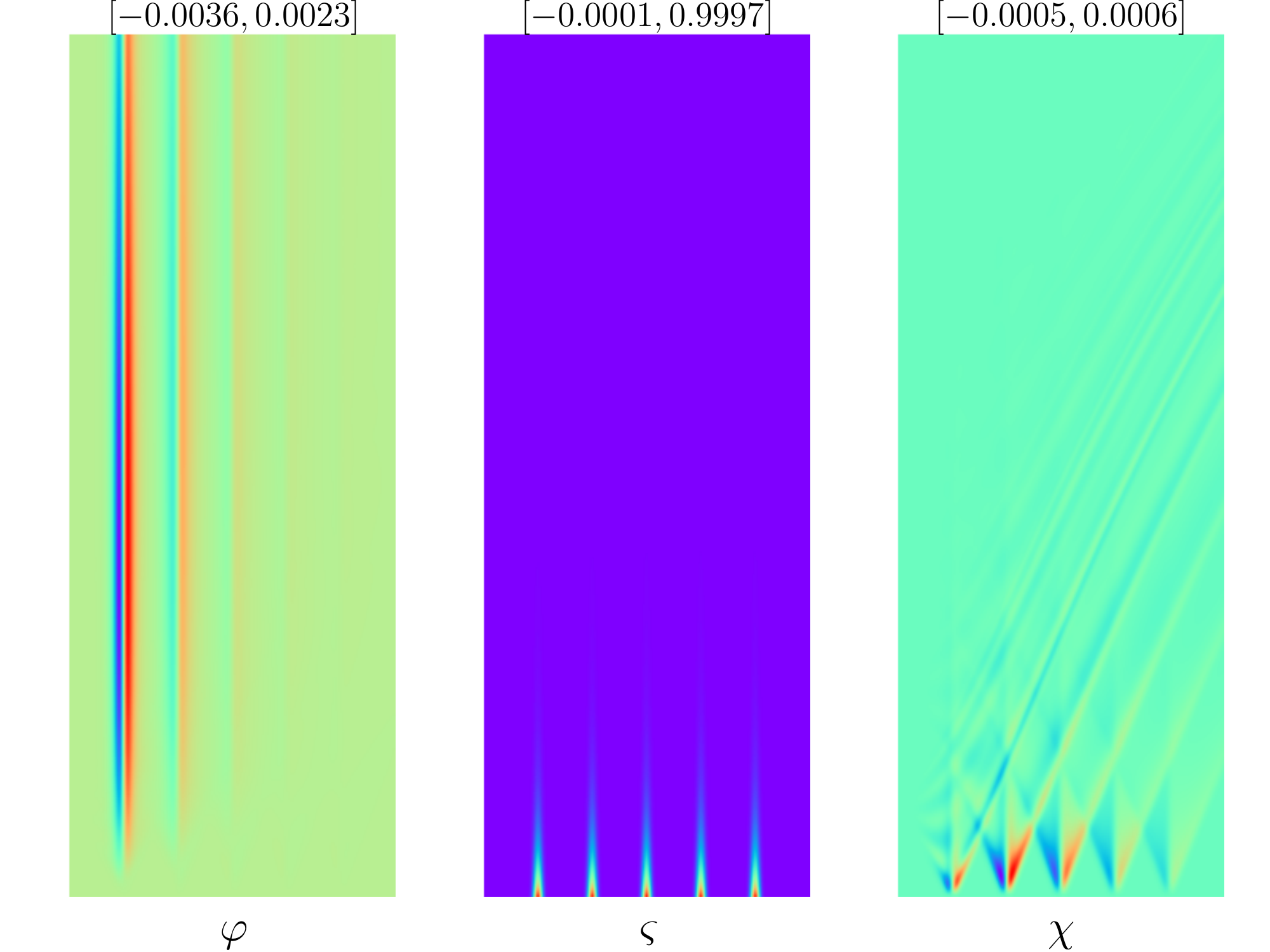} } \\
  \subfigure[Case 5: initialise $\chi$ for $\ell=2$]{ \includegraphics[width=0.48\hsize]{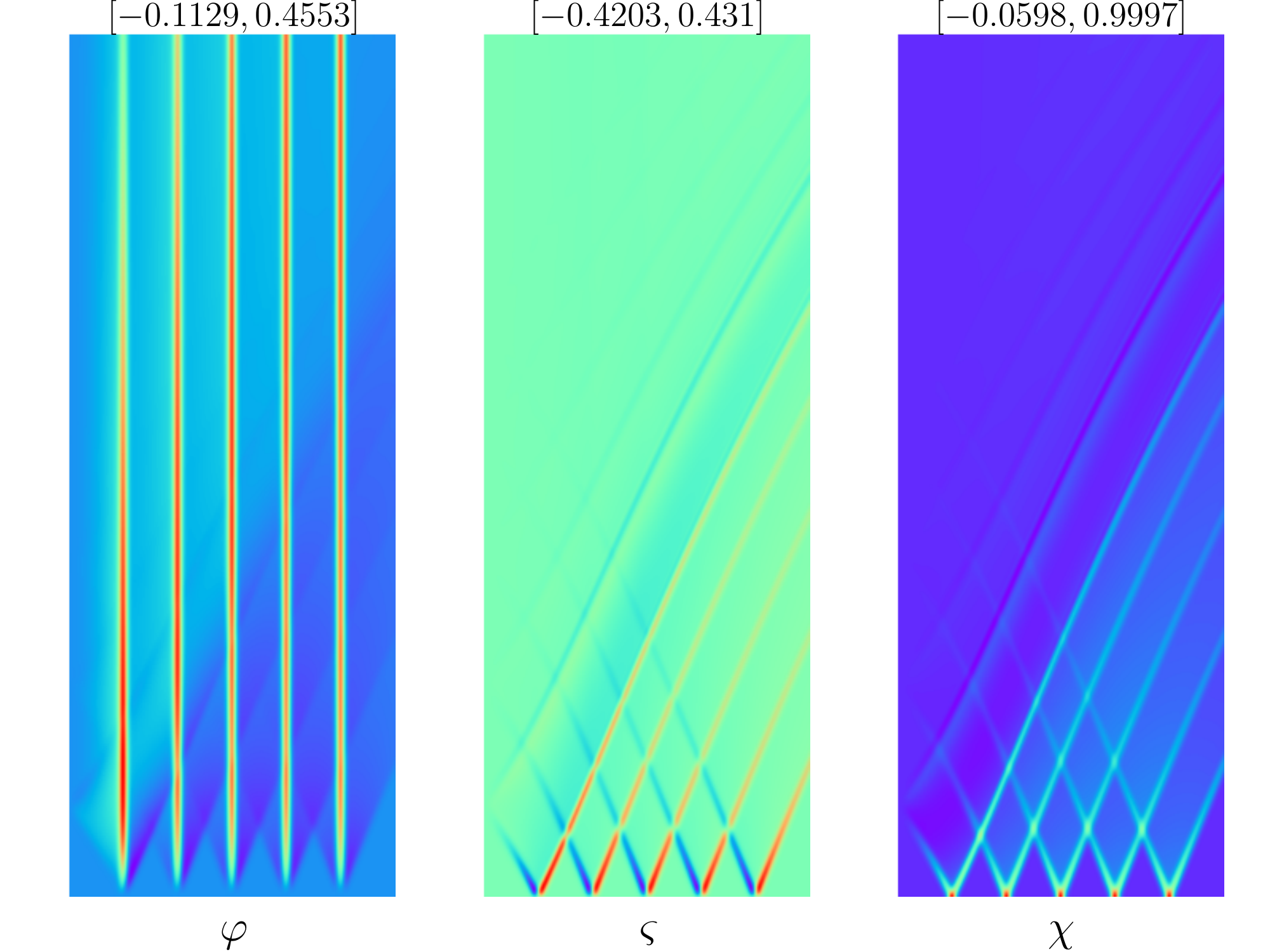} } 
  \subfigure[Case 6: initialise $\chi$ for $\ell=10$]{ \includegraphics[width=0.48\hsize]{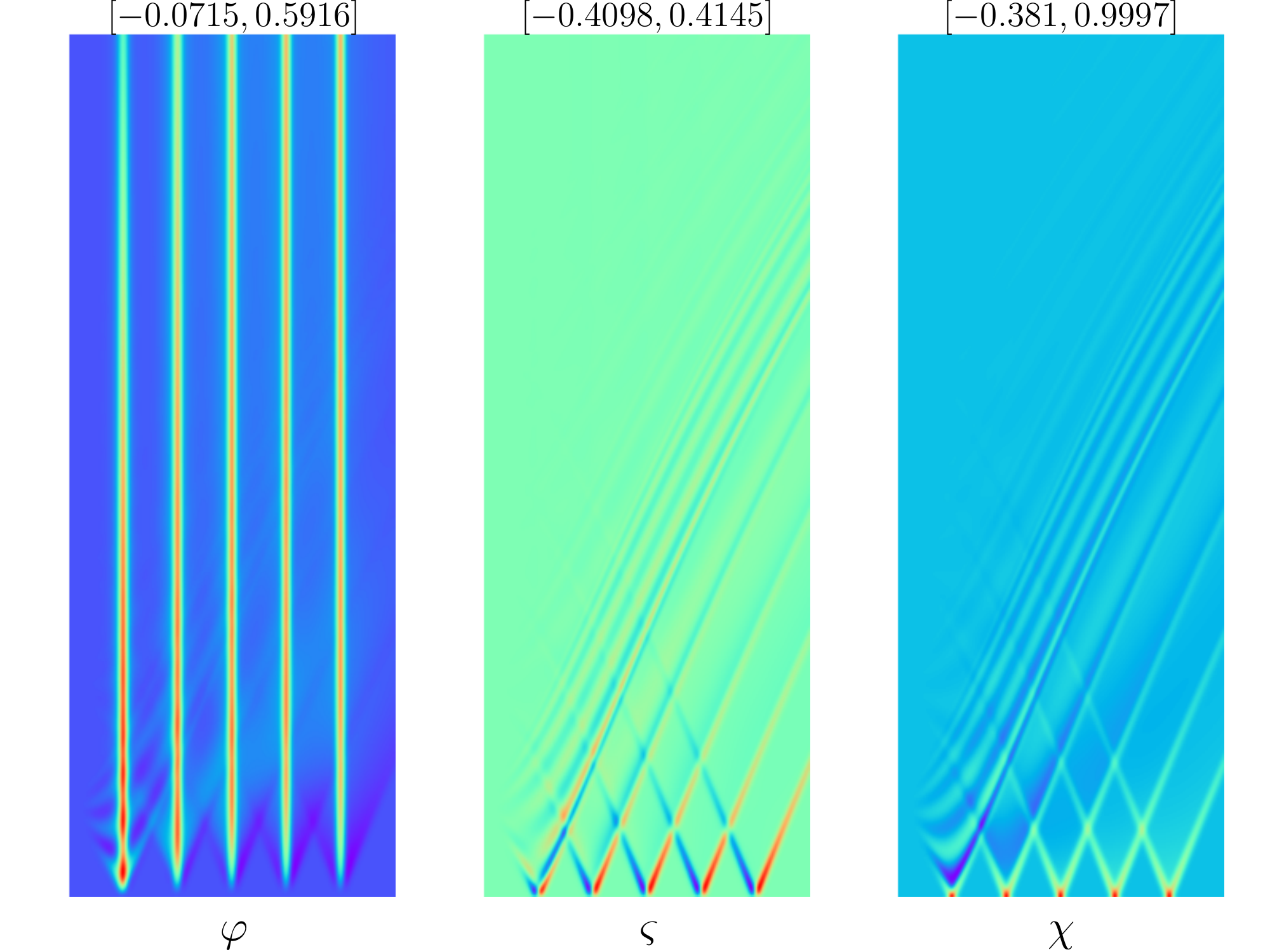} } 
  \caption{Space-time-evolution of LTB gauge-invariants $\{ \varphi, \varsigma, \chi \}$ with initial conditions according to Eqs. (\ref{ltb:comparison:1}) and (\ref{ltb:comparison:2}) for two exemplary modes $\ell=2$ and $\ell=10$. Minimum and maximum values of the full spacetime diagram are given in the headlines. We applied a Gaussian shaped void density profile $\rho(t_0,r)$ with density contrast $\Omega_\mathrm{in} = 0.2$ and $L = 2$ Gpc asymptotically embedded into an EdS model with $h=0.7$. These results have been obtained on an equidistant grid in radius with $512$ nodes and time steps dynamically computed by the CFL condition (Eq. (\ref{ltb:numerics:2})). }
 \label{ltb:comparison:fig:2}
 \end{figure}

 \begin{figure}[ht]
  \centering
  \subfigure[Case 1: initialise $\varphi$ for $\ell=2$]{ \includegraphics[width=0.48\hsize]{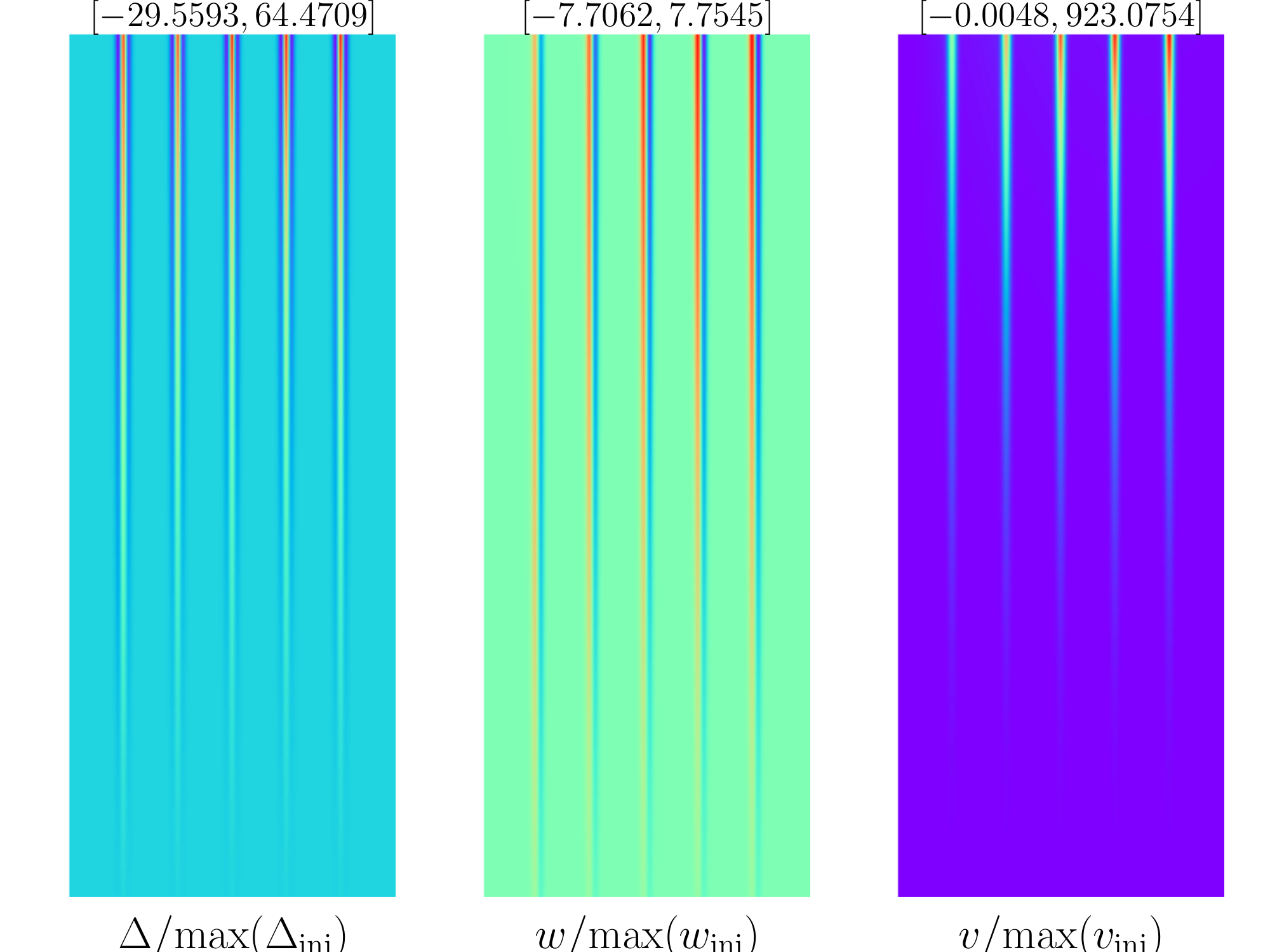} } 
  \subfigure[Case 2: initialise $\varphi$ for $\ell=10$]{ \includegraphics[width=0.48\hsize]{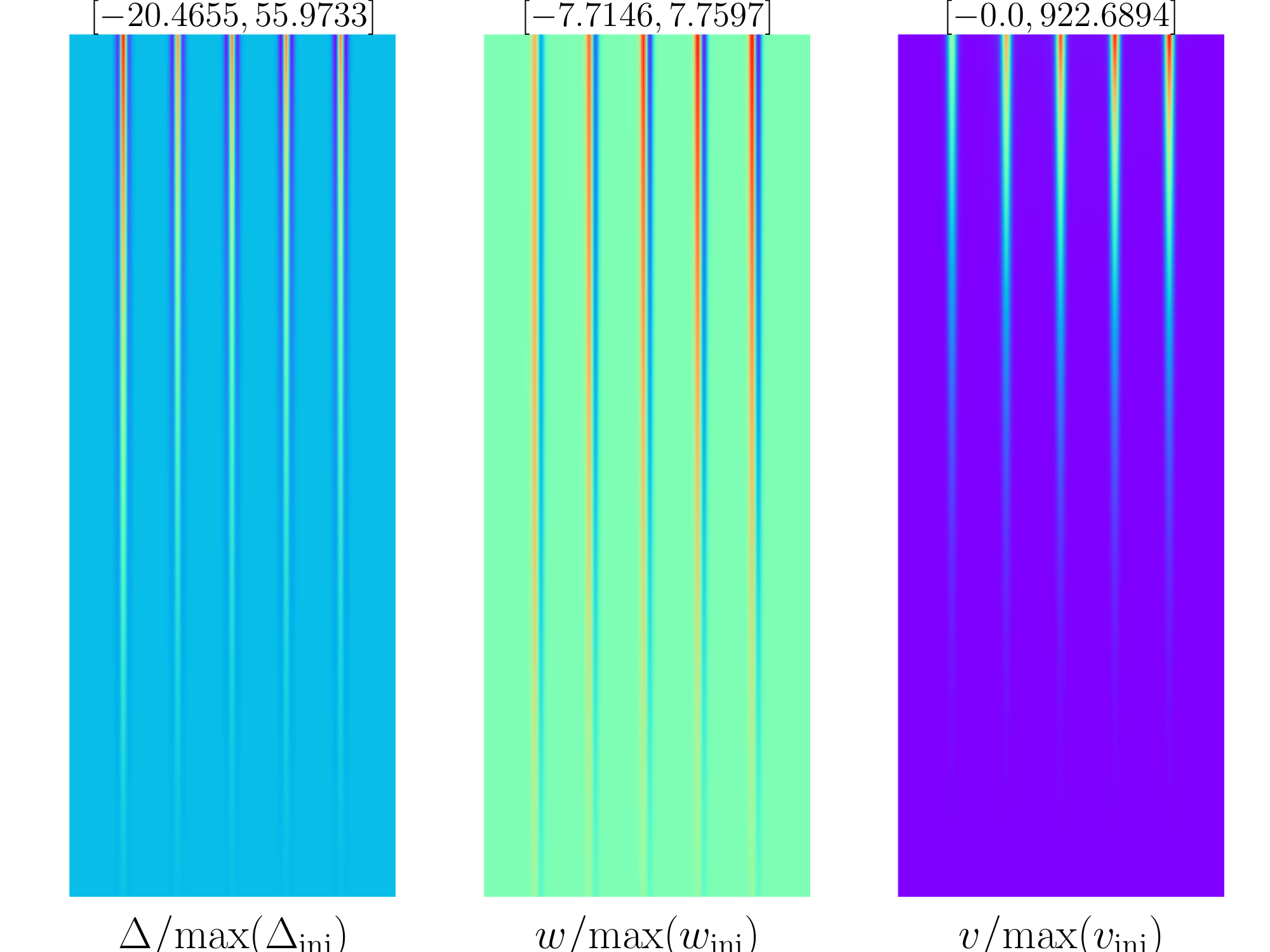} } \\
  \subfigure[Case 3: initialise $\varsigma$ for $\ell=2$]{ \includegraphics[width=0.48\hsize]{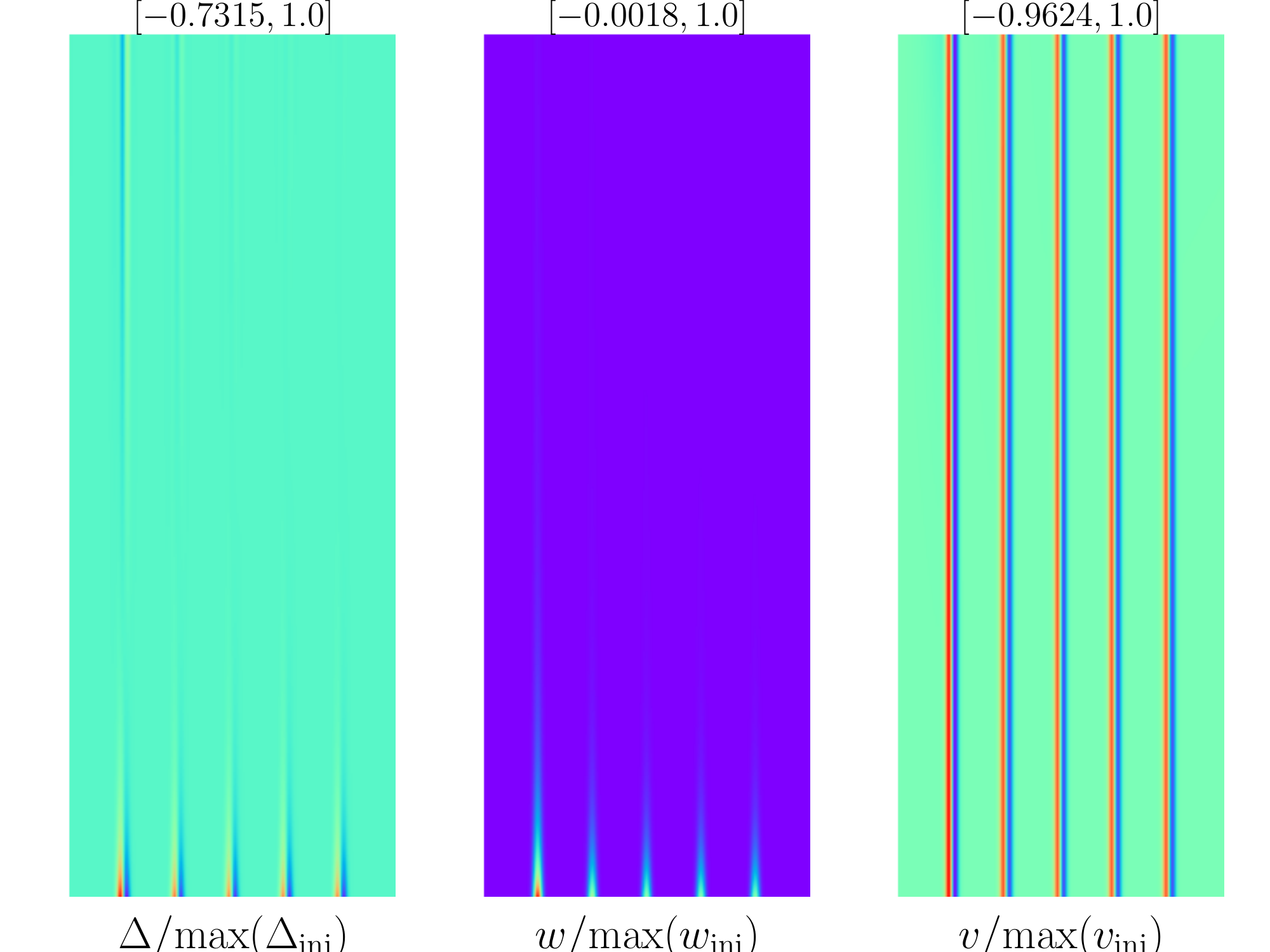} } 
  \subfigure[Case 4: initialise $\varsigma$ for $\ell=10$]{ \includegraphics[width=0.48\hsize]{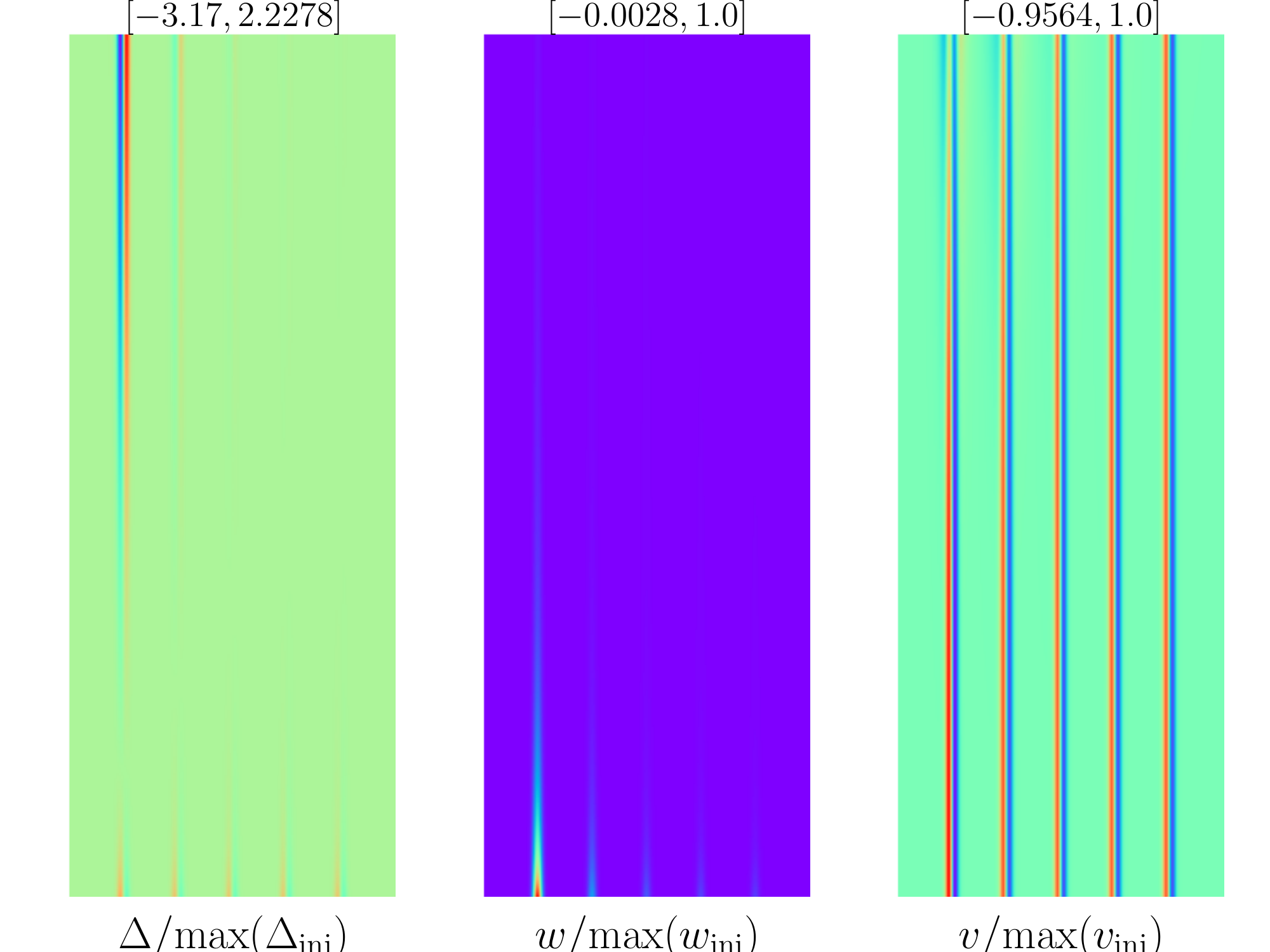} } \\
  \subfigure[Case 5: initialise $\chi$ for $\ell=2$]{ \includegraphics[width=0.48\hsize]{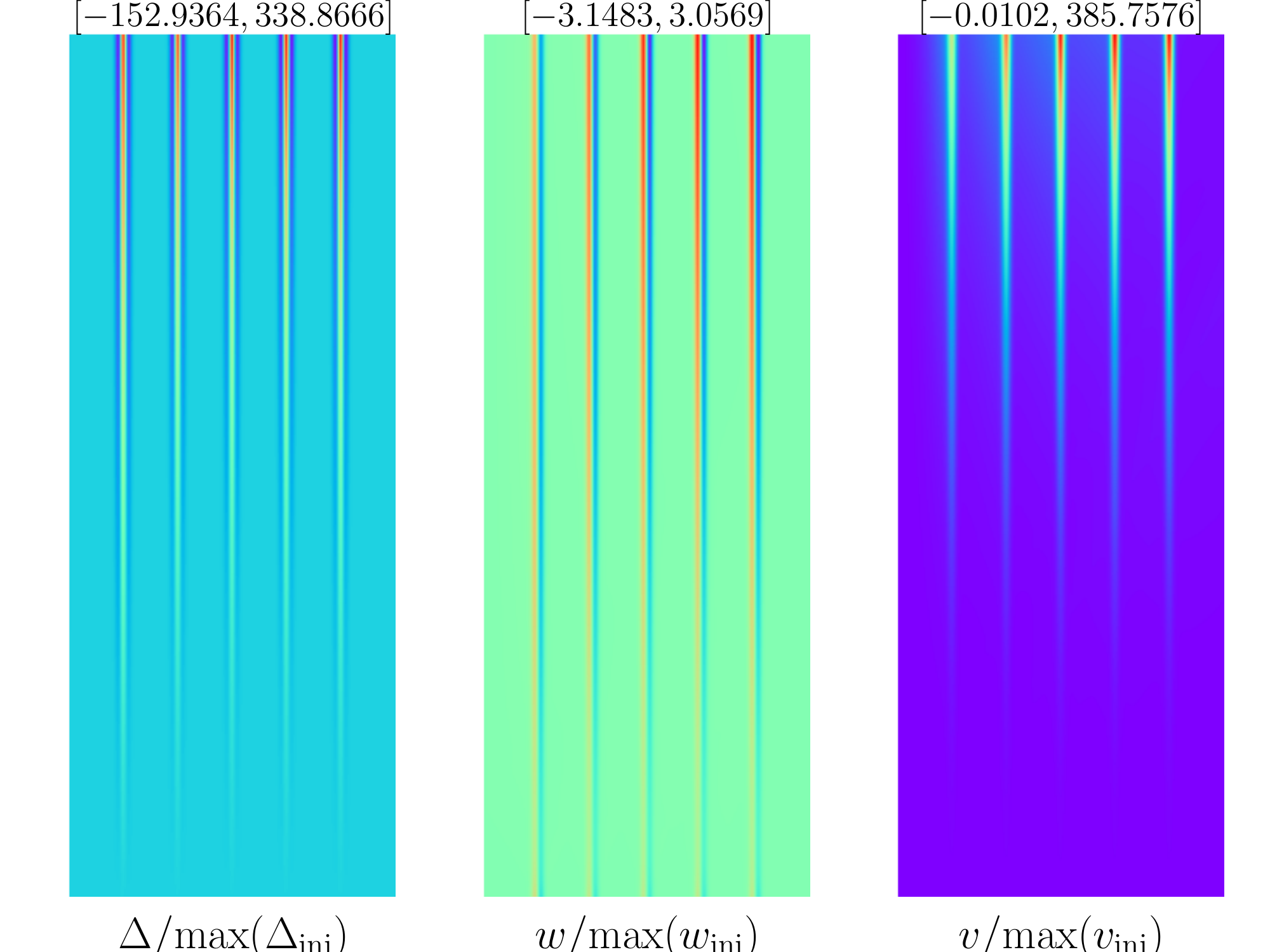} } 
  \subfigure[Case 6: initialise $\chi$ for $\ell=10$]{ \includegraphics[width=0.48\hsize]{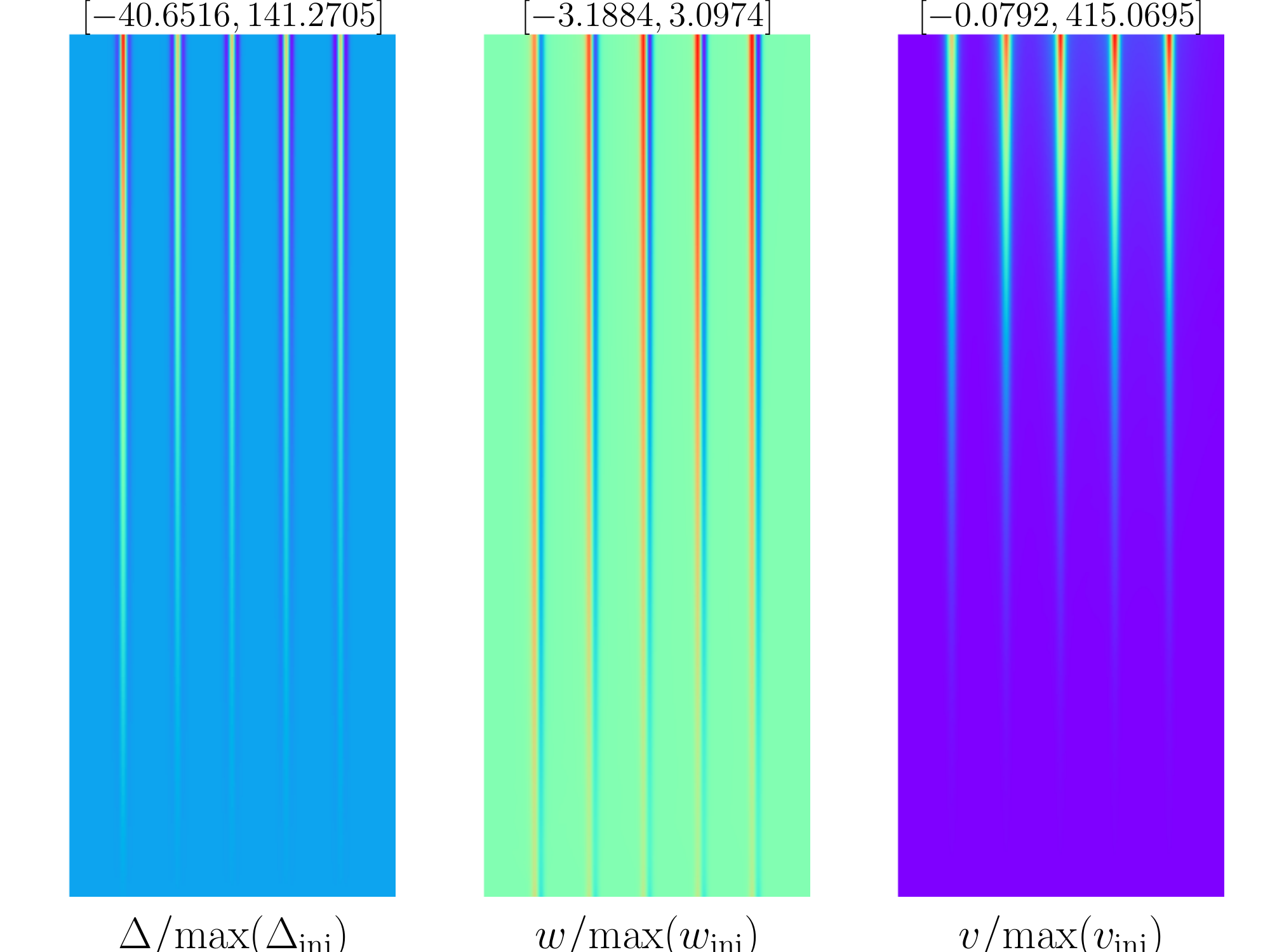} } 
  \caption{Space-time-evolution of corresponding gauge-invariant fluid perturbations $\{ \Delta, w, v \}$ with initial conditions according to Eqs. (\ref{ltb:comparison:1}) and (\ref{ltb:comparison:2}). We have again two exemplary modes $\ell=2$ and $\ell=10$ as considered in \cite{february_evolution_2014}. Minimum and maximum values of the full spacetime diagram are given in headlines where we have normalised spacetime profiles using their maximum initial values. }
 \label{ltb:comparison:fig:3}
 \end{figure}
 
 We see from Fig. (\ref{ltb:comparison:fig:2}) that the initially non-zero modes, while evolving in time, excite the remaining initially zero modes at a partly significant level. In addition, we see the internal mode mixing of each gauge-invariant variable. Time evolution suggests that $\chi$, as a purely tensorial mode in FLRW limit, contains only propagating degrees of freedom, whereas $\varsigma$ and $\varphi$ are mixtures of SVT perturbation types. $\varsigma$ contains vectorial and tensorial degrees of freedom as can be seen by their decaying mode (Case 3 and 4) and propagating mode (Case 5 and 6) as well as modes similar to infall velocities (Case 1 and 2). $\varphi$ contains scalar, vector and tensor modes. Cases 1 and 2 illustrate the behavior of $\varphi$ as the scalar Bardeen potential decaying inside the void (as $\kappa(r)<0$) and staying constant in the EdS region. Nonetheless, it also contains propagating modes as can be clearly seen in cases where it is excited from an initially zero state. \\
 
 The corresponding fluid perturbations can be seen in Fig. (\ref{ltb:comparison:fig:3}). The behavior of the generalised density contrast $\Delta$ as well as the two gauge-invariant velocity perturbations $w$ and $v$ generally depends on the initialised metric perturbations. Strong decay of $\varsigma$ also causes the fluid perturbations to decay whereas $\varphi$ and $\chi$ source growing modes. Similar to the metric perturbations, these gauge-invariants are mixtures of scalar, vector and tensor degrees of freedom in the FLRW limit (see \cite{clarkson_perturbation_2009}).  \\
 
 We do not want to give a detailed analysis of the behavior of the time evolution of each perturbation variable here, since the results obtained are not new and have already been analyzed in great detail in \cite{february_evolution_2014}. The main intention of this section is to give an independent confirmation and to show that our numerical results are in agreement with previous works in the field. \\
 
  \FloatBarrier
 
 Indeed, compared with results obtained by February et al. (2014), the spacetime diagrams agree well in shape for each initial configuration and the scales of the excited perturbations are similar. We have to point out here that the background model implementation of February et al. is different from the one used in this work. We fix the void density profile today and the Hubble constant in the asymptotic FLRW regime. February et al. instead apply a functional shape of the matter density parameter profile $\Omega_\mathrm{m}(r)$ today and fix the Hubble constant at the spatial origin. We believe that this causes the minor differences in the values of the perturbation variables. Nonetheless, the basic features of the spacetime evolution are clearly reproduced.

%---------------------------------------------------------------------------------------------------------------------------
\subsection{Angular power spectra} 
\label{ltb:powerspectra}
%---------------------------------------------------------------------------------------------------------------------------

As a next step, we follow the approach outlined in section (\ref{ltb:initial}) to obtain a realisation of a Gaussian initial Bardeen potential field. We compute spherical harmonic coefficients that are going to be propagated independently in time. Since we deal with statistical information and want to explore general properties of the spacetime evolution of perturbations in void models, we compute estimates of the angular power spectra for each perturbation type that are evaluated on the unperturbed LTB-lightcone using Eqs. (\ref{ltb:background:11}) and (\ref{ltb:background:12}):

\begin{equation}
 C^\ell(z) = \sum_{m=-\ell}^\ell \left| a\lm(t(z), r(z))\right|^2.
\label{ltb:powerspectra:1}
\end{equation}

Since we only consider a single Gaussian realisation of the scalar potential field, we can just compute estimates of the angular power spectra and quantify the statistical error of these estimates to be (similar to cosmic variance)\footnote{Note that the evolution equations (\ref{ltb:perturbation:4})-(\ref{ltb:perturbation:10}) decouple into the spherical harmonic modes $(\ell m)$ due to spherical symmetry of the background model. First-order spacetime evolution does therefore not correlate different $(\ell m)$ modes and Gaussianity, as needed for the derivation of this error estimate, is preserved in each spherical shell. Consequently, we can safely compute the statistical error of the angular power spectra using this formula at all times and radii.}

\begin{equation}
\langle  (C^\ell  - C^\ell_\mathrm{theor})^2 \rangle = \frac{2}{2\ell+1} (C^\ell)^2
 \label{ltb:powerspectra:2}
\end{equation}

\begin{figure}[ht]
\subfigure[ $z=0.5$ ]{ \includegraphics[page=2, width=0.48\hsize]{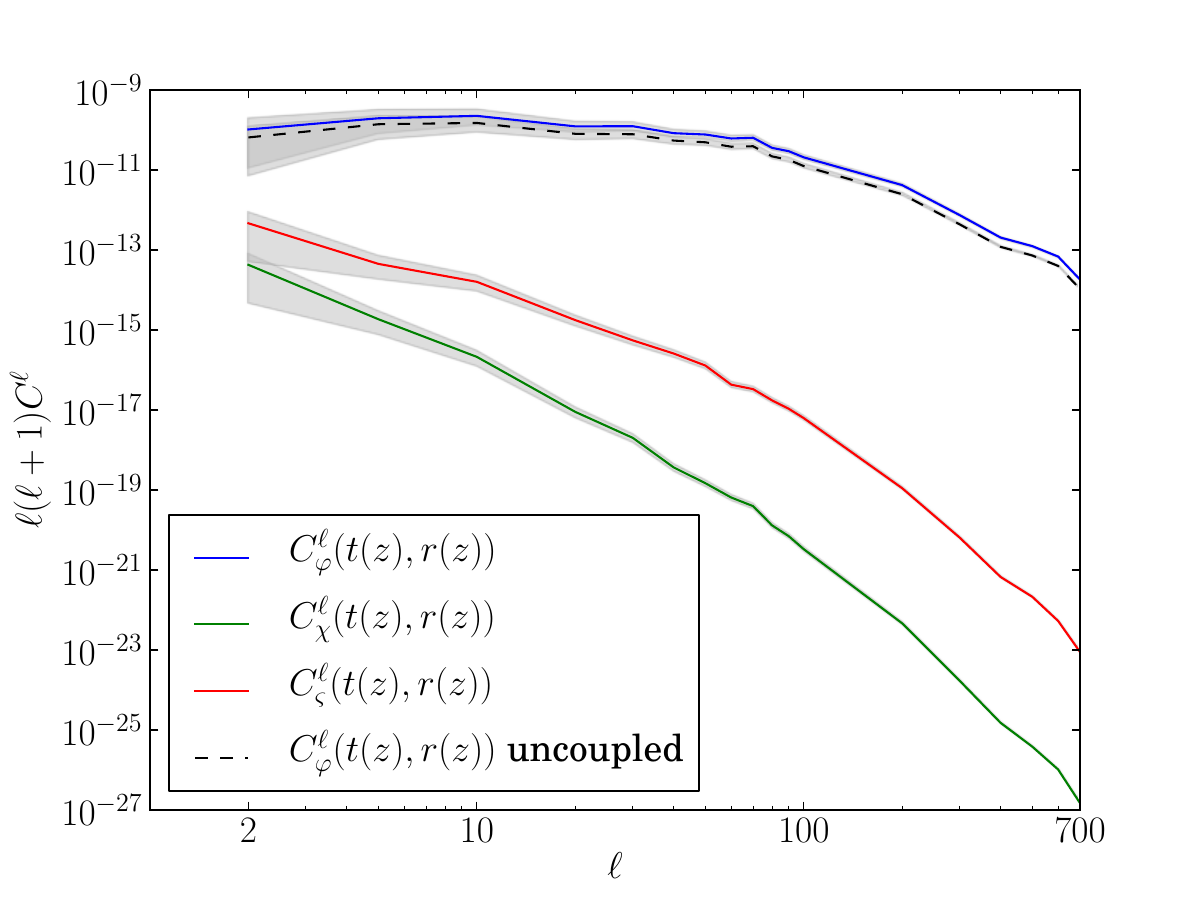} } 
\subfigure[ $z=1$ ]{ \includegraphics[page=3, width=0.48\hsize]{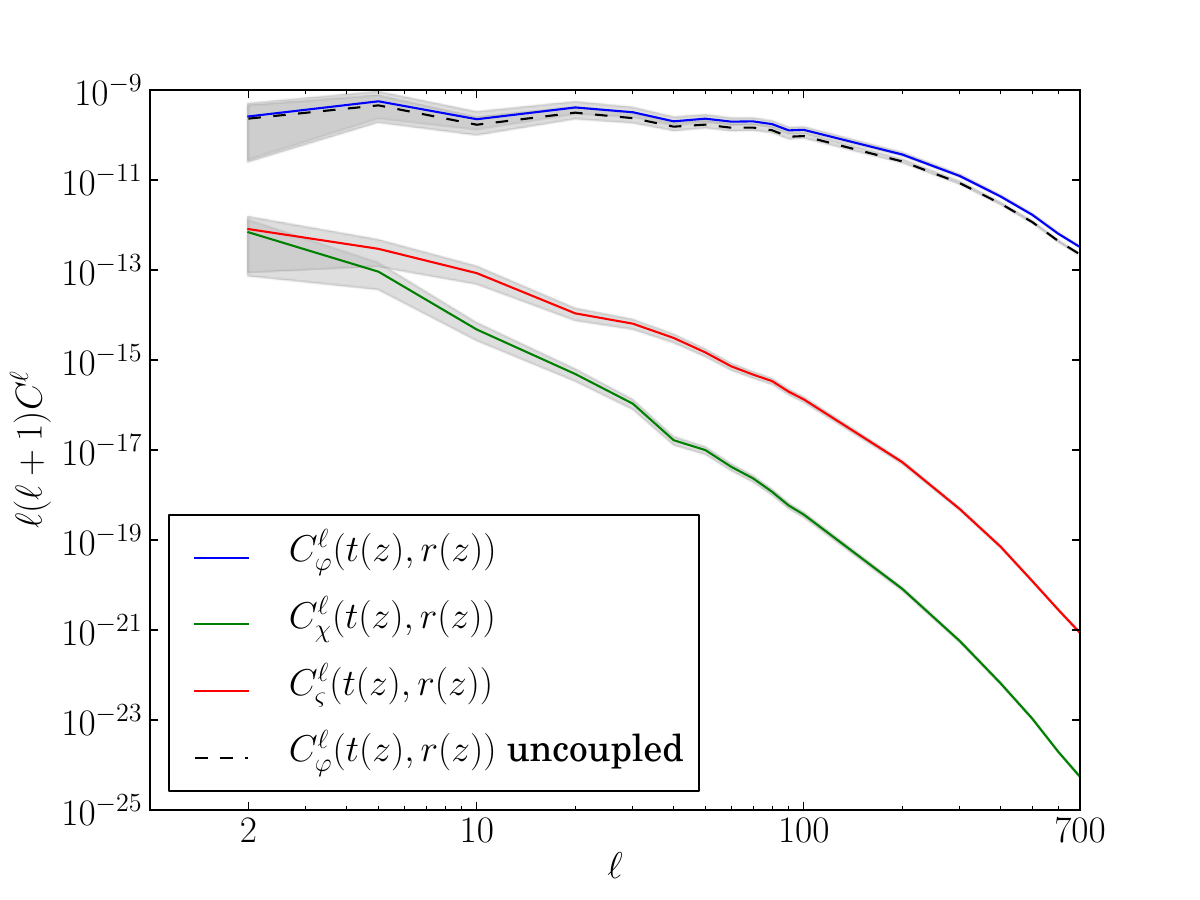} } \\
\subfigure[ $z=2$ ]{ \includegraphics[page=4, width=0.48\hsize]{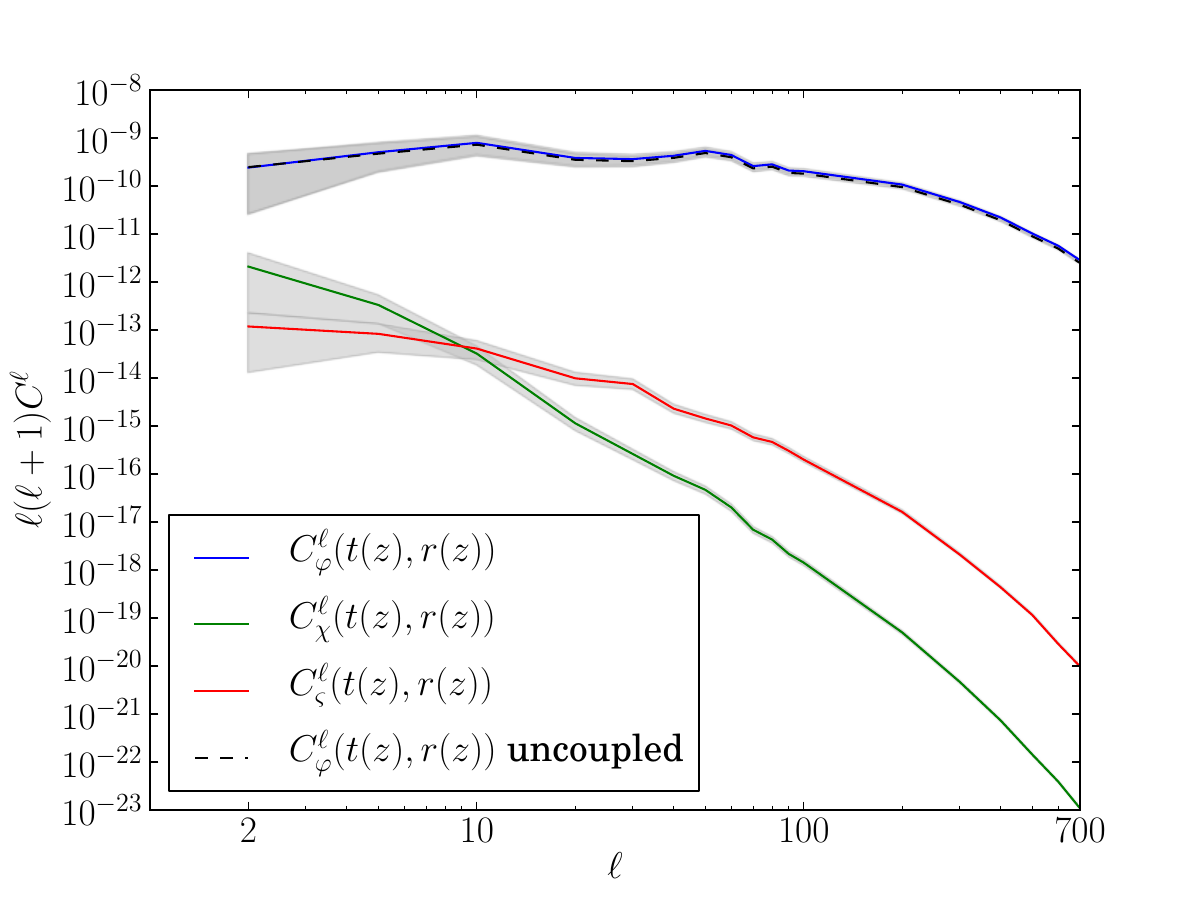} } 
\subfigure[ $z=5$ ]{ \includegraphics[page=7, width=0.48\hsize]{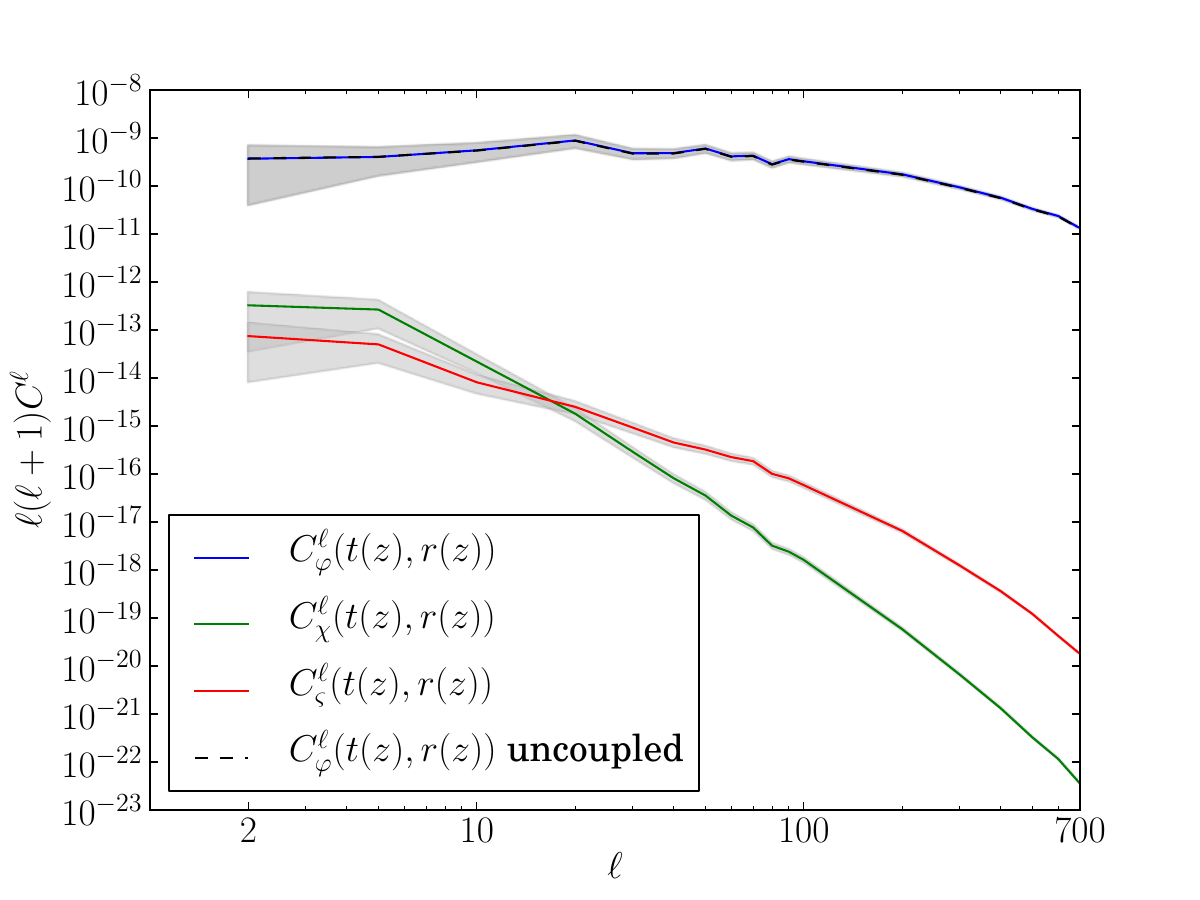} } \\
\caption{Angular power spectra of the metric perturbations $\left\{ \varphi, \varsigma, \chi \right\}$ evaluated on the LTB backward lightcone for different redshifts: The gray-shaded areas indicate the uncertainty in the estimation of the powerspectra due to statistical fluctuations in the finite sample. We consider a Gaussian shaped void according to Eq. (\ref{ltb:background:14}) with $\Omega_\mathrm{in}=0.2$ and $L=2$ Gpc. The solution to the uncoupled evolution equation for $\varphi$ is considered as well for completeness. It shows the difference of the coupled and uncoupled case for the spacetime evolution of $\varphi$. We see that the amplitudes decrease with decreasing angular scale which is expected due to the shape of the Bardeen potential power spectrum the initial conditions are sampled from. In addition, the difference between the fully coupled and uncoupled solution for $\varphi$ is increasing with decreasing redshift as the void depth increases significantly at small redshifts. As a consequence, non-vanishing contributions of the remaining two gauge-invariants are excited that slow down the decay of the $\varphi$ in the interior of the void compared to the uncoupled solution. The effects of coupling on the evolution of $\varphi$ will be analysed in detail in Sect. \ref{ltb:coupling}.}
\label{ltb:powerspectra:fig:1}
\end{figure}

In this way, we can investigate the behavior of the different perturbation types at different redshifts $z$ and angular scales $\ell$. \\

Fig. (\ref{ltb:powerspectra:1}) shows power spectra for each LTB gauge-invariant at redshifts $z = \left[ 0.5, \ 1.0, \ 2.0, \ 3.0, \ 5.0 \right]$. Starting from initial Bardeen potential perturbations leading to an initial $\varphi$ and vanishing $\chi$ and $\varsigma$, we clearly see excited modes of these metric perturbations at smaller redshifts. The angular power spectra basically agree with the shape of the initial Bardeen potential power spectrum. Since initial amplitudes of large k-modes are suppressed, we expect to see the same behavior in the angular power spectra on small angular scales. At small redshifts, we are in the regime in which discretization of the initial 3d cube plays a role causing higher $\ell$-modes to be strongly suppressed. \\

The angular scale at given spherical harmonic mode $\ell$ can be roughly estimated to be $ \delta \theta \approx \pi/\ell$. In order to estimate the maximum angular scale that can be resolved by our pixel size $\Delta x$, we can write   

\begin{equation}
 r \cdot \delta \theta \ge \Delta x = \frac{L_\mathrm{cube}}{(N_\mathrm{nodes})^{1/3} r}
 \label{ltb:powerspectra:3}
\end{equation}

where $L_\mathrm{cube}$ denotes the side length of the initial 3d cube and $N_\mathrm{nodes}$ the number of nodes in it. \\

Solving for $\ell$, we obtain

\begin{equation*}
 \ell \le \pi \frac{r (N_\mathrm{pixel})^{1/3}}{L_\mathrm{cube}}.
\end{equation*}

For our setup, the initial cube has a side length of $12$ Gpc with $1024^3$ nodes. At different redshifts, the maximum angular scale allowed by the discretization is shown in Tab. (\ref{ltb:powerspectra:tab:1}). At $z=0.1$ for instance, we have a maximum $\ell$ mode of only $57$ and all higher modes are strongly suppressed. We therefore choose a lower redshift boundary of $z\geq 0.5$ for our considerations. 

 \FloatBarrier

\begin{table}
\centering
\begin{tabular}{rrr}
  \hline \\
  $z$ & $r(z)\mathrm{[Mpc]}$ & $\ell_\mathrm{max}$ \\ 
  \hline \\
  0.1 & 306.495 & 57 \\
  0.5 & 1386.31 & 260 \\
  1.0 & 2440.38 & 457 \\
  2.0 & 3686.05 & 691 \\
  3.0 & 4403.68 & 826 \\
  4.0 & 4887.78 & 917 \\
  5.0 & 5243.44 & 983 \\
  \hline
 \end{tabular}
 \caption{Estimation of maximum $\ell$-modes on a Bardeen potential cube of $L_\mathrm{cube}=12000$ Mpc extension and $1024^3$ pixels. Results are taken on LTB lightcone positions for a background void model with $\Omega_\mathrm{in}=0.2$, $\Omega_\mathrm{out}=1$ and $L=2000$ Mpc.}
 \label{ltb:powerspectra:tab:1}
\end{table}

%---------------------------------------------------------------------------------------------------------------------------
\subsection{Influence of initial modifications} 
\label{ltb:influence}
%---------------------------------------------------------------------------------------------------------------------------
 
 \begin{figure}[ht]
  \includegraphics[page = 2, width=\hsize]{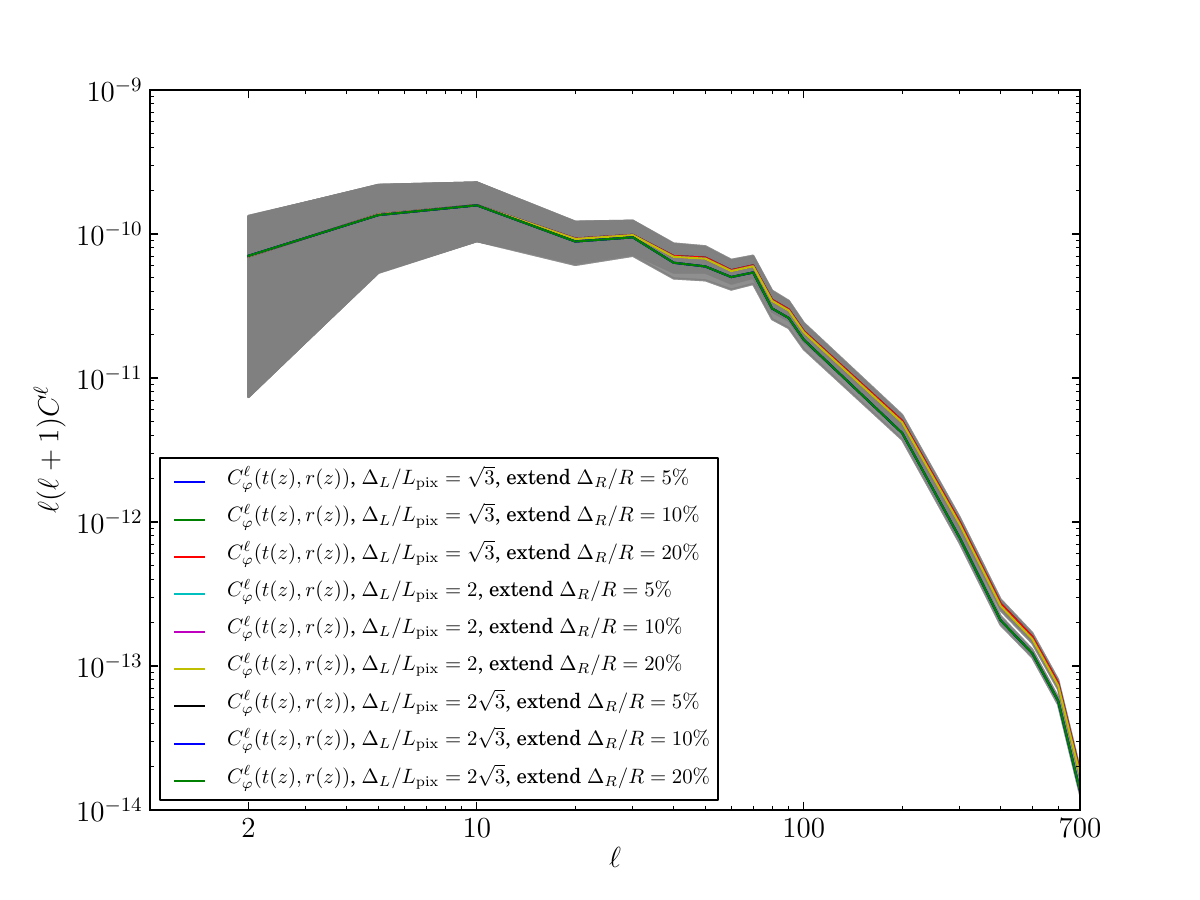}
  \caption{ Influence on two main modifications on the initial profile (smoothing scale factor $\Delta_L$ and extension scale factor $\Delta_R$) on the angular power spectra for $\varphi$ at redshift $z = 0.5$. }
  \label{ltb:influence:fig:1}
  \end{figure}
 
 As outlined in section (\ref{ltb:initial}), we had to modify the initial profiles by a smoothing scale and an extension region in which the profile is extended by a Gaussian dropping to zero. The smoothing scale itself and the Gaussian extension are additional parameters to the initial profiles and we need to investigate their influence on the final angular power spectra. We therefore performed several runs with smoothing scales $\Delta_L = \left[0, \sqrt{3}, 2\sqrt{3}\right] \times$ pixel size and $\Delta_R = \left[ 5\%, 10\%, 20 \% \right]$ of the size of the domain of interest. The results in Fig. (\ref{ltb:influence:fig:1}) indicate that the extension scale has no significant influence on the final results. Deviations caused by a smoothing scale up to 2 pixels are well within the statistical error of the final angular power spectra.  We can therefore safely apply a smoothing scale of one pixel diagonal and extension of $10\%$ of the domain of interest.
  
%---------------------------------------------------------------------------------------------------------------------------
\subsection{Coupling strength} 
\label{ltb:coupling}
%---------------------------------------------------------------------------------------------------------------------------
 
 We finally want to quantify the coupling of the perturbation variables in a statistical way. Starting from an initial scalar perturbation as Gaussian random field with vanishing mean, we want to compare its evolution on different angular scales for the coupled and uncoupled case. As the gauge invariant variables $\varphi$ and $\Delta$ represent the generalized gravitational potential and density contrast, we will restrict our analysis on these two variables in particular. Their uncoupled evolution is described by
 
 \begin{align}
  \label{ltb:coupling:1}
 \ddot{\varphi} &= - 4 H\o \dot{\varphi} + \frac{2 \kappa}{a\o^2} \varphi \\
  \label{ltb:coupling:2}
   \alpha  \Delta &= - \frac{1}{Z^2} \varphi'' + \frac{1}{Z^2} \left( C - 4 \frac{a\p}{r a\o} \right) \varphi' + \left( H\p + 2 H\o \right) \dot{\varphi} + \left[ \frac{\ell(\ell+1)}{r^2 a\o^2} + 2 D \right] \varphi 
 \end{align}

 The fully coupled system of master and constraint equations given by Eqs. (\ref{ltb:perturbation:4}) - (\ref{ltb:perturbation:7}) and Eqs. (\ref{ltb:perturbation:8}) - (\ref{ltb:perturbation:10}) will be evolved simultaneously in time  using the same initial Gaussian random seed.  We then compute the corresponding angular power spectra $C^\ell_\varphi$ and $C^\ell_{\varphi,\mathrm{uc}}$ on the PNC. In order to quantify the coupling strength, we compute the relative deviation $\epsilon^\ell$ of the square roots of the angular power spectra:
 
 \begin{equation}
  \epsilon^\ell(z) = \left| \frac{ \sqrt{C^\ell_x(t(z), r(z))} - \sqrt{C^\ell_{x,\mathrm{uc}}(t(z), r(z))} }{\sqrt{C^\ell_{x,\mathrm{uc}}(t(z), r(z))} } \right| \ \mathrm{with} \ x=\varphi, \Delta
  \label{ltb:coupling:3}
 \end{equation}

 with combined statistical errors given by

 \begin{equation}
  \Delta \epsilon^\ell =  \frac{\epsilon^\ell}{\sqrt{C^\ell_{x, \mathrm{uc}}}} \sqrt{ \left(\frac{\Delta C^\ell_{x, \mathrm{uc}}}{2C^\ell_{x, \mathrm{uc}}}\right) + \left(\frac{\Delta C^\ell_{x}}{2C^\ell_{x}}\right) + \frac{1}{2}\left( \frac{\Delta C^\ell_{x, \mathrm{uc}}}{C^\ell_{x, \mathrm{uc}}} \right)^2 \left|C^\ell_{x} - C^\ell_{x, \mathrm{uc}} \right| }
  \label{ltb:coupling:4}
 \end{equation}

 The coupling strength can now be investigated as function of void size and void depth. For this purpose, we consider two exemplary redshifts at $z=0.5$ and $z=5$ and perform two runs with different configurations:
 
 \begin{enumerate}
   \item Case 1: fixed void size $L = 2000$ Mpc and modified void depth $\Omega_\mathrm{in} = 0.2, \ 0.4, \ 0.6, \ 0.8$, and $1.0$ (FLRW limit)
   \item Case 2: fixed void depth $\Omega_\mathrm{in} = 0.2$ and modified void size $L = 1000, \ 1200, \ 1500, \ 1800, \ 2000$ Mpc
 \end{enumerate}

  \begin{figure}[ht]
  \subfigure[ $z=0.5$ ]{ \includegraphics[page=2, width=0.48\hsize]{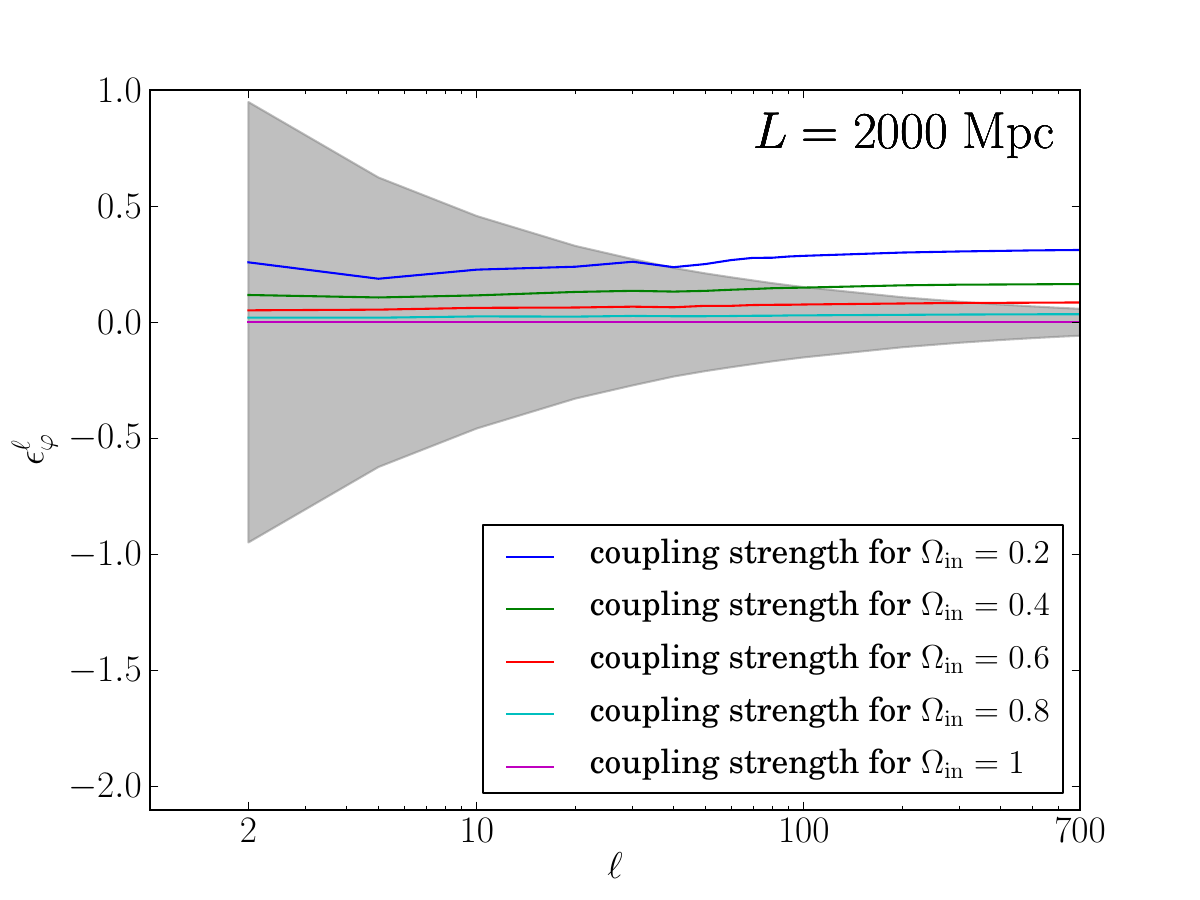} } 
  \subfigure[ $z=5$ ]{ \includegraphics[page=7, width=0.48\hsize]{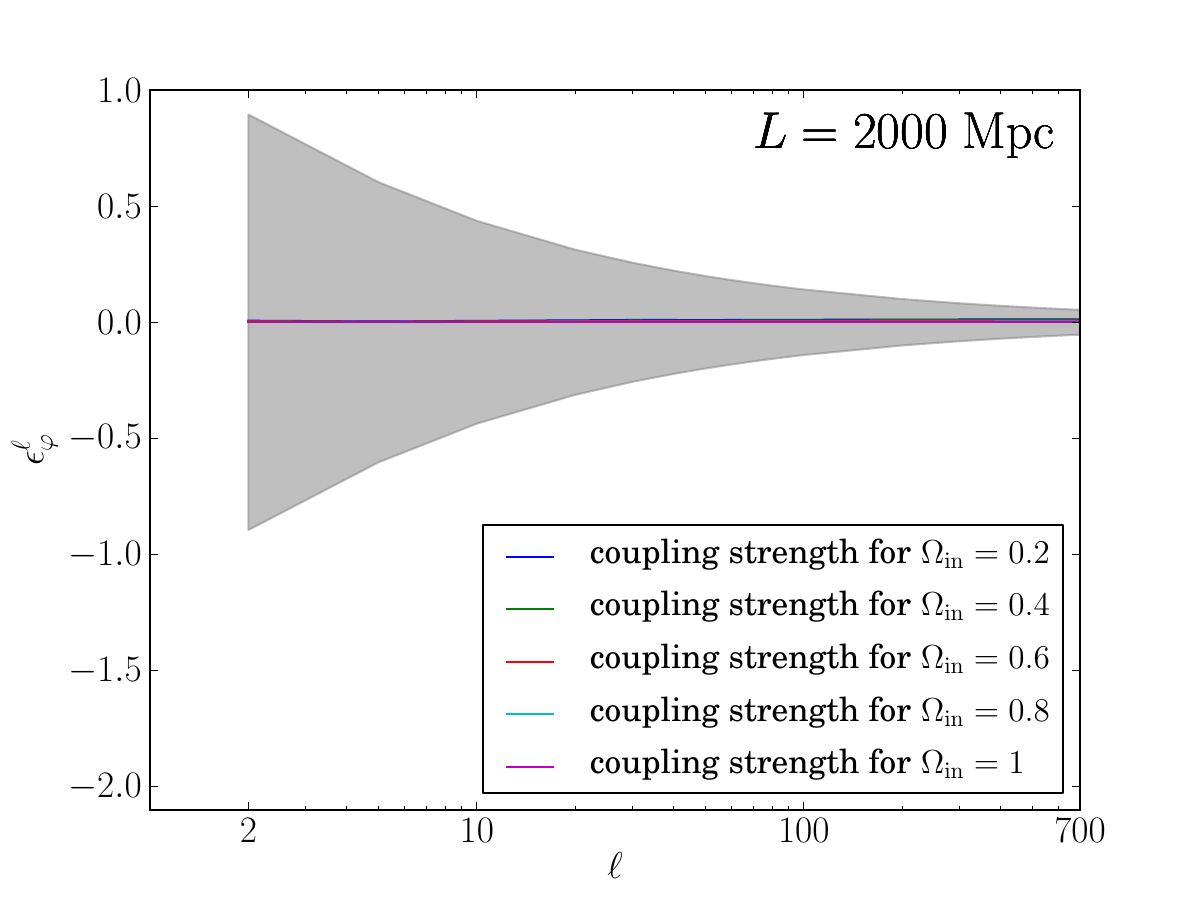} } \\
  \subfigure[ $z=0.5$ ]{ \includegraphics[page=2, width=0.48\hsize]{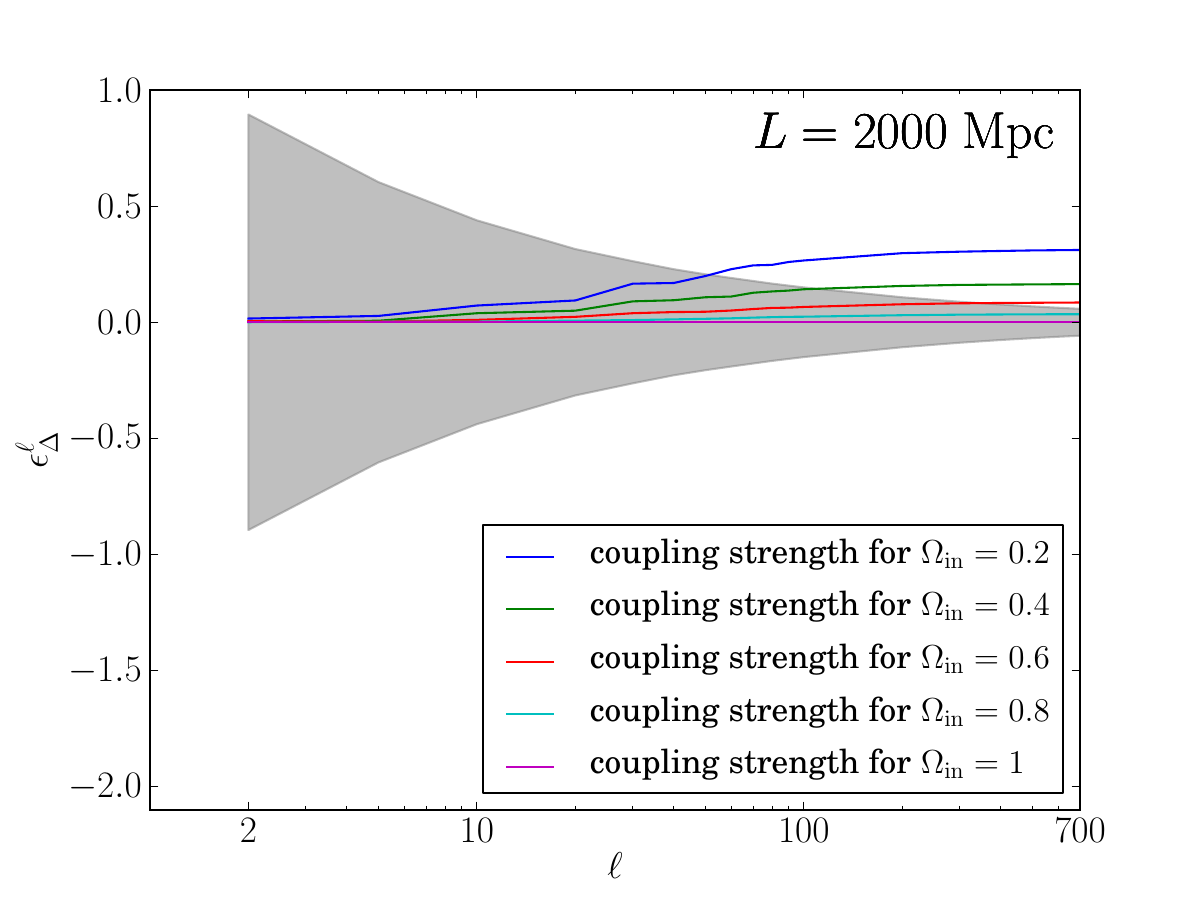} } 
  \subfigure[ $z=5$ ]{ \includegraphics[page=7, width=0.48\hsize]{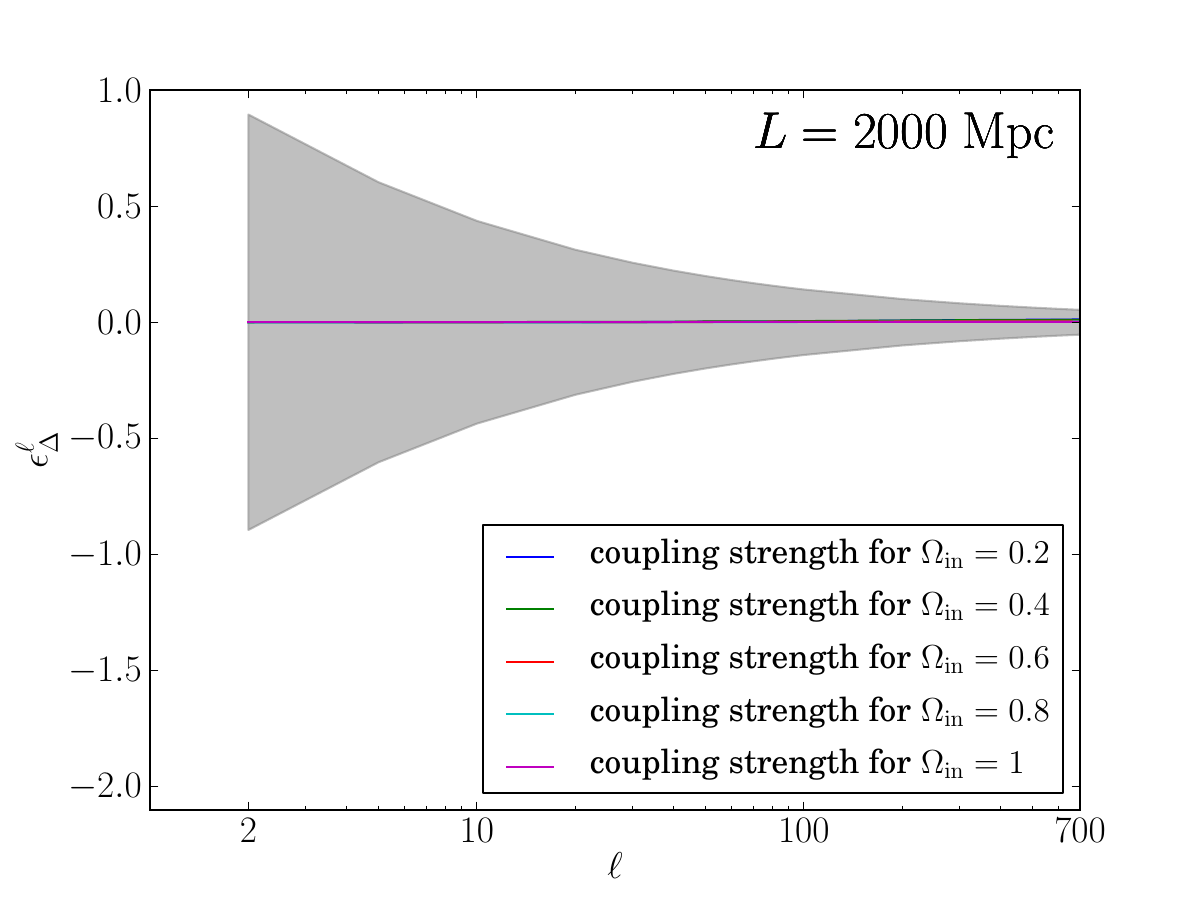} } \\
  \caption{ Coupling strength $\epsilon^\ell$ for the two gauge invariants $\varphi$ and $\Delta$ (as defined in Eq. (\ref{ltb:coupling:3})) for different void depth $\Omega_\mathrm{in} = 0.2, 0.4, 0.6, 0.8, 1.0$ at fixed void size of $L=2$ Gpc. The results are evaluated on the corresponding LTB past lightcones for two exemplary redshifts $z=0.5$ and $z=5$. The grey-shaded area marks the statistical error due to fluctuations in the finite sample estimated due to Eq. (\ref{ltb:coupling:4}) which would be expected for an EdS model. At redshift $z=0.5$, we see the coupling strength increasing with increasing void depth up to nearly $30$ \% for the strongest deviation of $\Omega_\mathrm{in}=0.2$ from the EdS solution the void is embedded in. This is also significant regarding the statistical fluctuation in the finite sample . At higher redshifts, the coupling strength quickly drops as soon as the lightcone becomes FLRW-like. Coupling increases with void depth as expected, since the space-time induced anisotropy (quantified by the shear) and radial dependence of the curvature profile which are the main sources of perturbation coupling are also significantly increasing with void depth. We also see an increase with multipole order $\ell$ which can also be expected, since one of the coupling terms in Eq. (\ref{ltb:perturbation:5}) as well as Eq. (\ref{ltb:perturbation:9}) scales quadratically with $\ell$ and quickly dominates the coupling strength on small angular scales. The two gauge invariants mainly behave similar and we reach similar coupling strength, despite the increase of coupling strength with decreasing angular scales is more prominent in case of $\Delta$. The FLRW limit of vanishing coupling is nicely reproduced in both cases.}
  \label{ltb:coupling:fig:1}
 \end{figure}

 \begin{figure}[ht]
  \subfigure[ $z=0.5$ ]{ \includegraphics[page=2, width=0.48\hsize]{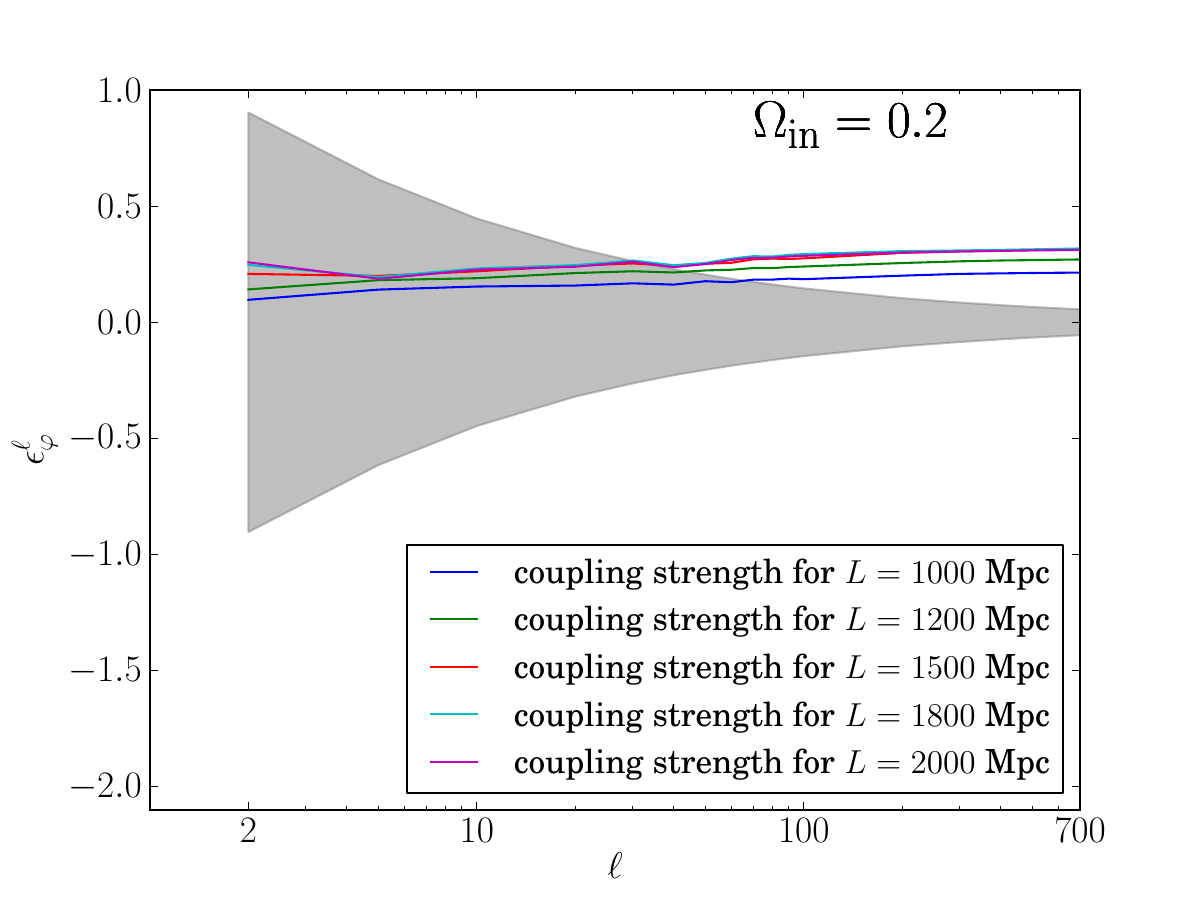} } 
  \subfigure[ $z=5$ ]{ \includegraphics[page=7, width=0.48\hsize]{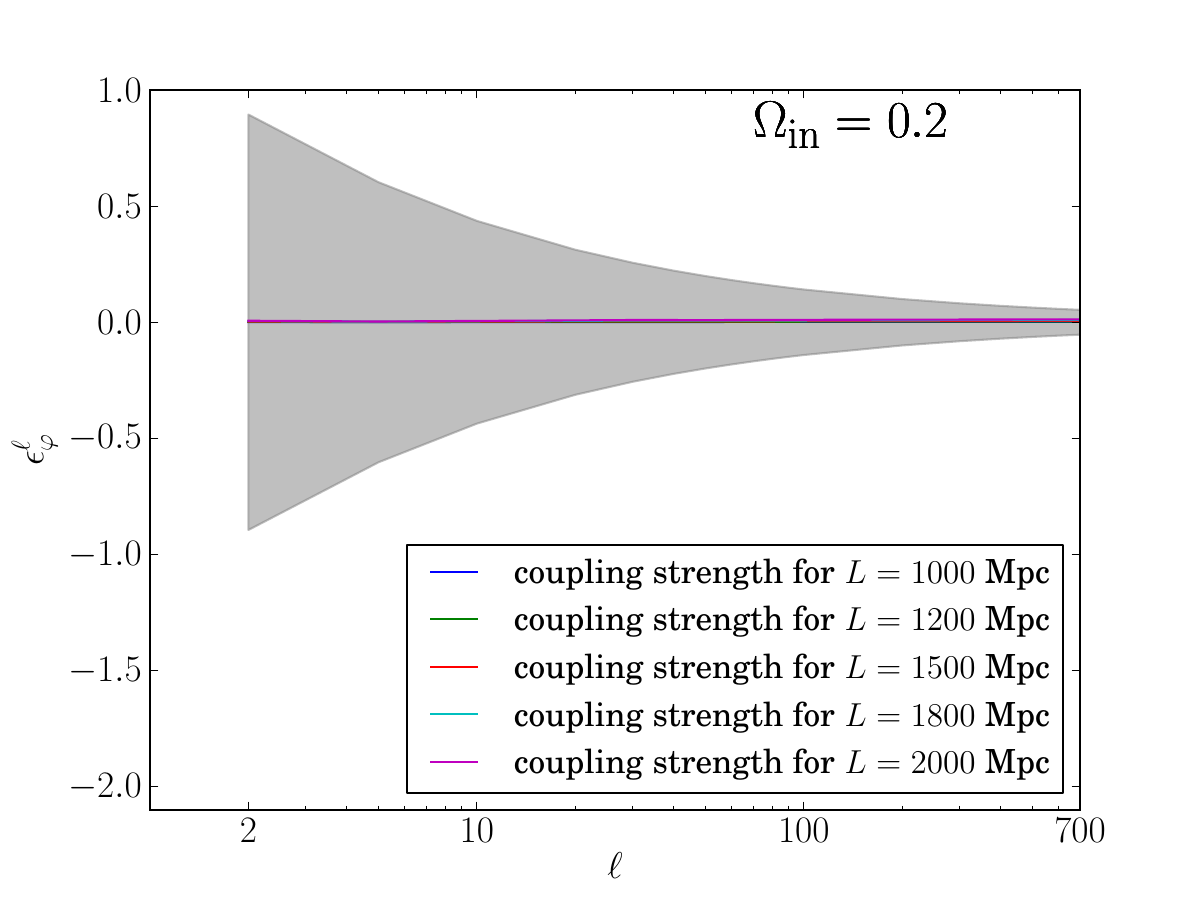} } \\
  \subfigure[ $z=0.5$ ]{ \includegraphics[page=2, width=0.48\hsize]{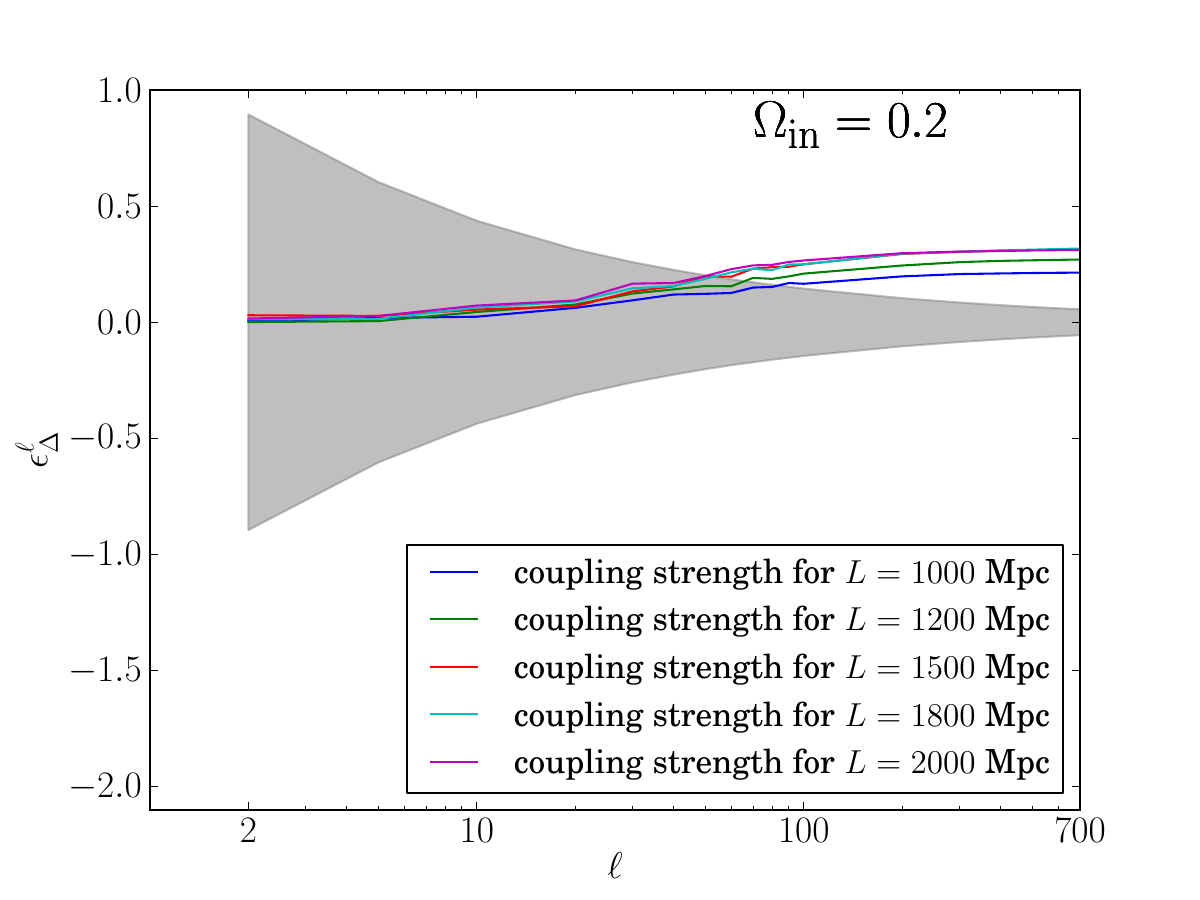} } 
  \subfigure[ $z=5$ ]{ \includegraphics[page=7, width=0.48\hsize]{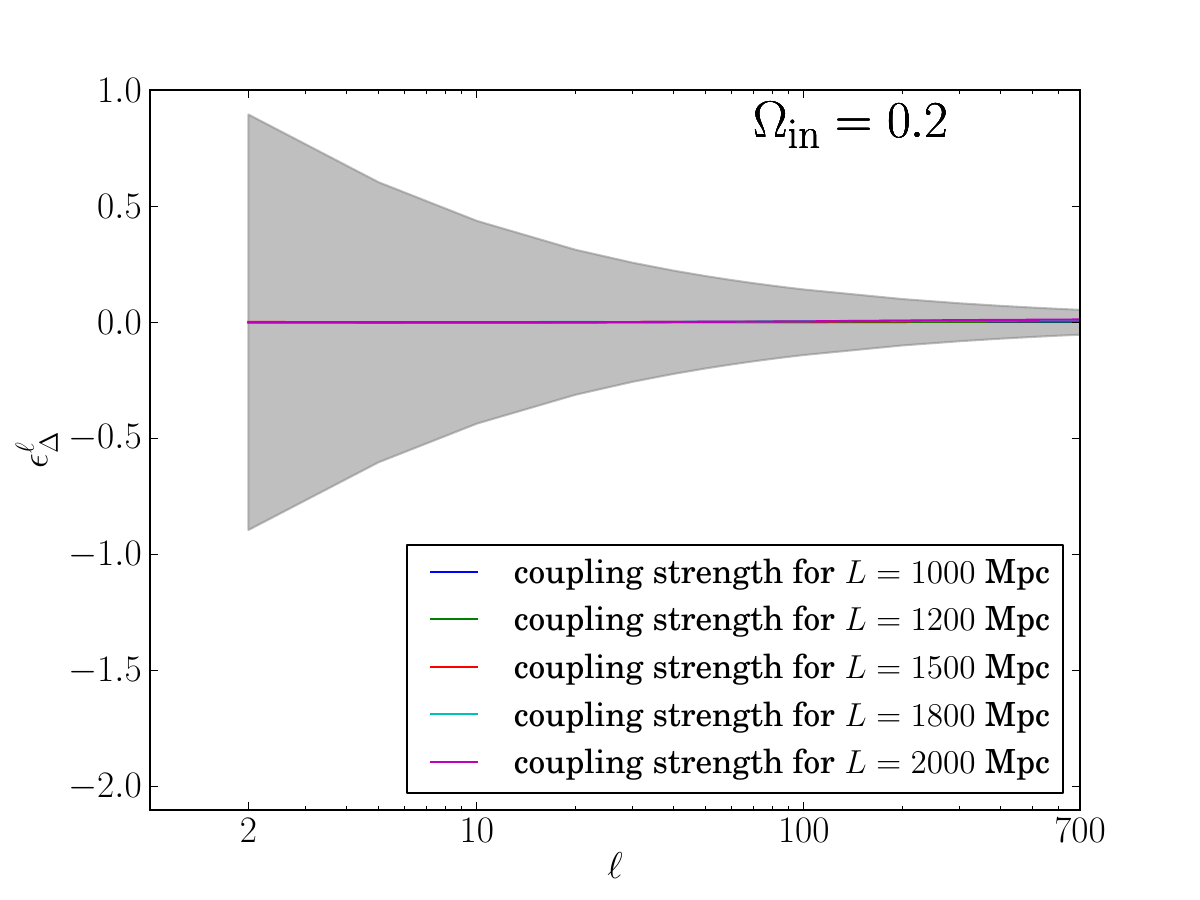} } \\
  \caption{ Coupling strength $\epsilon^\ell$ for the two gauge invariants $\varphi$ and $\Delta$ for different void sizes $L = 300, \ 500 , \ 1000, \ 1200, \ 1500, \ 1800, \text{and} \ 2000$ Mpc at fixed void depth. The results are again evaluated on the LTB past lightcones for the two exemplary redshifts  $z=0.5$ and $z=5$. The grey shaded area marks again the uncertainty due to statistical fluctuations for EdS. As already seen in Fig. (\ref{ltb:coupling:fig:1}), we see again couplings up to $30 \%$ at redshift $z=0.5$ which drops to insignificant values at $z=5$ as the lightcone is no longer strongly affected by the void. Nonetheless, these plots have to be interpreted with care since, due to discretization limits, we can only analyse the coupling reliably for redshifts $z \geq 0.5$ which corresponds to a radial coordinate of already $\sim 1500$ Mpc. This does not mean that we do not expect significant couplings in smaller sub-Gpc scale voids. A drop in the coupling strength at smaller void sizes can therefore partially be caused by the decreasing void depth at $r(z=0.5)$.}
  \label{ltb:coupling:fig:2}
 \end{figure}
 
 The coupling strength in case of varying void depth (Case 1) with fixed void size of $2$ Gpc is shown in Fig. (\ref{ltb:coupling:fig:1}). As expected, coupling increases with void depth as the shear (a measure for off-centre anisotropy) and curvature gradients decrease as well and the FLRW limit of vanishing coupling is correctly reproduced. Coupling decreases with increasing redshift as the void itself evolves in time and has significant depth only in the non-linear FLRW regime ($z \lesssim 1.0$) . The huge deviations of the uncoupled and coupled case are nonetheless surprising as it rises up to nearly $\sim 30 \%$ for a deep void with a density contrast of $\Omega_\mathrm{out} - \Omega_\mathrm{in} = 0.8$ with respect to the background EdS model. Deep voids cause large off-center anisotropies quantified by the difference in radial and tangential Hubble rate (shear $\sigma  \sim H\p - H\o$). This background shear can be identified as one of the main contributions to the coupling. In case of big voids of Gpc scale, this coupling is also present at large distances from the center. Coupling increases with angular scale as the coupling terms scale quadratically with $\ell$ which was also found in \cite{february_evolution_2014}. \\

 Results on Case 2 are shown in Fig. (\ref{ltb:coupling:fig:2}). We see prominent coupling up to  $30\%$ for a large $2$ Gpc void at redshift $z=0.5$. In case of smaller voids the coupling is not significant, since the radial coordinates on the PNC are $r(z=0.5) \sim 1500$ Mpc and therefore the density contrast and the corresponding anisotropy are already small there. Note that this does not mean that coupling is small in the interior of small voids. Since we are limited in the discretization scale, we are only able to make reliable statements about the coupling strength at redshifts smaller than $z = 0.5$ (see Tab (\ref{ltb:powerspectra:tab:1})). This issue will be addressed in more detail in the discussion section. In fact, one can assume coupling to be stronger at very small redshifts as the void depth increases nonlinearly in this regime. In the interior of small voids a large spacetime anisotropy is generated at already small distances from the centre due to strong gradients in the curvature and mass profiles. Nonetheless, as the redshift increases, coupling decreases until it is not significantly measureable any more at $z=5$ as the lightcone approaches FLRW shape. \\
 
\begin{figure}[ht]
 \subfigure[ $\varphi$ ]{ \includegraphics[width=0.48\hsize]{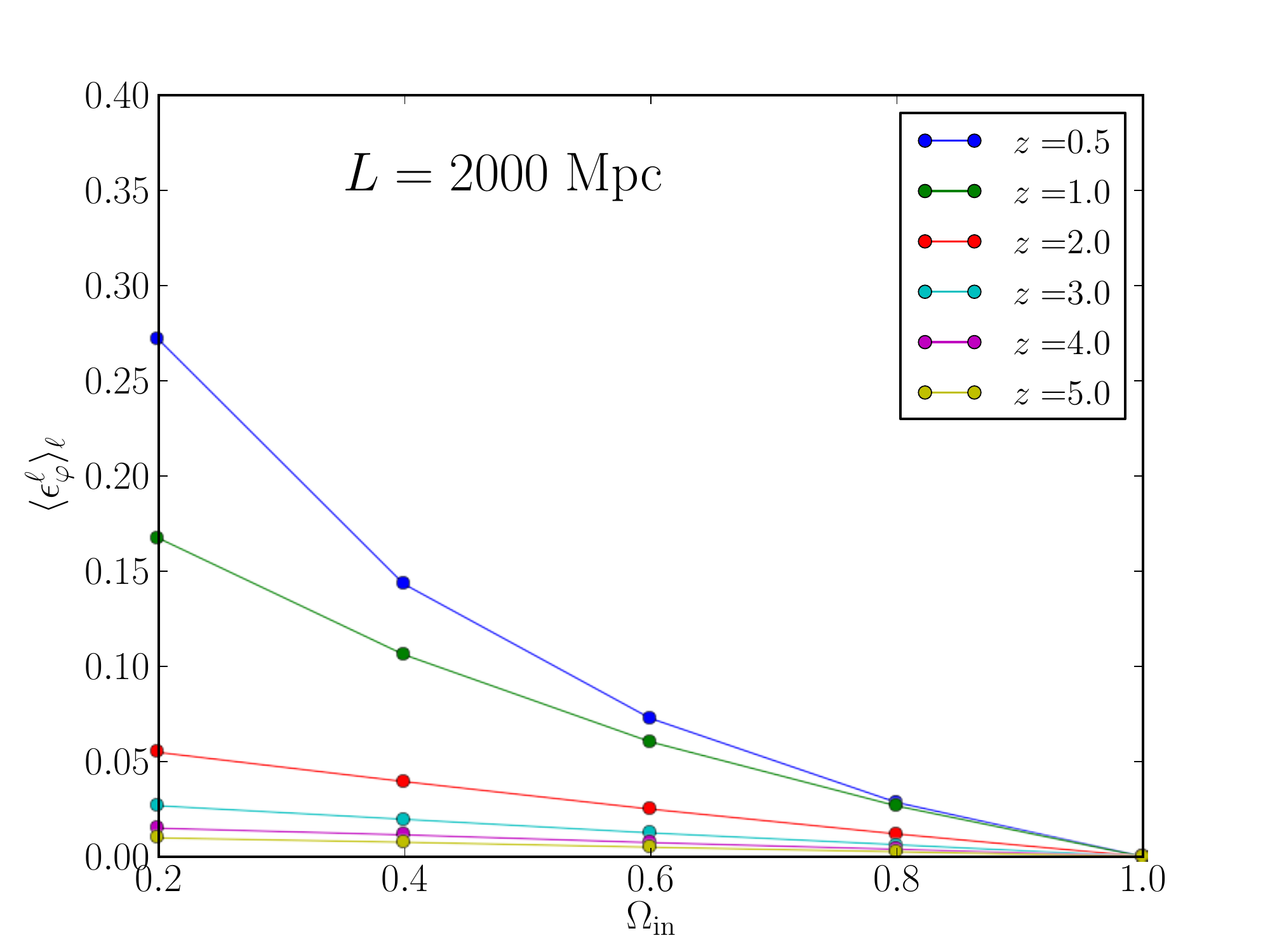} } 
 \subfigure[ $\varphi$ ]{ \includegraphics[width=0.48\hsize]{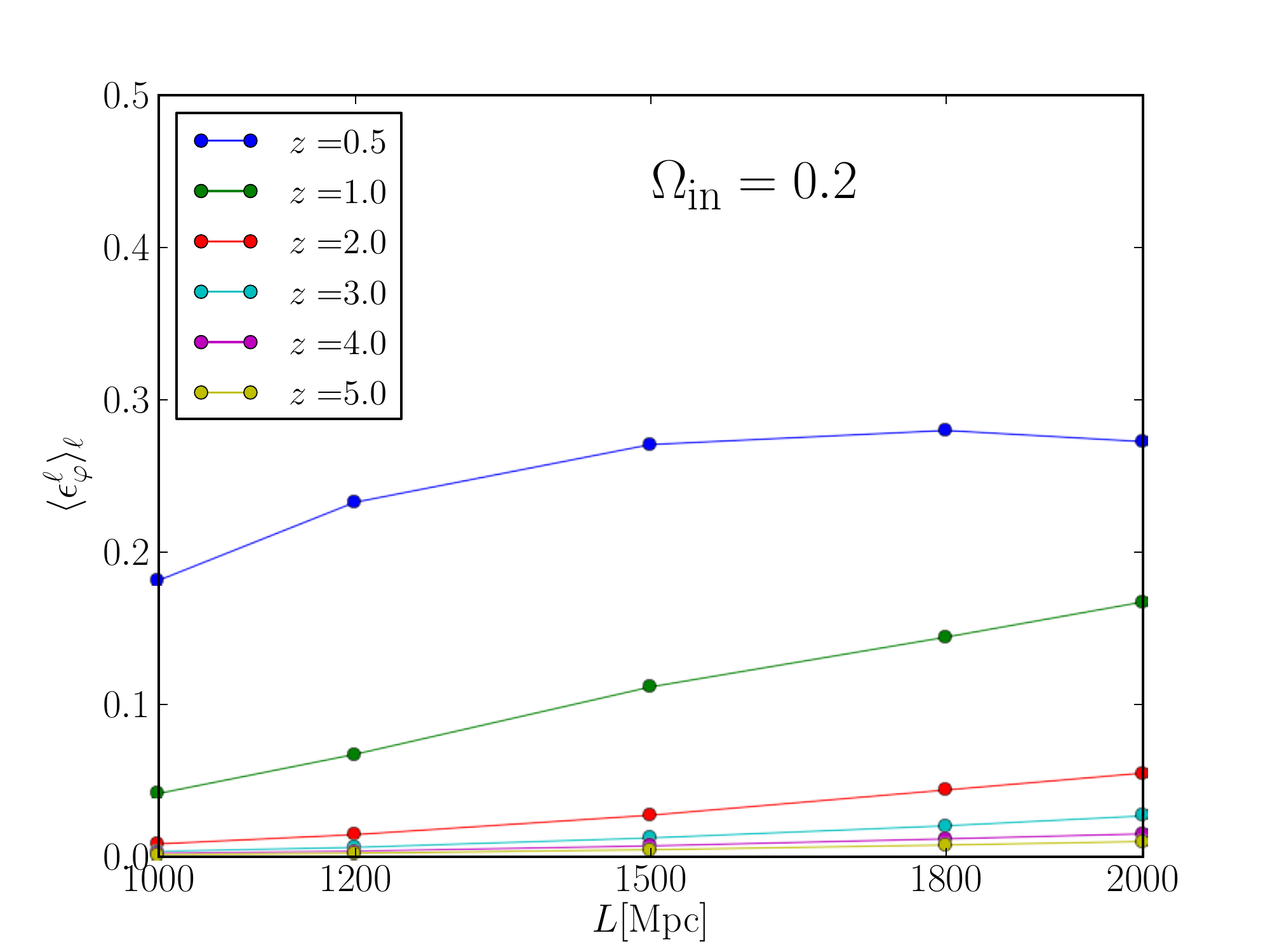} } \\
 \subfigure[ $\Delta$ ] { \includegraphics[width=0.48\hsize]{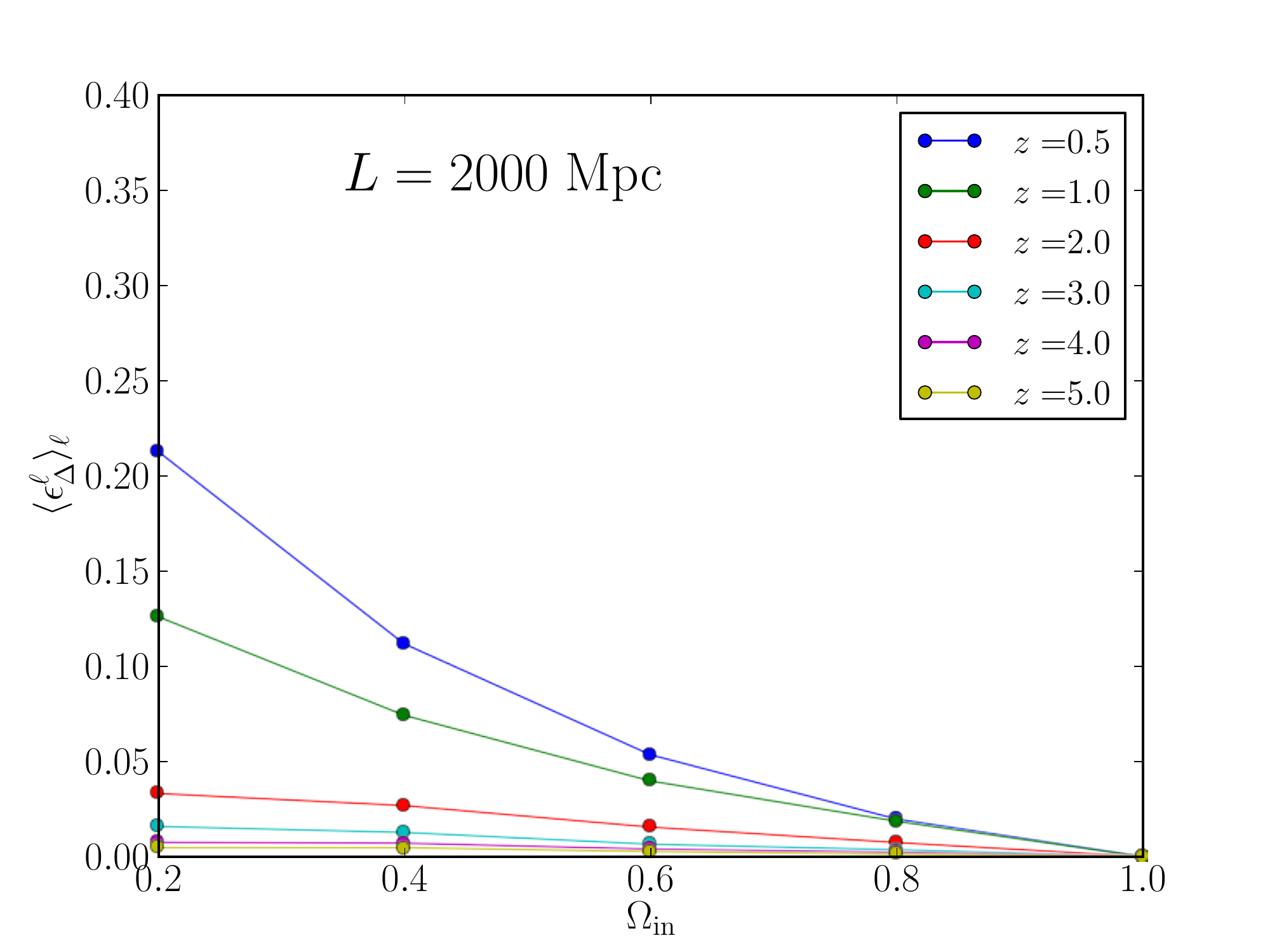} } 
 \subfigure[ $\Delta$ ] { \includegraphics[width=0.48\hsize]{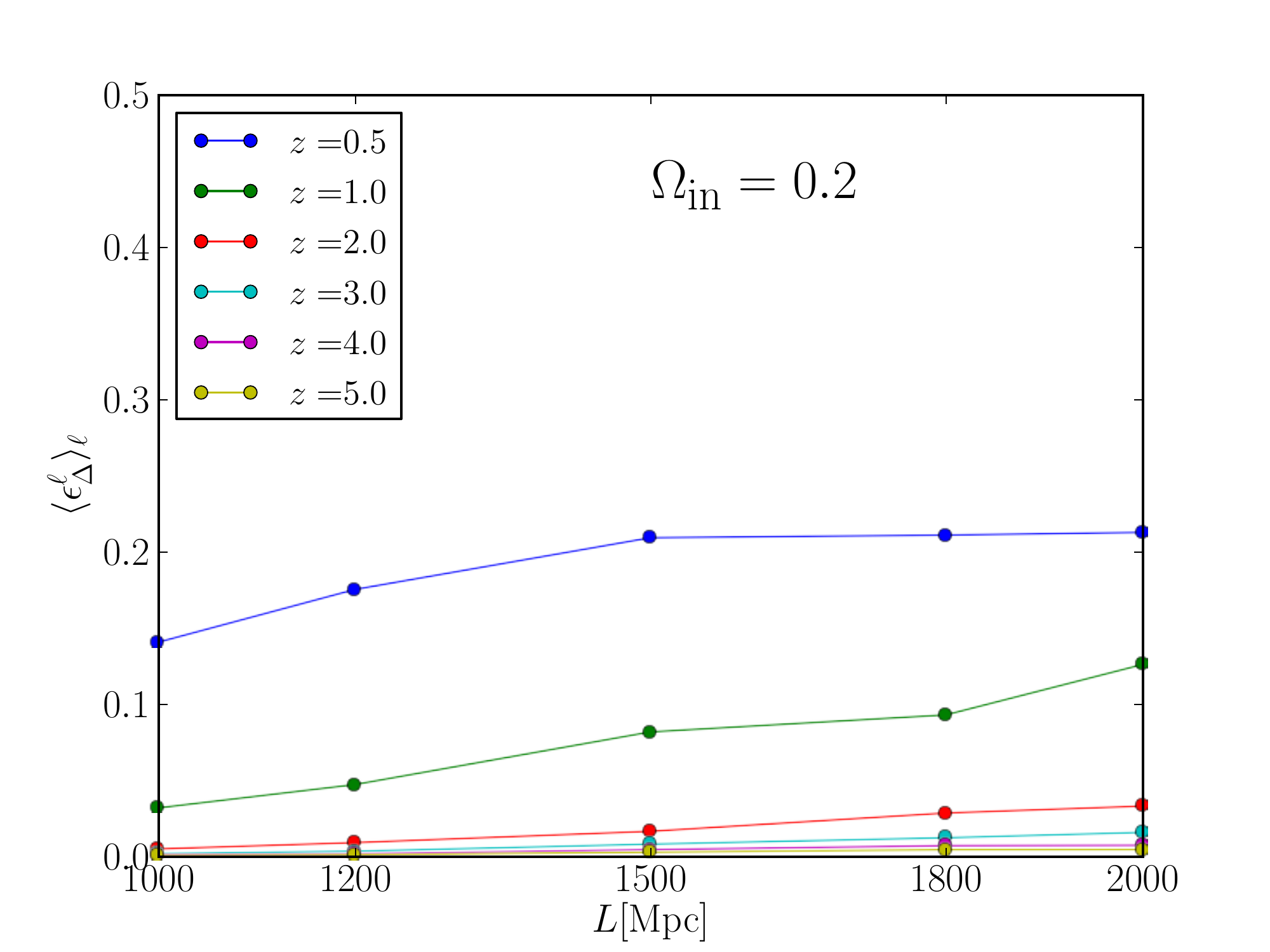} } 
 \caption{ Averaged coupling strength $\langle \epsilon^\ell \rangle$ as function of void depth and size for decreasing redshifts: The coupling strength according to the definition of Eq. (\ref{ltb:coupling:3}) is averaged over all angular scales and then plotted for the two different scenarios considered. (a) and (c) show the averaged coupling strengths as function of void depth $\Omega_\mathrm{in}$ at fixed void size. The behaviour as function of void size at fixed void depth is shown in (b) and (d). Note that due to the discretization of the initial Bardeen potential we are not yet able to study the coupling at small voids with $L < 1 \mathrm{Gpc}$ and small redshifts $z<0.5$.  As already observed in Figs. (\ref{ltb:coupling:fig:1}) and (\ref{ltb:coupling:fig:2}), coupling increases with void depth and also with decreasing redshift reaching averaged values of $27\%$ in case of $\varphi$ and $22\%$ in case of $\Delta$. As the void forms non-linearly at small redshifts, we can expect significant coupling for $z<1$. In case of a fixed void depth and varied void size, we see an increase in the coupling strength with void size. This is caused by the fact that large voids still have significant depth at $z<1$ whereas at larger redshifts we essentially probe the coupling in the asymptotic FLRW regime. }
 \label{ltb:coupling:fig:3}
 \end{figure}

 Fig. (\ref{ltb:coupling:fig:3}) show the coupling strength averaged over all angular scales as function of void depth and void size according to the two cases considered here. The results are shown for several redshifts on the LTB backward lightcone such that different stages of the void evolution can be traced. In case of varying void depth, we see significant couplings of $25\%$ for deep voids of $80 \%$ density contrast which is decreasing

 \FloatBarrier 
 
%***************************************************************************************************************************
\section{Discussion} 
\label{ltb:discussion}
%***************************************************************************************************************************

Perturbation theory in inhomogeneous backgrounds is inherently more complicated than in standard FLRW models. Not only the coupling of the gauge-invariant perturbations (necessarily described by partial instead of ordinary differential equations), but also their physical interpretation are very challenging. Nonetheless, it is necessary to perform this analysis in order to add further constraints on LTB models and, eventually, to rule them out on a solid ground. The current work does not yet address the physical interpretation of these results and therefore does not enable a direct comparison to FLRW models. Carrying out an intermediate step, we analyse the coupling strength of the gauge-invariants in a cosmological environment starting from a dark matter power spectrum in the FLRW limit. As we are able to distinguish scalar, vector and tensor perturbations on initial homogeneous backgrounds, we started from an initially scalar gravitational potential $\Psi$. By comparing perturbed FLRW metrics and perturbed LTB metrics in the FLRW limit in a general gauge, the FLRW Bardeen potential $\Psi$ can be related to the LTB gauge-invariants (see \cite{clarkson_perturbation_2009}). It turns out that only $\varphi = -2\Psi$ is remaining which yields a simple initial configuration for our setup. We can then study the coupling effect on the evolution of this generalized gravitational potential and the corresponding generalized density contrast $\Delta$. For particularly deep and large voids (as needed to recover the distance redshift relation of type Ia SNe and measurements of the local Hubble rate (\cite{february_rendering_2010, celerier_we_2000, redlich_probing_2014})), we find an averaged coupling strength $26\%$ for the generalized gravitational potential and $22\%$ for the generalised density contrast at redshifts $z<1$. These results are in agreement with previous considerations of February et al. (2014) (\cite{february_evolution_2014}) as they find maximum deviations of the amplitudes of $\varphi$ of around $30\%$ well within deep voids at late times. Analytical treatments in the framework of second order FLRW perturbations (see \cite{nishikawa_comparison_2014}) predict a non-negligible coupling as well. \\

In fact, considering second order perturbations of a background EdS model in RW gauge yields very similar evolution equations for the corresponding second order quantities. The void itself is modelled by isotropic first-order perturbations and general second order perturbations are placed ``on top" of these. However, a full analysis of this formalism with cosmological initial conditions has not yet been performed in RW gauge and therefore only general arguments about the growing and decaying modes of second order perturbations are known (see \cite{nishikawa_comparison_2014} for details). Although one can argue that, at second order, the influence of non-scalar perturbations on the scalar potential or density contrast are present and even growing in time, this does not allow any direct quantitative comparison with the exact treatment of linear perturbations in voids so far. Nonetheless, with the current numerical scheme at hand, such comparisons can, in principle, be made and the accuracy of the second order FLRW approach can directly be tested.\\

So far, coupling effects of perturbation variables or even linear structure growth in LTB models have been neglected as corrections due to the background shear have been estimated to be of percent level. In fact, we found percent level corrections for voids that are not significantly deep, but those models do not account for a sufficiently high local Hubble rate to describe the observed local universe without an exotic fluid assumption. However, for deep voids, coupling is indeed prominent and affects the evolution of the gauge invariants in a significant way which should be taken into account when studying observational effects on linear structure formation. \\

Although a careful analysis and comparison to FLRW models is still left to be done, coupling in the generalized metric potential $\varphi$ would yield corrections in weak gravitational lensing measurements and the integrated Sachs Wolfe effect that probe the metric perturbations on different angular scales. Previous studies of weak gravitational lensing in perturbed LTB spacetimes have been performed by Dunsby et al (2010) (\cite{dunsby_how_2010}) using the approach of \cite{zibin_scalar_2008}). In this framework, a model comparison is possible under the assumption of a negligible magnetic part of the Weyl tensor. Our results suggest that weak gravitational lensing based on the fully coupled perturbation equations should be investigated to assess possible corrections. In case of the integrated Sachs Wolfe effect, there is, to our knowledge, no direct observational study in perturbed LTB spacetimes. However, Tomita (2010) (see \cite{tomita_gauge-invariant_2009}) worked out a gauge-invariant treatment in LTB spacetimes based on the formalism of \cite{gerlach_relativistic_1978}. \\   

Corrections in the generalised density contrast $\Delta$ are expected to be seen in the two-point correlation function of the galaxy distibution and to affect the BAO scales. This could be relevant for analysing typical galaxy surveys like BOSS-SDSS-III (\cite{dawson_baryon_2013}) or WhiggleZ (\cite{blake_wigglez_2008}) in the context of large void models. The BAO have been used as an additional probe to constrain properties of void models observationally (see \cite{garcia-bellido_radial_2009, biswas_testing_2010, zumalacarregui_tension_2012, zibin_can_2008}) by taking the geometric distortion of the sound horizon due to the background shear into account, but not the effect of linear structure formation on inhomogeneous backgrounds. February et al. (2013) (\cite{february_galaxy_2013}) modeled this effect using the uncoupled evolution described by Eqs. (\ref{ltb:coupling:1}) and (\ref{ltb:coupling:2}) with the assumption that the background shear is only of percent level and therefore causes negligible coupling. By direct comparison to the geometric approximation, the authors found that the influence of the structure formation model is, in fact, subdominant, but nonetheless at percent level (see Figs. (7) and (8) in \cite{february_galaxy_2013}) such that future large volume surveys like Euclid (\cite{scaramella_euclid_2015}) and SKA (\cite{schwarz_testing_2015}) might be sensitive to it. Corrections to BAO scales due to a full treatment of perturbations in LTB models are expected to be of similar size and might therefore also be relevant for constraining void models with future surveys. Regarding our findings that coupling has a significant effect on the generalized density contrast $\Delta$ for deep voids, the effect on BAO measurements is worth to be investigated. However, any statements about observable predictions have to be considered with great caution in this context, as the notion of a physical density contrast in fully perturbed LTB models is poorly understood and a rigorous analysis is still left to be done.\\

Since we have taken the complete master equations and constraint equations into account, we are not straightforwardly able to compute transfer functions for the theoretical angular power spectra obtained from an initial potential power spectrum (as done in \cite{february_galaxy_2013} for the uncoupled case). We therefore have to draw a finite 3d realisation of a Gaussian random field, evolve it forward in time, and compute estimates of the angular power spectra. However, this approach has its limitations. We are resolution limited at small redshifts and high $\ell$-modes (see again Table (\ref{ltb:powerspectra:tab:1})) leading to reasonable results only at $z \ge 0.5$. Within our current technique of sampling initial conditions, improvement can only be achieved with higher resolutions in the initial 3d cube that becomes very memory-consuming. In fact, this issue needs to be resolved for the future, since we are now technically forced to crop the most interesting region of small redshifts and small voids. As the void grows non-linearly at late times causing also depth and slope of the density profile to grow considerably, we should see the strongest coupling effects in this spacetime region. On the contrary, spacetime anisotropy decreases in the vicinity of the void centre. This poses the question whether low shear regions in the void can approximately be described as open FLRW models. If negligible coupling strength is found close to the void centre, the validity and accuracy of this approximation can directly be investigated. However, the present technique for sampling initial conditions for the Bardeen potentials results in sparsely resolved spherical harmonic coefficients in the central region, which prevents us from making any reliable statement here. We are in the progress of developing an advanced sampling technique that allows us to solve this resolution problem. We will address this issue in a subsequent paper and analyse the low shear limit in detail. A second limitation is, of course, the resulting statistical error due to the finite realisation of the initial sample. This prevents us from giving reliable statements about coupling strengths at small $\ell$-modes similar to the cosmic variance limit in the CMB analysis. However, the statistical error can be reduced by considering several realisations in a row which is computationally very costly and the corresponding reduction would only scale with the square root of the number of trials. \\

The master variables $\varphi$, $\varsigma$, and $\chi$ are constructed to be gauge-invariant quantities and therefore contain only physical (in principle observable) degrees of freedom. Nonetheless, reducing them to ``usable" observable constraints will be necessary for direct comparison to homogeneous models and for tests of the Copernican principle. There are several promising approaches:

\begin{itemize}
 \item Clarkson et al. (2009) (\cite{clarkson_perturbation_2009}) constructed perturbation variables that reduce to pure Scalar-Vector-Tensor variables in FLRW limit. Using these variables, we might be able to directly compare large void models with best fit $\Lambda$CDM models and study the effects of coupling strengths.
 \item  A second approach would be looking at conserved gauge-invariant quantities in LTB spacetimes that can be compared in initial FLRW states and in the LTB final state. Leithes \& Malik (2014) (see \cite{leithes_conserved_2014}) identified the spatial metric trace perturbation $\xi_\mathrm{SMTP}$ to be conserved over spacetime evolution in spherically symmetric dust spacetimes.
 \item A third possibility would be weak gravitational lensing and the integrated Sachs Wolfe effect. By tracing null geodesics in perturbed LTB spacetimes, we account for combined effects of all gauge-invariant quantities summarized in first order Ricci and Weyl focusing terms in the optical tidal matrix (see \cite{bartelmann_topical_2010, clarkson_misinterpreting_2012} for promising approaches that can be adapted to perturbed LTB spacetimes). 
\end{itemize}

 Following one on these possible approaches, we would be able to add additional constraints to the void density profile from structure formation which would be an extension of recent work by Redlich et al. (2014) (\cite{redlich_probing_2014}). However, it can also help to systematically rule out local void models for the description of the late-time universe. Combined multi-probe analyses of large void models with homogeneous big bang (\cite{moss_precision_2011, zumalacarregui_tension_2012, redlich_probing_2014}) showed strong tension with observations as even very flexible void profiles are not able to fit local (SNe, local $H_0$) and global measurements (CMB) of the Hubble rate simultaneously (see \cite{redlich_probing_2014} and references therein). \\

 In addition, deviations from isotropy can directly be probed using the kinetic Sunyaev-Zel'dovich (kSZ) effect. The basic idea is to use rescattering of CMB photons by hot electrons in galaxy clusters to access information from the interior of the observer's backward lightcone. As the corresponding galaxy cluster is placed at considerable radial distance from the void centre, it is exposed to a large spacetime anisotropy and therefore should see an anisotropic CMB signal (see \cite{bull_kinematic_2012, garcia-bellido_looking_2008} for details). A related approach is the so-called linear kSZ effect that takes scattering of CMB photons at all structures (linear density fluctuations) in the LTB patch into account. This effect allows to estimate corrections to the CMB angular power spectrum (see \cite{zhang_confirmation_2011} for the original and \cite{zibin_linear_2011} for a fully relativistic treatment). In fact, the prediced linear kSZ power and corresponding corrections to the CMB power spectrum are much larger than actually observed. Although considerable effort has been made to approximate or circumvent linear structure formation in void models to describe this effect, a rigorous treatment requires proper modelling of the evolution of linear perturbations on the LTB background. In fact, deriving the angular power spectrum of the linear kSZ effect will be possible without scale dependent approximations for the growth factor of perturbations (as done in \cite{zibin_linear_2011}) since spherical harmonic coefficients can be propagated in time directly. We would like to stress that any statement about the influence of evolution effects on inhomogeneous backgrounds goes beyond the scope of this work, as no direct expression for the scalar density contrast is available so far in gauge invariant LTB perturbation theory. Nonetheless, it is worth being investigated and should be addressed in the future. \\
 
 Following the recent trends in the literature, we cannot expect linear structure formation to alleviate the problems, but rather to strengthen arguments against  void models. One can therefore also think of a second application in testing the Copernican Principle with LTB models in combination with a cosmological constant $\Lambda$. As shown in \cite{february_evolution_2014}, Eqs. (\ref{ltb:perturbation:4})-(\ref{ltb:perturbation:10}) can be augmented by a cosmological constant and are therefore also valid in, so-called, $\Lambda$LTB models. Marra \& Pääkonen (2010) (\cite{marra_observational_2010}), Valkenburg et al. (2012) (\cite{valkenburg_testing_2014}), and  Redlich et al. (2014) \cite{redlich_probing_2014} performed detailed analyses of this kind of models to constrain local deviations in the radial density profile using combinations of several observational probes. In this context, a full treatment of linear structure growth in $\Lambda$LTB models would be a valuable extension of current methods to test the Copernican principle, although we would expect coupling effects to be less important due to smaller deviations of the density profile from the homogeneous $\Lambda$CDM model.

%***************************************************************************************************************************
\section{Conclusion} 
\label{ltb:conclusion}
%***************************************************************************************************************************

We have investigated the evolution of polar perturbations in LTB spacetimes in a cosmological setup starting from realistically sampled scalar initial conditions. In extension to previous numerical studies of \cite{february_evolution_2014}, we are now able to characterise basic properties of the spacetime evolution and the coupling strength in a statistical way. Though limited by resolution effects and statistical errors, we are confident that these issues can be resolved with higher computational effort. We see statistically a coupling strength (as defined in Eq. (\ref{ltb:coupling:3})) of nearly $30\%$ for large and deep voids needed to recover the distance redshift relation of SNe. For the moment, only the polar branch of the perturbations has been considered, since interesting physics mainly happens in this branch that contains the generalised Bardeen potential $\varphi$ and density contrast $\Delta$. For simplicity, we have taken only initial scalar perturbations into account, but we plan to generalise this by starting with initial scalar and tensor perturbations. The background void model can easily be interchanged by, for example, universal void profiles found in simulations (\cite{ricciardelli_universality_2014}) or observationally constrained best-fit void models (\cite{redlich_probing_2014}). \\

However, for confronting results with observational data, a proper conversion of the LTB gauge-invariants to observable quantities will be needed which is still left to be done. The overall goal will be to approach a concrete comparison of the evolution of perturbations in FLRW models and LTB models in order to test the Copernican principle reliably by using information on structure formation. For future work, we hope to successfully follow one of the approaches outlined in the discussion to compare structure formation in the best fit $\Lambda$CDM model and the best fit void model of \cite{redlich_probing_2014}.     

\appendix

%***************************************************************************************************************************
\section{Appendix: DUNE and local operators} 
\label{ltb:dune}
%***************************************************************************************************************************

Originating from developments in structural mechanics in the 1950s, finite element methods became a well-established technique to solve partial differential equations on very flexible and even unstructured grids. The basic idea consists of writing the model problem (PDE and boundary conditions) into weak formulation and converting it to a variational problem. The variational problem can then be solved by discretizing the domain of interest into finite elements and approximating the solution by suitable basis functions (polynomials in particular) on each element with coefficients to be determined. This typically leads to large, but sparsely populated, linear equation systems for the expansion coefficients. These methods are based on a solid mathematical background using techniques developed in functional analysis like weak derivatives and corresponding Sobolev spaces. \\ 

The \textit{Distributed Unified Numerics Environment} (DUNE) is a template-based, multi-purpose C++-library for numerical solution of partial differential equations on arbitrary grids using finite element methods. The basic structure is divided into modules performing different tasks like grid setup (\texttt{dune-grid}), iterative solvers for equation systems (\texttt{dune-istl}) or proper PDE setup (\texttt{dune-pdelab}) (see \cite{bastian_dune_2011} for details). It provides preimplemented basis polynomials, grid-based spatial solvers and time integrators designed to be very flexible and applicable to a wide range of PDE problems in basically arbitrary dimensions. For time-dependent problems, DUNE uses the \textit{method of lines}, solving a spatial discretization problem in each timestep and propagating the resulting coefficients in time. For our purposes, we have a one dimensional spatial problem to be solved with finite elements that is integrated in time using an Alexander S-stable method (see \cite{alexander_diagonally_1977}). For proper numerical treatment, the PDE system has to be formulated in, so called, \textit{weak residual formulation} which we are going to outline in the following\footnote{We basically adapt the derivations given in the documentation of the \texttt{dune-pdelab-howto} module edited by the DUNE project team. (see \textit{http://www.dune-project.org/pdelab/pdelab-howto-2.0.0.pdf)}}:\\   

We consider Eqs. (\ref{ltb:perturbation:4}) - (\ref{ltb:perturbation:6}) in the domain $\Omega = (0, r\st)$. In order to approach the problem numerically, we introduce auxiliary variables $\tilde{\varphi}$ and $\tilde{\chi}$ to convert the system into 5 first order equations in time. For the spatial finite element problem, we then rewrite it in \textit{weak formulation} by multiplying with an ansatz function $v$ and integrating over the domain $\Omega$. The ansatz function is chosen to vanish at the boundary $\partial \Omega$ such that terms with second spatial derivatives are replaced by first spatial derivatives using partial integration.\\ 

For a test function  $v \in \left(H^1(\Omega)\right)^5$ (5 dimensional first-order Sobolev space, called test space), we have:

 \begin{align} 
  \label{ltb:dune:1}
 \I { \dot{\chi} v_1 \d r} &=  \I { \tilde{\chi} v_1 \d r} \\
 \label{ltb:dune:2}
 \begin{split}
 \I { \dot{\tilde{\chi} } v_2 \d r} &= - \I { \chi' \left(\frac{v_2}{Z^2} \right)' \d r} + \I { \left[  - \frac{C}{Z^2} \chi' - 3 H\p \tilde{\chi} + \left[A - \frac{(\ell-1) (\ell+2)}{r^2 a\o^2}\right] \chi \right. }\\
                                & \quad { \left. + \frac{2\sigma}{Z} \varsigma' + \frac{2}{Z} \left[H\p - 2 H\o\right]' \varsigma - 4 \sigma \tilde{\varphi} + A \varphi \right] v_2 \d r}
 \end{split}
 \end{align}

\begin{align}
 \label{ltb:dune:3}
 \I { \dot{\varphi} v_3 \d r} &=  \I { \tilde{\varphi} v_3 \d r} \\
 \label{ltb:dune:4}
  \begin{split}
 \I { \dot{\tilde{\varphi} } v_4 \d r} &=  \I { \left[ - 4 H\o \tilde{\varphi} + \frac{2 \kappa}{a\o^2} \varphi  - H\o \tilde{\chi} + Z^{-2} \frac{a\p}{r a\o} \chi' -  \left[ \frac{1 - 2\kappa r^2}{r^2 a\o^2} - \frac{\ell (\ell +1)}{2 r^2 a\o^2}\right] \chi  \right. }\\
                                   & \quad { \left. + \frac{2}{Z} \frac{a\p}{r a\o} \sigma \varsigma  \right] v_4 \d r}
 \end{split}
 \end{align}

 \begin{equation}
   \I {\dot{\varsigma} v_5 \d r} = \I { -2 H\p \varsigma v_5 \d r}   -  \I {  \frac{\chi'}{Z} v_5  \d r }
  \label{ltb:dune:5}
  \end{equation} \\
 
 The index $i \in \{1,2,3,4,5\}$ of the test function $v$ expresses the components of the tuple $\left(v_1, v_2, v_3, v_4, v_5\right) \in \left(H^1(\Omega)\right)^5$. \\
 
 We shorten notation by writing all terms in the LHS as temporal residuals $m_i$ and RHS terms as spatial residuals $r_i$:
 
\begin{align}
\label{ltb:dune:6}
\frac{d}{\d t} m_1 (\chi, v_1, t) &= r_1(\tilde{\chi}, v_1, t) \\
\label{ltb:dune:7}
\frac{d}{\d t} m_2 (\tilde{\chi}, v_2, t) &= r_2(\chi, \tilde{\chi}, \varphi, \tilde{\varphi}, \varsigma,v_2, t)
\end{align}

\begin{align}
\label{ltb:dune:8}
\frac{d}{\d t} m_3 (\varphi, v_3, t) &= r_3(\tilde{\varphi}, v_3, t) \\
\label{ltb:dune:9}
\frac{d}{\d t} m_4 (\tilde{\varphi}, v_4, t) &= r_4(\chi, \tilde{\chi}, \varphi, \tilde{\varphi}, \varsigma, v_4, t)
\end{align}
 
\begin{align}
\frac{d}{\d t} m_5 (\tilde{\varsigma}, v_5, t) &= r_5(\chi, \varsigma, v_5, t)
\label{ltb:dune:10}
\end{align}
 
We can then define the full temporal and spatial residuals by taking the sum of each

\begin{equation}
 m = \sum_{i=1}^5 m_i , \qquad r = \sum_{i=1}^5 r_i  
\label{ltb:dune:11}
\end{equation}

The problem of solving the polar master equations can then be reformulated in terms of a residual (variational) problem \\

\par
\begingroup
\leftskip=1cm 
\noindent
\textit{Find a solution $u = (\chi, \tilde{\chi}, \varphi, \tilde{\varphi}, \varsigma)^T \in \left(H^1(\Omega)\right)^5$ (trial space) such that } 

\begin{equation*}
 \frac{d}{\d t} m(u, v, t) - r(u,v,t) = 0
\end{equation*}

\textit{holds for each test function  $v \in \left(H^1(\Omega)\right)^5$ and for each time $t \in \left[ t_ {min}, t_{max}\right]$.} \\

\par
\endgroup

In order to approach this problem numerically, we discretize the spatial domain of interest $\Omega$ and approximate the solution $u$ and test function $v$ by polynomials of given degree on each spatial element. Therefore, we define general conforming finite elements  $\Omega_e$ of the full domain $\Omega$ with $ e \in E^0_h = \{ e_0, \ldots, e_{N^0_h - 1}  \}$. The corresponding conformal finite-element-space is given by

\begin{equation*}
 U^k_h = \left\{ u \in \mathcal{C}^0(\bar{\Omega}) \left| \right. u_{\left|\Omega_e \right. }  \in P_k(\Omega_e) \ \forall e \in E^0_h  \right\}
\end{equation*}

By construction, it is the function space of continuous, element-wise polynomial functions of degree $k$. Since this function space is the same for each variable $\chi_h, \tilde{\chi}_h, \varphi_h, \tilde{\varphi}_h \  \text{and} \ \varsigma_h$, we have an overall 5-dimensional trial- and test space $\left(U_h^k\right)^5$. We assume test and trial space to be equal (Galerkin approach).\\  

Since \texttt{dune-pdelab} requires the local residual contributions of each element as input, the local contributions of Eqs. (\ref{ltb:dune:1}) - (\ref{ltb:dune:5}) on each finite element of the given residuals need to be determined. These contributions are called \textit{local operators} and will be computed for our model problem in the following.  For practical reasons, we define a reference finite element $\hat{\Omega}_e$ for all computations and an element transformation $\mu_e$ that maps the result to the actual element considered. We then fix a local polynomial basis $\{ \pvec \}$ on the reference element. Thus, each function of the conforming finite-element-subspace $U^k_h$ can be expanded into this local basis 

\begin{equation*}
 u^{(i)}_h = \sum_{e \in E^0_h} \sum_{l=0}^{n(e)-1}{ {\mathbf{u^{(i)}}_{g(e,l)}} \hpel{(i)} \chi_e(x) } \in U^k_h  
 \end{equation*}

with the following expressions used:

\begin{itemize}
 \item $n(e)$: number of polynomial basis functions on reference element $\hat{\Omega}_e$ 
 \item $\hat{\Omega}_e$: reference element of finite element $e$ 
 \item $\mu_e: \hat{\Omega}_e \longrightarrow \Omega_e$ : element transformation
 \item $g: E^0_h \times \mathbb{N}_0  \longrightarrow \mathcal{I}_{U^k_h} = \{0, \ldots, N_{U^k_h-1} \}$: local to global index map
 \item $\mathbf{u} \in \mathbb{R}^{\mathcal{I}_{U^k_h}}$: global coefficient vector
 \item $\chi_e$: Characteristic function of finite element $e$ ($\chi_e \equiv 1$ for our purposes)
\end{itemize}

Thus, the components of the full trial and test space read

\begin{align*}
  \chi_h &= \sum_{e \in E^0_h} \sum_{l=0}^{n(e)-1}{\boldsymbol{\chi}_{g(e,l)} \hpel{(1)} \chi_e(x) }, \\
  \tilde{\chi}_h &= \sum_{e \in E^0_h} \sum_{l=0}^{n(e)-1}{ {\boldsymbol{\tilde{\chi}}_{g(e,l)}} \hpel{(2)} \chi_e(x) }, \\
  \varphi_h &= \sum_{e \in E^0_h} \sum_{l=0}^{n(e)-1}{ {\boldsymbol{\varphi}_{g(e,l)}} \hpel{(3)} \chi_e(x) }, \\
  \tilde{\varphi}_h &= \sum_{e \in E^0_h} \sum_{l=0}^{n(e)-1}{ {\boldsymbol{\tilde{\varphi}}_{g(e,l)}} \hpel{(4)} \chi_e(x) }, \\
  \varsigma_h &= \sum_{e \in E^0_h} \sum_{l=0}^{n(e)-1}{ {\boldsymbol{\varsigma}_{g(e,l)}} \hpel{(5)} \chi_e(x) }, \\
  \text{and} \\
  v^{(i)}_h &= \sum_{e \in E^0_h} \sum_{l=0}^{n(e)-1}{ {\boldsymbol{v^{(i)}}_{g(e,l)}} \hpel{(i)} \chi_e(x) }. 
\end{align*}

%(We will drop the index $i$ for simplicity and refer to trial space components in the following.) \\

We construct a global basis $\{\phi_{j}\}$ from the local reference basis $\{\hat{\phi}_{e,i}\}$ by defining 

\begin{equation*}
 \Phi_{U^k_h} = \left\{ \phi_j(x) = \sum_{(e,l):g(e,l)=j} \hpel{} \chi_e(x) \ : \ j \in  \mathcal{I}_{U^k_h} \right\}
\end{equation*}

such that $u_h \in U^k_h$ can be expanded into this global basis 

\begin{equation*}
 u_h = \sum_{j \in \mathcal{I}_{U^k_h} } {u_j \phi_j} = \mathrm{FE}_{\Phi_{U^k_h}} (\mathbf{u})
\end{equation*}

with $\mathrm{FE}_{\Phi_{U^k_h}}: \ \mathbf{U} \subset \mathbb{R}^{\mathcal{I}_{U^k_h}} \longrightarrow U^k_h$ being the Finite-Element-Isomorphism. \\
 
 Let us consider again the residual formulation of the problem
 
 \begin{align}
  \label{ltb:dune:12}
  &\frac{d}{\d t} m(u,v,t) - r(u,v,t) = 0 \\
  \Rightarrow & \sum_{i=0}^{5} \left[ \frac{d}{\d t} m_i(u_i, v_i,t) - r_i(u_i, v_i, t) \right] = 0
 \end{align}

 We take $v_i = \phi^{(i)}_j$ with $j \in \mathcal{I}_{U^k_h}$ which allows to write the residual as component of residual vectors $\mathcal{M}^{(i)}$ and $\mathcal{R}^{(i)}$. 
 
 \begin{align*}
  m_i(u^{(i)}_h, \phi^{(i)}_j) &= m_i(\mathrm{FE}(\mathbf{u^{(i)}}), \phi^{(i)}_j) = \mathcal{M}^{(i)}(\mathbf{u^{(i)}})_j \\
  r_i(u^{(i)}_h, \phi^{(i)}_j) &= r_i(\mathrm{FE}(\mathbf{u^{(i)}}), \phi^{(i)}_j) = \mathcal{R}^{(i)}(\mathbf{u^{(i)}})_j 
 \end{align*}

We want to restrict ourselves to the local degrees of freedom. Therefore, the residual contributions of each element need to be separated and formulated in terms of local operators. Therefore, we define the subset $\mathbf{U_e} \subset \mathbf{U}$ of local degrees of freedom (coefficients of the local expansion) and the (linear) reduction map

\begin{equation*}
  R_e: \mathbf{U} \longrightarrow \mathbf{U}_e \quad \text{with} \ R_e(\mathbf{u})_l = \sum_{j} (\mathbf{R_e})_{lj} (\mathbf{u})_j = \mathbf{u}_{g(e,l)}.  
\end{equation*}

The local operators can then be defined as maps 

\begin{equation*}
  \avol : \mathbf{U}_e \longrightarrow \mathbf{U}_e   
 \end{equation*}

 such that they contain the local weak formulation of the problem.\\
 
 In this way, the full residuals can be expressed in terms of a sum over the local residual contributions of each finite element:

\begin{align*}
 \mathcal{R}^{(i)}(\mathbf{u^{(i)}})_j &= \sum_{e \in E^0_h} R_e^{-1} \left[ \avolspat{i}{R_e(\mathbf{u^{(i)}})} \right] \\
                                     &= \sum_{(e,l):g(e,l)=j} (R_e)_{lj} \left[ \avolspat{i}{R_e(\mathbf{u^{(i)}})} \right]_l \\
                                     &= \sum_{e \in E^0_h} \mathbf{R}_e^{T} \left[ \avolspat{i}{R_e(\mathbf{u^{(i)}})} \right] \\ \\
\mathcal{M}^{(i)}(\mathbf{u^{(i)}})_j &= \sum_{e \in E^0_h} R_e^{-1} \left[ \avoltemp{i}{R_e(\mathbf{u^{(i)}})} \right] \\
                                     &= \sum_{(e,l):g(e,l)=j} (R_e)_{lj} \left[ \avoltemp{i}{R_e(\mathbf{u^{(i)}})} \right]_l \\
                                     &= \sum_{e \in E^0_h} \mathbf{R}_e^{T} \left[ \avoltemp{i}{R_e(\mathbf{u^{(i)}})} \right] 
\end{align*}

By direct comparison to Eqs. (\ref{ltb:dune:1}) - (\ref{ltb:dune:5}), we obtain the following local operator expressions as integrals over reference elements in terms of the local reference basis $\hat{\phi}^{(i)}$: \\
 
 \textit{temporal part:}
 
 \begin{align}
\label{ltb:dune:13}  
 \avoltemp{1}{R_e(\boldsymbol{\chi})}_m &= \int_{\hat{\Omega}_e} \sum_{l=0}^{n(e)-1} \left[R_e(\boldsymbol{\chi})\right]_l \ \hpelh{(1)} \ \hpemh{(1)} \ \det\left[{\nabla{\mu_e}}\right] \d \hat{x} \\
\label{ltb:dune:14}   
 \avoltemp{2}{R_e(\boldsymbol{\tilde{\chi}})}_m &= \int_{\hat{\Omega}_e} \sum_{l=0}^{n(e)-1} \left[R_e(\boldsymbol{\tilde{\chi}})\right]_l \ \hpelh{(2)} \ \hpemh{(2)} \ \det\left[{\nabla{\mu_e}}\right] \d \hat{x} \\
\label{ltb:dune:15}  
 \avoltemp{3}{R_e(\boldsymbol{\varphi})}_m &= \int_{\hat{\Omega}_e} \sum_{l=0}^{n(e)-1} \left[R_e(\boldsymbol{\varphi})\right]_l \ \hpelh{(3)} \ \hpemh{(3)} \ \det\left[{\nabla{\mu_e}}\right] \d \hat{x} \\
\label{ltb:dune:16}   
 \avoltemp{4}{R_e(\boldsymbol{\tilde{\varphi}})}_m &= \int_{\hat{\Omega}_e} \sum_{l=0}^{n(e)-1} \left[R_e(\boldsymbol{\tilde{\varphi}})\right]_l \ \hpelh{(4)} \ \hpemh{(4)} \ \det\left[{\nabla{\mu_e}}\right] \d \hat{x} \\
\label{ltb:dune:17}  
 \avoltemp{5}{R_e(\boldsymbol{\varsigma})}_m &= \int_{\hat{\Omega}_e} \sum_{l=0}^{n(e)-1} \left[R_e(\boldsymbol{\varsigma})\right]_l \ \hpelh{(5)} \ \hpemh{(5)} \ \det\left[{\nabla{\mu_e}}\right] \d \hat{x} 
  \end{align}\\

 \textit{spatial part:}
 \begin{align}
\label{ltb:dune:18}
 \begin{split}
  &\avolspat{1}{R_e(\boldsymbol{\tilde{\chi}})}_m \\
  &= \int_{\hat{\Omega}_e} \sum_{l=0}^{n(e)-1} \left[R_e(\boldsymbol{\tilde{\chi}})\right]_l \ \hpelh{(2)} \ \hpemh{(1)} \ \det\left[{\nabla{\mu_e}}\right] \d \hat{x}  
  \end{split} \\ \nonumber \\ 
 \label{ltb:dune:19}
   \begin{split}
     &\avolspat{2}{R_e(\boldsymbol{\chi}), R_e(\boldsymbol{\tilde{\chi}}), R_e(\boldsymbol{\varphi}), R_e(\boldsymbol{\tilde{\varphi}}), R_e(\boldsymbol{\varsigma})}_m \\
     &=  -\int_{\hat{\Omega}_e} \sum_{l=0}^{n(e)-1} R_e(\boldsymbol{\chi})_l \ \partial_{\hat{x}}\hpelh{(1)} \ \partial_{\hat{x}} \left( \frac{\hpemh{(2)} }{Z^2} \right) \ \left( \partial_{\hat{x}}\mu_e \right)^{-2} \ \det\left[{\nabla{\mu_e}}\right] \d \hat{x}  \\
     &+  \int_{\hat{\Omega}_e} \sum_{l=0}^{n(e)-1} \left[ \left( -\frac{C}{Z^2} R_e(\boldsymbol{\chi})_l \ \partial_{\hat{x}} \hpelh{(1)} + \frac{2 \sigma}{Z} R_e(\boldsymbol{\varsigma})_l \ \partial_{\hat{x}}\hpelh{(5)} \right)  \left(\partial_{\hat{x}}\mu_e \right)^{-1}  \right. \\ 
     &-   3 H\p R_e(\boldsymbol{\tilde{\chi}})_l \ \hpelh{(2)} + \left(A - \frac{(\ell-1)(\ell+2)}{\mu_e(\hat{x})^2 a\o^2}\right) R_e(\boldsymbol{\chi})_l \ \hpelh{(1)} \\
     &+  \left.  \frac{2}{Z} \left[H\p - 2 H\o\right]' R_e(\boldsymbol{\varsigma})_l \ \hpelh{(5)} - 4 \sigma R_e(\boldsymbol{\tilde{\varphi}}) \ \hpelh{(4)} + A R_e(\boldsymbol{\varphi})_l \ \hpelh{(3)} \right] \hpemh{(2)} \\
     &\det\left[{\nabla{\mu_e}}\right] \d \hat{x} 
   \end{split}\\ \nonumber \\ 
  \label{ltb:dune:20}
   \begin{split}
  &\avolspat{3}{R_e(\boldsymbol{\tilde{\varphi}})}_m \\
  &= \int_{\hat{\Omega}_e} \sum_{l=0}^{n(e)-1} \left[R_e(\boldsymbol{\tilde{\varphi}})\right]_l \ \hpelh{(4)} \ \hpemh{(3)} \ \det\left[{\nabla{\mu_e}}\right] \d \hat{x}     
   \end{split} \\  \nonumber \\ 
   \label{ltb:dune:21}
   \begin{split}
     &\avolspat{4}{R_e(\boldsymbol{\chi}), R_e(\boldsymbol{\tilde{\chi}}), R_e(\boldsymbol{\varphi}), R_e(\boldsymbol{\tilde{\varphi}}), R_e(\boldsymbol{\varsigma})}_m \\
     &=  \int_{\hat{\Omega}_e} \sum_{l=0}^{n(e)-1} \left[  -4 H\o R_e(\boldsymbol{\tilde{\varphi}})_l \ \hpelh{(4)} + \frac{2 \kappa}{a\o^2} R_e(\boldsymbol{\varphi})_l \ \hpelh{(3)} - H\o R_e(\boldsymbol{\tilde{\chi}})_l \ \hpelh{(2)}  \right.\\
     &- \left. \left(  \frac{1-2\kappa \mu_e(\hat{x})^2}{\mu_e(\hat{x})^2 a\o^2} - \frac{\ell (\ell +1)}{2 \mu_e(\hat{x})^2 a\o^2} \right)  R_e(\boldsymbol{\chi})_l  \ \hpelh{(1)} + \frac{1}{Z^2} \frac{a\p}{\mu_e(\hat{x}) a\o} R_e(\boldsymbol{\chi})_l  \partial_{\hat{x}} \hpelh{(1)} \left(\partial_{\hat{x}} \mu_e \right)^{-1} \right. \\
     &\left. + \frac{2}{Z} \sigma \frac{a\p}{\mu_e(\hat{x}) a\o} R_e(\boldsymbol{\varsigma}) \hpelh{(5)}  \right] \hpemh {(4)} \ \det{\left[\nabla{\mu_e}\right]} \d \hat{x} 
   \end{split} \\ \nonumber \\
   \label{ltb:dune:22}
   \begin{split}
     &\avolspat{5}{R_e(\boldsymbol{\chi}), R_e(\boldsymbol{\tilde{\chi}}), R_e(\boldsymbol{\varphi}), R_e(\boldsymbol{\tilde{\varphi}}), R_e(\boldsymbol{\varsigma})}_m
     \\
     &= \int_{\hat{\Omega}_e} \sum_{l=0}^{n(e)-1} \left[ -2 H\p R_e(\boldsymbol{\varsigma})_l  \ \hpelh{(5)} - \frac{1}{Z} R_e(\boldsymbol{\chi})_l \ \partial_{\hat{x}} \hpelh{(1)}  \right] \ \hpemh{(5)} \\
     & \ \det{\left[\nabla{\mu_e}\right]} \d \hat{x}
   \end{split}
  \end{align}

 These operators can directly be passed to the \texttt{dune-pdelab}-framework. The local operator formulation of the fluid system (given by Eqs. (\ref{ltb:perturbation:8})-(\ref{ltb:perturbation:10})) can be done analogously. If the spatial discretization is completed and residual contributions of all finite elements are known, Eq. (\ref{ltb:dune:12}) can be integrated as large scale ODE problem in time. All results in this work are obtained by using basis polynomials for degree 2 in space combined with a third order time integrator.\footnote{ Since the solution can be expected to be sufficiently smooth, the approximation using polynomials of degree 2 lead to second order convergence with respect to the $H^1$-Norm and even third order convergence with respect to the $L^2$-Norm.} Initial and boundary conditions are fixed according to the methods outlined in Sects. \ref{ltb:numerics} and \ref{ltb:initial}.\\ 
 
 In the one dimensional case we want to consider, the general formulation is simplified considerably:
 
 \begin{itemize}
  \item The domain of interest is a 1d open interval $\Omega = \left(0, r\st \right)$. 
  \item The finite elements are subintervals $\Omega_e = \left( r_j, r_{j+1}\right)$ with $e=j, \ r_0 = 0, \ r_{N_{U^k_h}} = r\st$.  
  \item The reference element is the unit interval $(0,1)$ and the transformation map $\mu_e$ is given by
\begin{align*}
  \mu_e: \hat{\Omega}_e &\longrightarrow \Omega_e \\
		     \hat{x} &\longmapsto \left(r_{j+1} - r_{j}\right) \hat{x} + r_{j}
 \end{align*}

  Therefore, the Jacobian is just a constant $\partial_{\hat{x}} \mu_e = r_{j+1} - r_j = \det\left[{\nabla{\mu_e}}\right]$.
 
 \begin{comment}
 \item We use a linear conforming finite-element-space $U^1_h = \left\{ u \in \mathcal{C}^0(\bar{\Omega}) \left| \right. u_{\left|\Omega_e \right. }  \in P_1(\Omega_e) \ \forall e \in E^0_h  \right\}$ and linear global basis functions (Lagrange polynomials, "hat"-functions): 
  
    \begin{equation*}
   \phi_j(x) = \left\{ \begin{array}{ll}
                        \dfrac{x-r_{j-1}}{r_j-r_{j-1}} &, \ x \in \Omega_{e-1} \\ 
                        \\
                        \dfrac{r_{j+1}-x}{r_{j+1}-r_j } &, \   x \in \Omega_e
                       \end{array}
                        \right.
  \end{equation*}
  
  \item On $\Omega_e$ we define the local polynomial basis
  
    \begin{align*}
   \psi_1(x) &= \frac{r_{j} - x}{r_{j} - r_{j-1}} = \phi_j(x) , \\
   \psi_2(x) &= \frac{x - r_j}{r_{j+1} - r_j} = \phi_{j+1}(x)    
  \end{align*}
  
  \item On the reference element $\hat{\Omega}_e$ (there exists just one for all elements $e$), we have the local reference basis
  
   \begin{align*}
   \hat{\phi}^{(i)}_{e,0}(\hat{x}) &=  \hat{\psi}_1(\hat{x}) = 1-\hat{x} \\
   \hat{\phi}^{(i)}_{e,1}(\hat{x}) &= \hat{\psi}_2(\hat{x}) = \hat{x}
 \end{align*}
  
  Thus, $n(e)=2$ on each finite element and $g(e,l)$ has the same value just for two local indices $l$ (at fixed $e$). This also fixes the reduction map $R_e$.  
  
  \end{comment}
  
 \end{itemize}

\acknowledgments

We thank Peter Bastian for long discussions and his great helpfulness and support with the DUNE framework. SM wants to thank the DUNE project team for a very informative and well-organised DUNE/PDELab workshop on March 24-28, 2014. Furthermore, we want to thank Martin Reinecke for help and advice with the Healpix C++ implementation.  Part of this work was supported by the German
\emph{Deut\-sche For\-schungs\-ge\-mein\-schaft, DFG\/} project
number BA 1359 / 20-1. Most simulations required for this work were performed on the bwGRiD cluster (http://www.bw-grid.de), member of the German D-Grid initiative, funded by the Ministry for Education and Research (Bundesministerium für Bildung und Forschung) and the Ministry for Science, Research and Arts Baden-Württemberg (Ministerium für Wissenschaft, Forschung und Kunst Baden-Württemberg). We finally want to thank the anonymous referee for very detailed and constructive comments on the initial manuscript.

%*************************************************************************************************************************************
% References
%*************************************************************************************************************************************

\bibliographystyle{abbrv}
\bibliography{references/references}

\end{document}